\documentclass[a4paper,11pt]{article}
\usepackage{jcappub} 
\usepackage{lineno}
\usepackage{aas_macros}
\usepackage{multirow}

\arxivnumber{2411.12025} 
\title{\boldmath Characterization of DESI fiber assignment incompleteness effect on 2-point clustering and mitigation methods for DR1 analysis}

\usepackage{orcidlink}

\author[1,2]{{D.~Bianchi}\orcidlink{0000-0001-9712-0006},}
\author[3]{{M.~M.~S~Hanif}\orcidlink{0009-0006-2583-5006},}
\author[4,5]{{A.~Carnero Rosell}\orcidlink{0000-0003-3044-5150},}
\author[6]{{J.~Lasker}\orcidlink{0000-0003-2999-4873},}
\author[7,8,9]{{A.~J.~Ross}\orcidlink{0000-0002-7522-9083},}
\author[10]{{M.~Pinon}\orcidlink{0009-0009-3228-7126},}
\author[10]{{A.~de~Mattia}\orcidlink{0000-0003-0920-2947},}
\author[11,12]{{M.~White}\orcidlink{0000-0001-9912-5070},}
\author[13]{{S.~Ahlen}\orcidlink{0000-0001-6098-7247},}
\author[14]{{S.~Bailey}\orcidlink{0000-0003-4162-6619},}
\author[15]{{D.~Brooks},}
\author[10]{{E.~Burtin},}
\author[14]{{E.~Chaussidon}\orcidlink{0000-0001-8996-4874},}
\author[14]{{T.~Claybaugh},}
\author[16]{{S.~Cole}\orcidlink{0000-0002-5954-7903},}
\author[17]{{A.~de la Macorra}\orcidlink{0000-0002-1769-1640},}
\author[14,12]{{S.~Ferraro}\orcidlink{0000-0003-4992-7854},}
\author[18]{{A.~Font-Ribera}\orcidlink{0000-0002-3033-7312},}
\author[19,20]{{J.~E.~Forero-Romero}\orcidlink{0000-0002-2890-3725},}
\author[21,22,23]{{E.~Gaztañaga},}
\author[14]{{S.~Gontcho A Gontcho}\orcidlink{0000-0003-3142-233X},}
\author[24]{{G.~Gutierrez},}
\author[14]{{J.~Guy}\orcidlink{0000-0001-9822-6793},}
\author[25,26]{{C.~Hahn}\orcidlink{0000-0003-1197-0902},}
\author[7,27,9]{{K.~Honscheid}\orcidlink{0000-0002-6550-2023},}
\author[28]{{C.~Howlett}\orcidlink{0000-0002-1081-9410},}
\author[29]{{S.~Juneau}\orcidlink{0000-0002-0000-2394},}
\author[30]{{D.~Kirkby}\orcidlink{0000-0002-8828-5463},}
\author[14]{{T.~Kisner}\orcidlink{0000-0003-3510-7134},}
\author[14]{{A.~Kremin}\orcidlink{0000-0001-6356-7424},}
\author[14]{{M.~Landriau}\orcidlink{0000-0003-1838-8528},}
\author[31]{{L.~Le~Guillou}\orcidlink{0000-0001-7178-8868},}
\author[14]{{M.~E.~Levi}\orcidlink{0000-0003-1887-1018},}
\author[14]{{P.~McDonald}\orcidlink{0000-0001-8346-8394},}
\author[29]{{A.~Meisner}\orcidlink{0000-0002-1125-7384},}
\author[32,18]{{R.~Miquel},}
\author[33]{{J.~Moustakas}\orcidlink{0000-0002-2733-4559},}
\author[10,14]{{N.~Palanque-Delabrouille}\orcidlink{0000-0003-3188-784X},}
\author[34,35,36]{{W.~J.~Percival}\orcidlink{0000-0002-0644-5727},}
\author[37]{{F.~Prada}\orcidlink{0000-0001-7145-8674},}
\author[38]{{I.~P\'erez-R\`afols}\orcidlink{0000-0001-6979-0125},}
\author[14]{{A.~Raichoor}\orcidlink{0000-0001-5999-7923},}
\author[39]{{G.~Rossi},}
\author[40]{{E.~Sanchez}\orcidlink{0000-0002-9646-8198},}
\author[14]{{D.~Schlegel},}
\author[41,3]{{M.~Schubnell},}
\author[42,16]{{R.~Sharples}\orcidlink{0000-0003-3449-8583},}
\author[14]{{J.~Silber}\orcidlink{0000-0002-3461-0320},}
\author[29]{{D.~Sprayberry},}
\author[3]{{G.~Tarl\'{e}}\orcidlink{0000-0003-1704-0781},}
\author[17]{{M.~Vargas-Maga\~na}\orcidlink{0000-0003-3841-1836},}
\author[29]{{B.~A.~Weaver},}
\author[31]{{P.~Zarrouk}\orcidlink{0000-0002-7305-9578},}
\author[14]{{R.~Zhou}\orcidlink{0000-0001-5381-4372},}
\author[43]{{H.~Zou}\orcidlink{0000-0002-6684-3997}}

\affiliation[1]{Dipartimento di Fisica ``Aldo Pontremoli'', Universit\`a degli Studi di Milano, Via Celoria 16, I-20133 Milano, Italy}
\affiliation[2]{INAF-Osservatorio Astronomico di Brera, Via Brera 28, 20122 Milano, Italy}
\affiliation[3]{University of Michigan, 500 S. State Street, Ann Arbor, MI 48109, USA}
\affiliation[4]{Departamento de Astrof\'{\i}sica, Universidad de La Laguna (ULL), E-38206, La Laguna, Tenerife, Spain}
\affiliation[5]{Instituto de Astrof\'{\i}sica de Canarias, C/ V\'{\i}a L\'{a}ctea, s/n, E-38205 La Laguna, Tenerife, Spain}
\affiliation[6]{Astrophysics \& Space Institute, Schmidt Sciences, New York, NY 10011, USA}
\affiliation[7]{Center for Cosmology and AstroParticle Physics, The Ohio State University, 191 West Woodruff Avenue, Columbus, OH 43210, USA}
\affiliation[8]{Department of Astronomy, The Ohio State University, 4055 McPherson Laboratory, 140 W 18th Avenue, Columbus, OH 43210, USA}
\affiliation[9]{The Ohio State University, Columbus, 43210 OH, USA}
\affiliation[10]{IRFU, CEA, Universit\'{e} Paris-Saclay, F-91191 Gif-sur-Yvette, France}
\affiliation[11]{Department of Physics, University of California, Berkeley, 366 LeConte Hall MC 7300, Berkeley, CA 94720-7300, USA}
\affiliation[12]{University of California, Berkeley, 110 Sproul Hall \#5800 Berkeley, CA 94720, USA}
\affiliation[13]{Physics Dept., Boston University, 590 Commonwealth Avenue, Boston, MA 02215, USA}
\affiliation[14]{Lawrence Berkeley National Laboratory, 1 Cyclotron Road, Berkeley, CA 94720, USA}
\affiliation[15]{Department of Physics \& Astronomy, University College London, Gower Street, London, WC1E 6BT, UK}
\affiliation[16]{Institute for Computational Cosmology, Department of Physics, Durham University, South Road, Durham DH1 3LE, UK}
\affiliation[17]{Instituto de F\'{\i}sica, Universidad Nacional Aut\'{o}noma de M\'{e}xico,  Circuito de la Investigaci\'{o}n Cient\'{\i}fica, Ciudad Universitaria, Cd. de M\'{e}xico  C.~P.~04510,  M\'{e}xico}
\affiliation[18]{Institut de F\'{i}sica d’Altes Energies (IFAE), The Barcelona Institute of Science and Technology, Edifici Cn, Campus UAB, 08193, Bellaterra (Barcelona), Spain}
\affiliation[19]{Departamento de F\'isica, Universidad de los Andes, Cra. 1 No. 18A-10, Edificio Ip, CP 111711, Bogot\'a, Colombia}
\affiliation[20]{Observatorio Astron\'omico, Universidad de los Andes, Cra. 1 No. 18A-10, Edificio H, CP 111711 Bogot\'a, Colombia}
\affiliation[21]{Institut d'Estudis Espacials de Catalunya (IEEC), c/ Esteve Terradas 1, Edifici RDIT, Campus PMT-UPC, 08860 Castelldefels, Spain}
\affiliation[22]{Institute of Cosmology and Gravitation, University of Portsmouth, Dennis Sciama Building, Portsmouth, PO1 3FX, UK}
\affiliation[23]{Institute of Space Sciences, ICE-CSIC, Campus UAB, Carrer de Can Magrans s/n, 08913 Bellaterra, Barcelona, Spain}
\affiliation[24]{Fermi National Accelerator Laboratory, PO Box 500, Batavia, IL 60510, USA}
\affiliation[25]{Steward Observatory, University of Arizona, 933 N, Cherry Ave, Tucson, AZ 85721, USA}
\affiliation[26]{Steward Observatory, University of Arizona, 933 N. Cherry Avenue, Tucson, AZ 85721, USA}
\affiliation[27]{Department of Physics, The Ohio State University, 191 West Woodruff Avenue, Columbus, OH 43210, USA}
\affiliation[28]{School of Mathematics and Physics, University of Queensland, Brisbane, QLD 4072, Australia}
\affiliation[29]{NSF NOIRLab, 950 N. Cherry Ave., Tucson, AZ 85719, USA}
\affiliation[30]{Department of Physics and Astronomy, University of California, Irvine, 92697, USA}
\affiliation[31]{Sorbonne Universit\'{e}, CNRS/IN2P3, Laboratoire de Physique Nucl\'{e}aire et de Hautes Energies (LPNHE), FR-75005 Paris, France}
\affiliation[32]{Instituci\'{o} Catalana de Recerca i Estudis Avan\c{c}ats, Passeig de Llu\'{\i}s Companys, 23, 08010 Barcelona, Spain}
\affiliation[33]{Department of Physics and Astronomy, Siena College, 515 Loudon Road, Loudonville, NY 12211, USA}
\affiliation[34]{Department of Physics and Astronomy, University of Waterloo, 200 University Ave W, Waterloo, ON N2L 3G1, Canada}
\affiliation[35]{Perimeter Institute for Theoretical Physics, 31 Caroline St. North, Waterloo, ON N2L 2Y5, Canada}
\affiliation[36]{Waterloo Centre for Astrophysics, University of Waterloo, 200 University Ave W, Waterloo, ON N2L 3G1, Canada}
\affiliation[37]{Instituto de Astrof\'{i}sica de Andaluc\'{i}a (CSIC), Glorieta de la Astronom\'{i}a, s/n, E-18008 Granada, Spain}
\affiliation[38]{Departament de F\'isica, EEBE, Universitat Polit\`ecnica de Catalunya, c/Eduard Maristany 10, 08930 Barcelona, Spain}
\affiliation[39]{Department of Physics and Astronomy, Sejong University, 209 Neungdong-ro, Gwangjin-gu, Seoul 05006, Republic of Korea}
\affiliation[40]{CIEMAT, Avenida Complutense 40, E-28040 Madrid, Spain}
\affiliation[41]{Department of Physics, University of Michigan, 450 Church Street, Ann Arbor, MI 48109, USA}
\affiliation[42]{Centre for Advanced Instrumentation, Department of Physics, Durham University, South Road, Durham DH1 3LE, UK}
\affiliation[43]{National Astronomical Observatories, Chinese Academy of Sciences, A20 Datun Rd., Chaoyang District, Beijing, 100012, P.R. China}

\emailAdd{davide.bianchi1@unimi.it}

\abstract{We present an in-depth analysis of the fiber assignment incompleteness in the Dark Energy Spectroscopic Instrument (DESI) Data Release 1 (DR1).
This incompleteness is caused by the restricted mobility of the robotic fiber positioner in the DESI focal plane, which limits the number of galaxies that can be observed at the same time, especially at small angular separations. 
As a result, the observed clustering amplitude is suppressed in a scale-dependent manner, which, if not addressed, can severely impact the inference of cosmological parameters.
We discuss the methods adopted for simulating fiber assignment on mocks and data.
In particular, we introduce the fast fiber assignment (FFA) emulator, which was employed to obtain the  power spectrum covariance adopted for the DR1 full-shape analysis.
We present the mitigation techniques, organised in two classes: measurement stage and model stage.
We then use high fidelity mocks as a reference to quantify both the accuracy of the FFA emulator and the effectiveness of the different measurement-stage mitigation techniques.
This complements the studies conducted in a parallel paper for the model-stage techniques, namely the $\theta$-cut approach.
We find that pairwise inverse probability (PIP) weights with angular upweighting recover the ``true" clustering in all the cases considered, in both Fourier and configuration space.
Notably, we present the first ever power spectrum measurement with PIP weights from real data.
}

\begin{document}
\maketitle
\flushbottom

\section{Introduction}
\label{sec:intro}

The three-dimensional distribution of galaxies encodes a wealth of cosmological information about the evolution of our Universe and the nature of its constituents.
Over the past twenty-five years, large spectroscopic surveys have been essential in extracting this information by gathering millions of galaxy redshifts across increasingly larger volumes \cite{SDSS2000, dawson2013, dawson2016}.

The Dark Energy Spectroscopic Instrument (DESI) survey \cite{Snowmass2013.Levi} relies on multiplexed spectroscopic observations achieved through a multi-object spectrograph \cite{DESIinstrument}. 5020 robotic positioners move optical fibers into place on the focal plane \cite{DESI2016a.Science, FocalPlane.Silber.2023, Corrector.Miller.2023} and allow 5000 spectra\footnote{20 of the fibers send their light to a camera that monitors sky conditions.} to be measured at once \cite{Spectro.Pipeline.Guy.2023}.
DESI thus achieves a multiplexing that is approximately five times greater than its predecessor, the Sloan Digital Sky Survey (SDSS).
The improved multiplexing, combined with the ability to dynamically adapt exposure times and tiling strategy to the observing conditions \cite{SurveyOps.Schlafly.2023} and an approximately $2.5\times$ increase in mirror area, enabled DESI to obtain more than four times as many extragalactic redshifts in its first year of main survey operation compared to what SDSS collected during its entire twenty-year history.

However, to fully benefit of such an increase in statistical power, it is essential to have precise control over all the potential sources of systematic error. 
In other words, is it crucial to identify and remove or model all the spurious fluctuations in the galaxy density that stem from non-cosmological factors.
Like for eBOSS and SDSS, in DESI one significant source of these clustering artifacts comes from missing observations, particularly those due to fiber assignment (FA) issues.

There are two main instrumental limitations at play.
First, each of the 5000 positioners is limited to a ``patrol radius" which covers only a small fraction of the focal plane.
Although each patrol region overlaps with those of adjacent fibers (approximately $15\%$ of the total focal plane area is reachable by two fibers), in densely populated regions of the sky the total number of galaxies may surpass the number of available fibers.
Second, due the physical size of the positioner, it is impossible to place two optical fibers closer than a minimum angular separation, on the focal plane, which prevents the observation of closely spaced galaxy pairs with a single pass of the instrument.
For historical reasons, in the following we will often refer to the combination of these effects as {\it fiber collisions}, even though, as just explained, it would be more accurate to describe it as a competition (between galaxies) for a fiber.

The overall effect of fiber collisions is a complex pattern of missing observations unevenly distributed across the survey area, with a primary concentration in high-density regions.
This results in a non trivial, scale dependent reduction in the amplitude of the observed galaxy clustering, which, if not properly corrected, can significantly affect the cosmological constraints.

Through the years different countermeasures were put forward, which can be roughly grouped in two classes depending on whether the proposed correction is limited to the clustering measurements or rather it requires additional manipulation of the theory model:  
\begin{enumerate}
    \item 
    Estimator level, e.g. \cite{hawkins2003, guo2012, pinol2017, hahn2017, 2017Bianchi, bianchi2020, sunayama2020, makiya2022}
    \item
    Model level, e.g. \cite{reid2014, burden2017, hahn2017, paviot2022}
\end{enumerate}

A well known example of an estimator-level approach is angular upweighting \cite{hawkins2003}.
This technique involves weighting each pair, when estimating the three-dimensional correlation function, by the ratio of total pairs to observed pairs at that angular separation.
This correction alone is not sufficient in general because it assumes that the radial properties of the unobserved pairs are equivalent to those of the observed pairs.
Typically, an imperfect matching between observed and missing galaxies is the reason why many popular missing observations countermeasures fail to achieve the desired level of accuracy.
For example, this is why the method proposed by \cite{guo2012} and the first of the two methods proposed by \cite{hahn2017}, do not deliver unbiased results.
Both approaches utilize the regions of overlapping observations to gain insight into the missing data.
The first method uses this information to adjust the pair counts, while the second probabilistically assigns redshifts to the missing galaxies.

The standard approach used by the BOSS team \cite{anderson2012} is commonly known as the nearest neighbor method.
It works by assigning the weight of the missing galaxies to the nearest galaxy in angular separation that obtained a redshift.
This upweighting approximately corrects the total pair counts on large separations but fails at smaller scales, as the pair between the missed and nearest galaxy is lost.
More recently, the eBOSS collaboration improved on this method by introducing completeness weights \cite{bosslsscat}, which more accurately capture the nature of fiber incompleteness. 
However, they still struggle to recover the ``true" clustering on small scales.
A similar approach has been implemented for DESI and is discussed in detail below.

A formally unbiased estimator-level approach was introduced by \cite{2017Bianchi, percival2017} and further refined in \cite{bianchi2020}.
These studies demonstrated that the ``true" clustering can consistently be recovered using Pairwise Inverse Probability (PIP) weights.
As the name implies, this method involves weighting each galaxy pair by its inverse probability of being observed.
This ensures that the expectation value of both pair and individual galaxy counts aligns with their true values on all scales, as long as there is some area of the survey footprint which is observed multiple times.
In addition to real data applications, e.g. \cite{mohammad2018, mohammad2020}, the inverse probability approach was tested in preliminary DESI studies with approximate synthetic catalogues~\cite{bianchi2018, smith2019}.
Its application to DESI data is explored in detail throughout this paper.

Some model-level approaches focus on eliminating the (angular) modes impacted by fiber assignment \cite{burden2017, paviot2022} and adjusting the model accordingly.
However, this results in a significant loss of large-scale information.
Another popular model-level approach is the second method proposed in \cite{hahn2017}.
The small-scale effect of fiber collisions is described as a top-hat window in transverse separation, which is convolved with the theoretical power spectrum when performing parameter inference. 
The main limitation of this approach is the approximate description of the window.

A more traditional model-level approach consist of simply removing the small transverse separations or, similarly, the small angles between line of sight and separation, and adjusting the theory model accordingly.
Typically, this method have been used in configuration space, with relatively minor variations, e.g. \cite{reid2014, mohammad2016, zarrouk2018}.
As extensively discussed in \cite{KP3s5-Pinon} and summarized here in a dedicated section, this approach can be applied directly to the separation angle between galaxies, which more accurately capture the nature of the DESI fiber assignment, and can also be extended to Fourier space.
This is the strategy adopted for the DESI DR1 full-shape analysis \cite{DESI2024.V.KP5, DESI2024.VII.KP7B}.

In this paper, we conduct a comprehensive study of the impact of fiber-assignment incompleteness on 2-point statistics for DESI data release 1 (DR1) and we present the countermeasures put in place to address this issue, in both configuration and Fourier space.

To effectively study the impact of fiber collisions, it is essential to create realistic synthetic galaxy catalogs (``mocks'') that include this effect.
We discuss the strategies we employed to simulate the fiber-assignment process for the DESI-DR1 mock catalogs and to derive PIP weights.
In particular, we present the fast fiber assignment (FFA) emulator used to obtain the covariance matrix for the DR1 full shape analysis of the power spectrum \cite{DESI2024.V.KP5, DESI2024.VII.KP7B}.

The paper is structured as follows.
In Section \ref{sec:data}, we present the DESI data, along with an overview of the fundamental constituents of the fiber assignment algorithm and the effects of fiber collisions for DR1.
In section \ref{sec:simFA} we illustrate the two methodologies used to simulate the fiber assignment process on both real data and mock catalogs. 
In section \ref{sec:sims} we describe the mock catalogues and use them to validate the above assignment strategies.
In section \ref{sec:estimator_level} and \ref{sec:model_level} we introduce the different estimator-level and model-level techniques we used for fiber mitigation, respectively.
In section \ref{sec:results} we present configuration and Fourier space measurements of the 2-point statistics, for both mocks and data, obtained with the different mitigation techniques.
We conclude in section \ref{sec:conclusions}.

\section{Data}\label{sec:data}

DESI is a multi-object spectroscopic instrument \cite{DESIinstrument}: robotic positioners guide 5000 optical fibers \cite{FocalPlane.Silber.2023, DESIfa} to the celestial coordinates of the designated targets \cite{DESItarget}, with the fibers transmitting the collected light to the spectrographs.
The full coverage of the survey footprint is achieved trough multiple, overlapping, pointings of the instrument, over different regions of the sky.
We refer to such pointings or, equivalently, to the corresponding sets of 5000 targets, as tiles.
Specifically, we focus on the redshift catalogs assembled using the version named ``iron" of the spectroscopic processing \cite{Redrock.Bailey.2024}.

\begin{figure}[htbp]
    \centering
    \includegraphics[width=0.8\linewidth]{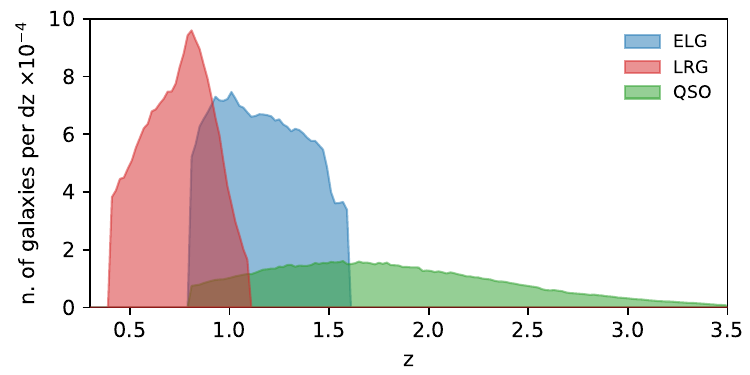}    
    \caption{Redshift distribution of the different DESI dark-time tracers, ELGs (blue), LRGs (red) and QSOs (green), with $dz=0.02$. For illustrative pourposes, the figure includes the QSO high-redshift tail up to $z=3.5$. However, note that objects with $z > 2.1$ are not considered for our clustering measurements.}
    \label{fig:zdist_data}    
\end{figure}

DESI large-scale observations are organised in two separate programs: ``bright" and ``dark" time, corresponding, for DR1, to 2275 and 2744 observed tiles, respectively.
In dark time DESI observes three classes of targets: Emission Line Galaxies (ELG) \cite{ELGtarget}, Luminous Red Galaxies (LRG) \cite{LRGtarget} and Quasars (QSO) \cite{QSOtarget}.
In addition, a low redshift Bright Galaxy Sample (BGS) \cite{BGStarget} is observed in bright time.
All the targets were selected based on photometry from the Legacy Survey Data Release 9 \cite{LS.Overview.Dey.2019,LS.dr9.Schegel.2024}.
For cosmological inference, the different target types are organised in redhsift bins as follows, 
\begin{itemize}
    \item ELG: \quad $0.8<z<1.1$; \quad $1.1<z<1.6$
    \item LRG: \quad $0.4<z<0.6$; \quad $0.6<z<0.8$; \quad $0.8<z<1.1$
    \item QSO: \quad $0.8<z<2.1$; \quad $2.1<z<3.5$
    \item BGS: \quad $0.1<z<0.4$
\end{itemize}
were the highest QSO bin is used for Lyman alpha forest analysis \cite{DESI2024.IV.KP6} but not for galaxy clustering studies, except for those on primordial non Gaussianity, which adopt $0.8 < z < 3.1$~\cite{chaussidon2024}.
In this work, for the sake of compactness, we instead collapse each tracer into a unique redshift bin (we only consider QSOs with $z < 2.1$) and we focus on the dark-program tracers: ELG, LRG and QSO.
In principle, all the studies and methods presented here can be applied to BGS tracers as well, as they are equipped with the same tools, such as weights and simulated catalogues, and exhibit a similar response to FA.
The raw, unweighted, redshift distributions of the observed galaxies are shown in figure~\ref{fig:zdist_data}.

We refer the reader to \cite{DESI2024.II.KP3}, and references therein, for an exhaustive description of all the properties of the different catalogues, the methodologies adopted to construct them and the corresponding nomenclature/conventions.
Here we focus on those aspects of the catalogues that are particularly relevant for fiber-assignment related phenomena.

The data can be organised into classes of catalogues, each playing a specific role in the FA process.
The mitigation methods we will present aim to recover the clustering of all galaxies, of a given tracer type, that had the opportunity to be targeted.
This is a binary true-or-false attribute that primarily depends on whether a galaxy entered the patrol radius of at least one fiber, but also takes into account additional effects, e.g. hardware failures, as described in \cite{DESI2024.II.KP3}.
We collectively refer to those galaxies as the {\it complete} sample, not to be confused with the parent sample from which they are extracted.
For DESI data the parent sample is simply given by the entire photometric catalogue.
Crucially, any difference between the two samples is not driven by the clustering itself, since, unlike the assignment probability, the chance of being targeted is independent of the local density, at least in principle. 
It thus can be handled with a proper random sample, tracing the effective volume of the complete sample.
We define the set of galaxies that were actually targeted and had their redshifts successfully measured as the {\it clustering} sample.
Lastly, we define the {\it full} sample, primarily used for angular upweighting, as the set of all galaxies of a given tracer type in the parent sample that meet the selection criteria (i.e. the veto mask, see \cite{DESI2024.II.KP3}) of the clustering sample, regardless of whether they were observed or not.
For consistency, the complete and clustering samples must be strictly defined within the same redshift bins, whereas the full sample does not include redshifts at all (it has by construction the same $z$-range of the parent sample).
Clearly, for real data, the complete sample is not available; it can only be accessed in simulated catalogs.
Each of these samples is paired with the correspondent random catalogue of unclustered tracers covering the same volume.
Table \ref{tab:cat} summarizes the relevant features of the different catalogues.
Further details can be found in appendix \ref{app:AUW}.

\begin{table}[htbp]
\centering
\begin{tabular}{c|ccccc}
\hline
catalogue & main purpose & FA & $z$-cut & mask & available for\\
\hline
\multirow{2}{2cm}{complete} & \multirow{2}{4cm}{``true'' 3-dim clustering} & \multirow{2}{0.5cm}{no} & \multirow{2}{0.5cm}{yes} & \multirow{2}{0.5cm}{no} & \multirow{2}{4cm}{altMTL mocks, \\ FFA mocks}
\\
&&&& \\
\multirow{2}{2cm}{full} & \multirow{2}{4cm}{angular upweighting} & \multirow{2}{0.5cm}{y/n} & \multirow{2}{0.5cm}{no} & \multirow{2}{0.5cm}{yes} & \multirow{2}{4cm}{data, altMTL mocks}
\\
&&&& \\
\multirow{2}{2cm}{clustering} & \multirow{2}{4cm}{3-dim clustering measurements} & \multirow{2}{0.5cm}{yes} & \multirow{2}{0.5cm}{yes} & \multirow{2}{0.5cm}{yes} & \multirow{2}{4cm}{data, altMTL mocks, FFA mocks}
\\
&&&& \\
\multirow{2}{2cm}{parent} & \multirow{2}{4cm}{running FA algorithm} & \multirow{2}{0.5cm}{no} & \multirow{2}{0.5cm}{no} & \multirow{2}{0.5cm}{no} & \multirow{2}{4cm}{data, altMTL mocks, FFA mocks}
\\
&&&& \\
\hline
\end{tabular}
\caption{Schematic description of the different catalogues used in this work. The redshift cuts are: $0.8<z_{\rm ELG}<1.6$;  $0.4<z_{\rm LRG}<1.1$; $0.8<z_{\rm QSO}<3.5$; $0.1<z_{\rm BGS}<0.4$. All catalogues except the parent are only defined within the DR1 angular footprint, corresponding to regions that entered the patrol radius of at least one (functioning) fiber.
The full and clustering catalogues also include an additional veto mask, as detailed in the table and discussed throughout the text. The ``y/n" symbol in the FA column signifies that the full catalogue includes a boolean entry, enabling the extraction of the subset of galaxies that were assigned, when needed.
The final column indicates whether a specific type of catalogue is available for real data, altMTL, or FFA, with the latter two representing the two classes of mocks considered in this work, as described in section \ref{sec:sims}.}
\label{tab:cat}
\end{table}

The allocation of fibers to designated targets is governed by the DESI targeting algorithm, \texttt{fiberassign}.
This code requires the merged-target-ledger (MTL) file as one of its inputs.
MTLs are lists of potential DESI targets, each with its own unique identifier ({\tt TARGETID}), sky coordinates (RA and Dec), {\tt PRIORITY}, {\tt SUBPRIORITY} and a survey/observing conditions dependent bitmask (see \cite{KP3s7-Lasker} for a detailed description).

When two or more targets are competing for the same fiber the conflict is settled by first looking at the {\tt PRIORITY} value, which is deterministic and solely depends on target type.
Specifically, for the dark time survey, $\text{\tt PRIORITY}_{\text{QSO}} > \text{\tt PRIORITY}_{\text{LRG}} > \text{\tt PRIORITY}_{\text{ELG}}$, except for a $10\%$ subsample of ELGs (ELG\_HIP class), whose {\tt PRIORITY} have been raised to match that of LRGs as a way to facilitate the extraction of the LRG-ELG cross-correlation signal. 
If the conflicting targets share the same {\tt PRIORITY}, then the assignment is determined by comparing their {\tt SUBPRIORITY} entry, which is simply a list of predetermined random numbers.

\begin{figure}[htbp]
    \centering
    \includegraphics[width=0.49\linewidth]{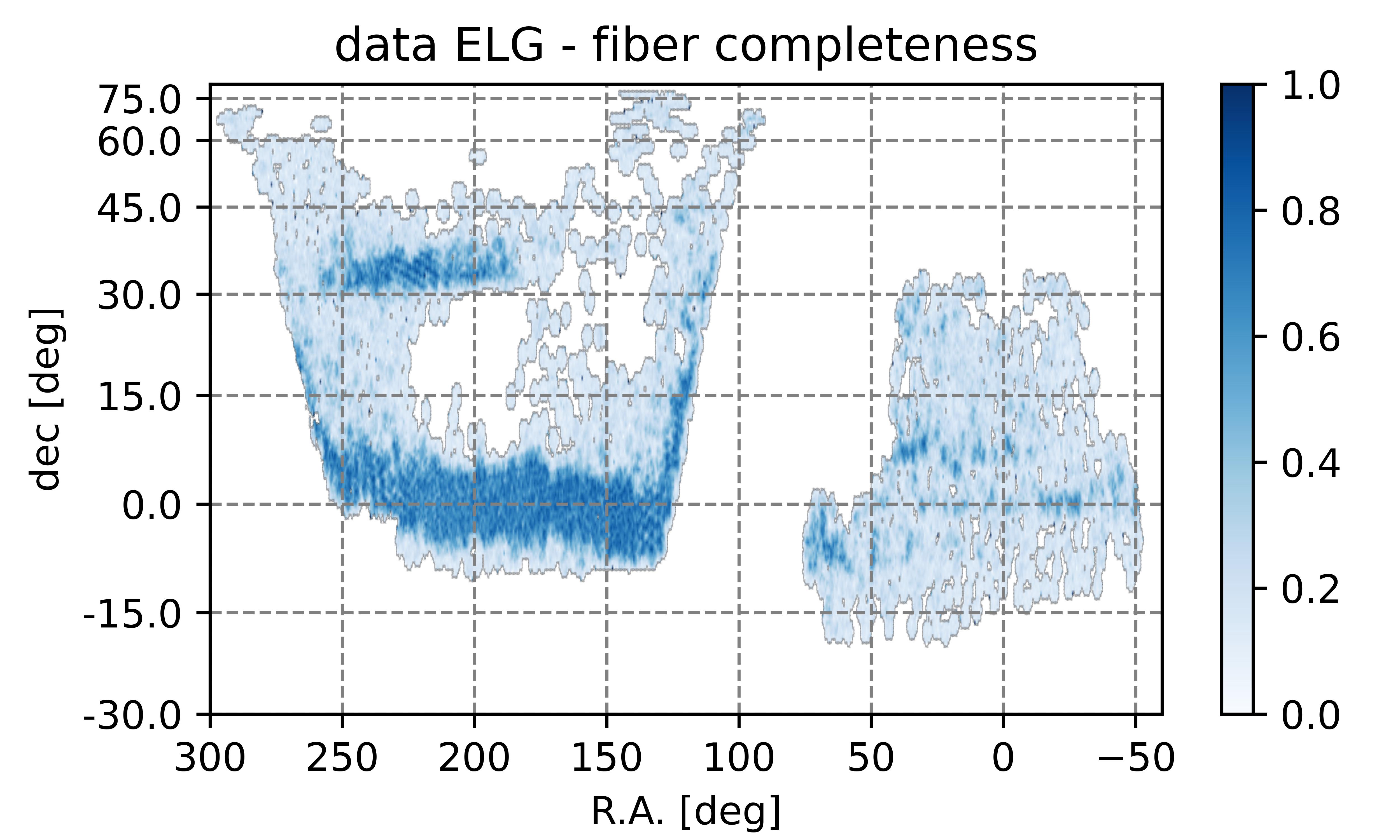}    \includegraphics[width=0.49\linewidth]{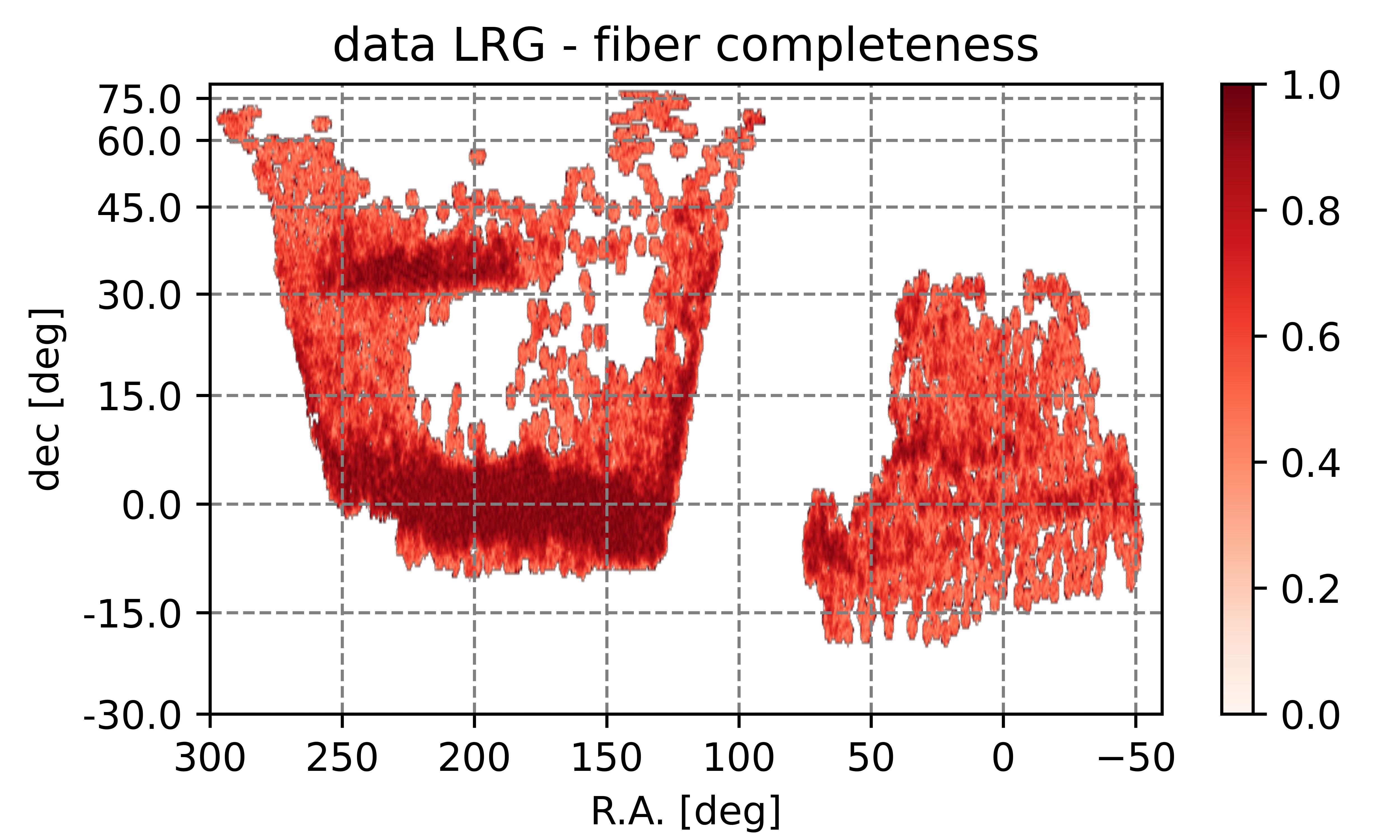}\\   \includegraphics[width=0.49\linewidth]{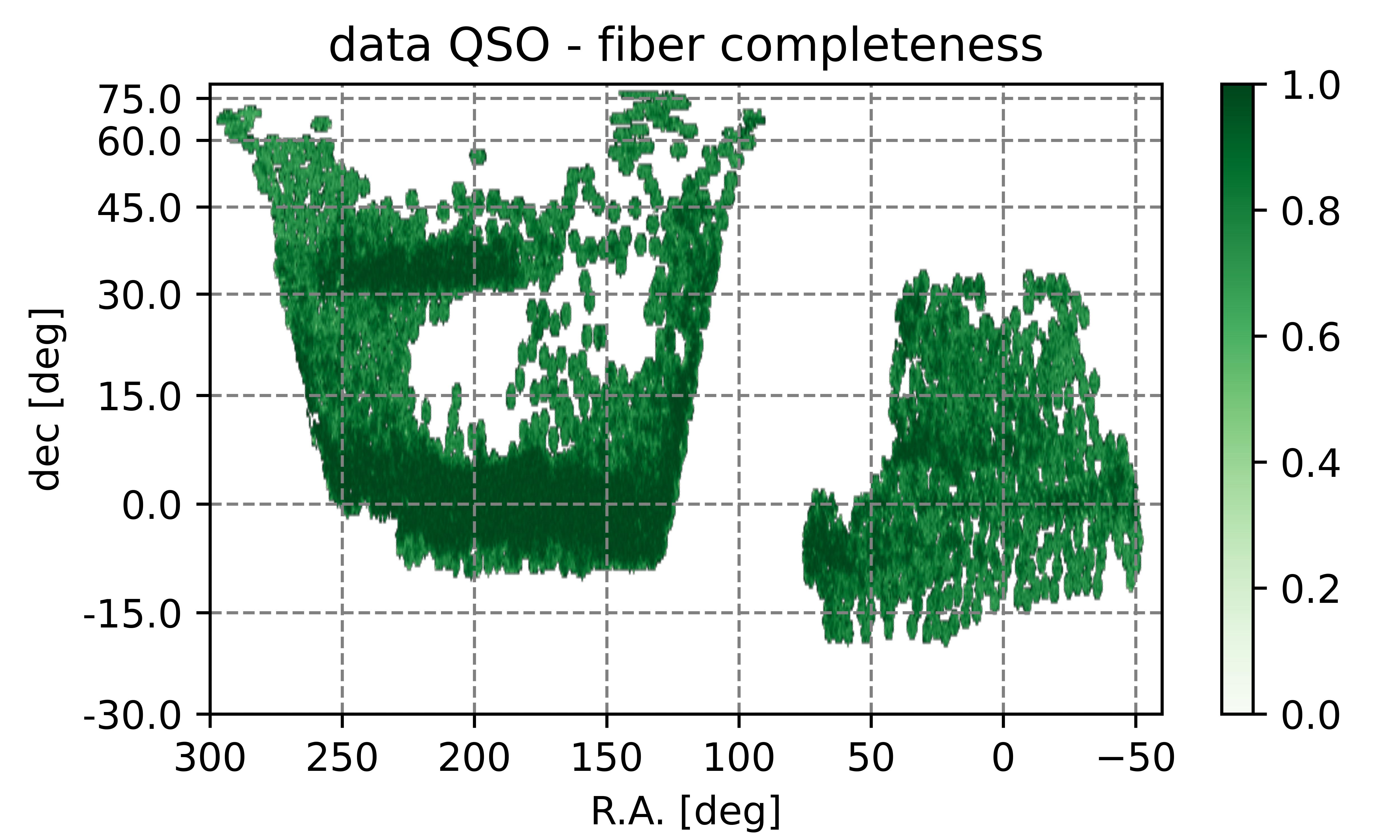}
    \caption{Angular assignment completeness of the galaxies observed by DESI in its first year (DR1), defined as the number of tracers assigned to a fiber divided by the total number of tracers in the corresponding full sample. 
    All three dark-time tracers, ELGs, LRGs and QSOs are shown, as labeled in the figure.}
    \label{fig:ang_2d_data}    
\end{figure}

In figure \ref{fig:ang_2d_data} we show the angular assignment completeness for the three dark-time tracers.
Unless otherwise stated, in this work we examine the North Galactic Cap (NGC) and the South Galactic Cap (SGC) separately.
These two regions are defined by their Galactic coordinates, not to be confused with the equatorial coordinates used for the figure.
More explicitly, NGC and SGC correspond to the two disconnected areas, from left to right, separated by the galactic plane (not shown in the figure).
One clear artefact created by fiber assignment immediately stands out: uneven large-scale distribution of galaxies inside the survey area.
These fluctuations are not of cosmological origin but rather they are artificially produced by the observing strategy.
In particular, as shown in \cite{DESI2024.II.KP3}, they correlate almost perfectly with the number of tile overlaps at each point in the survey footprint.
We refer to this number, which plays a crucial role for both fiber mitigation and emulation, as  $n_\text{tile}$. 

A second, even more important, artefact is not directly visible in the figure, as it manifests itself on much smaller scales and goes beyond 1-point statistics: fiber collisions.
As dicussed in section \ref{sec:intro}, for DESI this effect extends beyond mere physical collisions. 
In essence, in any area small enough to have a uniquely defined $n_\text{tile}$ value, we cannot observe more targets than $n_\text{tile}$ times the number of fibers in that area.

\begin{figure}
    \centering
    \includegraphics[width=0.32\linewidth]{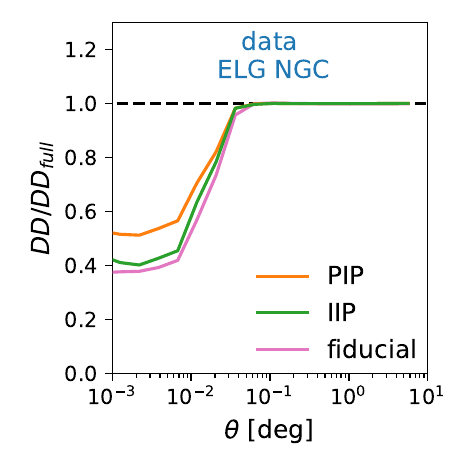}
    \includegraphics[width=0.32\linewidth]{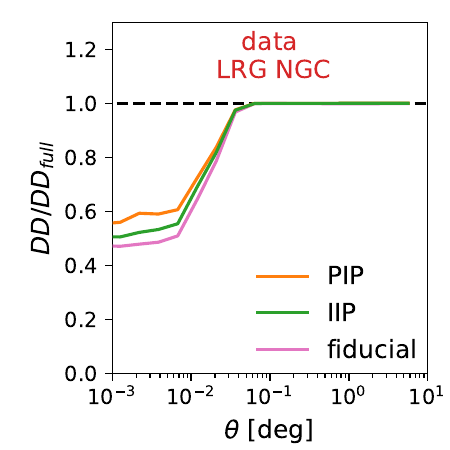}
    \includegraphics[width=0.32\linewidth]{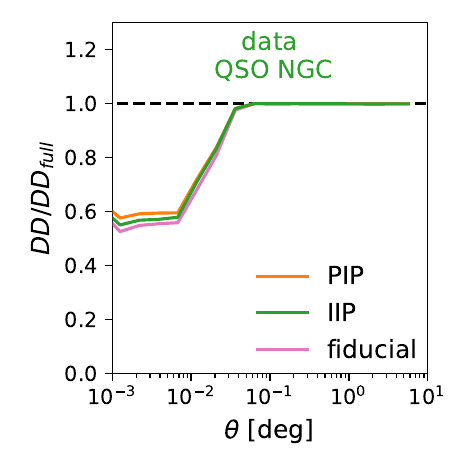}
    \caption{Normalised pair counts measured from DESI DR1, as a function of the separation angle $\theta$, with fiducial weights (solid pink), IIP weights (solid green) and PIP weights (solid orange), divided by the normalised pair counts from the full sample of potential targets. All three dark-time tracers are shown, as labeled in the figure.}
    \label{fig:ang_pair_counts_data}
\end{figure}

Figure \ref{fig:ang_pair_counts_data} illustrates the impact of fiber collisions on 2-point measurements, as a function of the separation angle $\theta$.
The solid pink curve represents the ratio of the pair counts from the sample of successfully observed galaxies in DR1 to the counts from the full sample of potential targets\footnote{The pair counts are divided by the total number of pairs in the respective sample.}.
In order to counteract the effect of fiber assignment, the observed galaxies are weighted with completeness weights\footnote{Here by completeness weights we mean $w_{comp}/f_{tile}$, as described in section \ref{sec:estimator_level}}, as detailed in \cite{DESI2024.II.KP3} and further discussed below in section \ref{sec:estimator_level}, which explains why the ratio nicely converges to 1 at large angles.
However, for separations $\lesssim 0.02 \ \rm deg$ there is a significant loss of power, up to $60\%$ for the ELGs, that clearly cannot be handled by the weights\footnote{ As discussed below, sections \ref{sec:estimator_level} and \ref{sec:results}, the fact that even PIP weights, which are inherently pairwise, fail to recover the ``true" clustering on small scale is direct evidence of the presence of zero-probability pairs, mostly coming from regions of the survey footprint covered by only a single pass of the instrument. 
In these regions, fiber collisions make it impossible to measure the redshifts of both members of a close pair, even when the two targets are QSOs and have the highest assignment priority.}.
Due to its lower completeness (figure \ref{fig:ang_2d_data}), the effect gets even stronger for the south galactic cap, not shown here for compactness, reaching about $80\%$.  
In the following we will often refer to this small-scale suppression as {\it collision window}.

The angular upweighting method introduced above and detailed in section \ref{sec:estimator_level} restores the ``true'' angular clustering by assigning to each galaxy pair a weight equal to the  inverse of the curve shown in the figure.
However, this alone does not guarantee that the corresponding 3-dim clustering is accurately recovered.

\section{Simulating fiber assignment}\label{sec:simFA}

Being able to understand and accurately reproduce the process of fiber assignment is of fundamental importance for, at least, two reasons.
First, it allows us to build realistic mock catalogues for covariance matrix estimations and validation of all the different inference models and estimators applied to the data, including
the fiber mitigation techniques discussed in this work.
Second, by building multiple realisations of the targeting, for both data and mocks, we can directly study the statistical properties of the fiber assignment process itself and use them to construct tailored countermeasures, as the inverse-probability weights.
In particular, we can associate to each galaxy a bitwise array, with as many bits as the number of realisations, where ones and zeros indicate whether the galaxy has been assigned or not in a given realisation.
These {\it bitweights} \cite{2017Bianchi}, usually compressed into integer numbers\footnote{The standard format adopted for DESI is 64 bit signed integers.}, are the building blocks that enable us to efficiently evaluate the assignment probability of any $t$-plet of galaxies in our samples (see section~\ref{sec:estimator_level}).

We developed two strategies for simulating fiber assignment: one follows the same FA processing as the data to create each replica and therefore it is computationally intensive, while the other relies on a simplified statistical representation of the process, enabling the generation of a much larger number of approximate replicas.

\subsection{Alternative merged target ledger (altMTL)}

As described in section \ref{sec:data}, the allocation of fibers relies on MTLs.
Any object in the MTL file gets a {\tt SUBPRIORITY} entry, i.e. a random value that controls the assignment when the object is competing against other targets with the same {\tt PRIORITY}.
The actual fiber assignment process can therefore be replicated, on data or mocks, by generating realistic alternative merged target ledgers (altMTLs). 
Since the only random component is given by the {\tt SUBPRIORITY} values, the assignment probability of the targets (and of the $t$-plets they form) can be estimated by rerunning \texttt{fiberassign} multiple times, each with a different collection of {\tt SUPRIORITY} values.

A detailed description of the process we have adopted to construct the altMTLs is provided in \cite{KP3s7-Lasker}.
For what concerns the clustering-analysis tools, altMTLs allowed us to obtain bitweighs for the DR1 samples and to build a set of realistic fiber-assigned mocks (see section \ref{sec:sims}), with the corresponding bitweights.   
Since these mocks have been processed through the actual targeting algorithm, the synthetic galaxies are delivered with the same entries as the real ones, including, e.g., completeness weights, $w_{comp}$, that we will explicitly define in section~\ref{sec:estimator_level}.
The downside is that the procedure is time consuming, limiting the number of samples we can create.

\subsection{Fast fiber assignment (FFA)}\label{sec:FFA}

As a way around the cpu-time limitations, we developed an emulator, which enables us to create thousands of fiber assigned samples, with an arbitrarily large number of targeting realisations each.  
We generically denote as emulator any surrogate algorithm that learns from a set of input-output pairs obtained with a reference algorithm.
Using machine-learning jargon, the inputs are the training sample and the outputs the corresponding target labels. 
Specifically, our reference algorithm is the \texttt{fiberassign} code.
The training sample can be either a set of accurate mocks or, notably, actual data.
The target label is simply the property of having been assigned or not, which each galaxy obtain through \texttt{fiberassign}.
As regards the architecture of the emulator, since we have a good understanding of the FA-algorithm average behaviour and a clear idea of which key features we need to emulate, we decided to adopt a shallow learning approach.
The advantage of our architecture, compared to approaches like deep learning with neural networks, is that it involves far fewer learnable parameters, all of which are easily interpretable, and is less sensitive to potential limitations of the training set, such as the number of samples available.
In brief, the emulator works as follows.
\begin{itemize}
    \item 
    As an input we need a catalogue of all the targets that had a chance of being assigned and a value of $n_{tile}$ for each of them.
    For the data this informations are obtained by concatenating the {\tt fiberassign} output files as the survey progresses and the code get repeatedly applied to the parent sample of all the DESI targets.
    For the synthetic data a mock parent sample is constructed with the same volume and number density as the real data and the mock potential assignments are obtained similarly as for the data. This concatenation of potential assignments provides all of the tiles on which a given target could have been assigned and $n_{tile}$ can thus be determined for every target.
    \item
    We run a friend-of-friend (FOF) algorithm, based on angular separation, on this sample of potential targets.
    The linking length, which is actually a linking angle, $\theta_{LL}$, is one of our free parameters. $\theta_{LL}$ can be roughly interpreted as the scale below which the assignment probabilities are not independent.
    In other words, below such scale, the assignment probability of a pair of galaxies is not the product of the two individual probabilities. See sections \ref{sec:estimator_level} and \ref{sec:model_level} for further discussions on the topic.
    In essence, this step serves to isolate groups of assignment-entangled galaxies. 
    \item
    Simultaneously, for each galaxy we count the number of companions within angular separation $\theta < \theta_{LL}$.
    We name this process close-friend count (CFC) and indicate the corresponding variable with $n_{CFC}$. Repetitions are allowed, i.e. a given galaxy can contribute to the CFC of more than one galaxy.
    In essence, each galaxy's $n_{CFC}$ roughly tells us how many other galaxies are competing with it for the same fiber.
    \item
    By comparing the pre- and post-assignment samples we obtain a two dimensional function, $\mathcal{F}$, which maps $n_{CFC}$ and $n_{tile}$ onto the fraction of assigned galaxies.
    Specifically, we count the number of galaxies with a given combination of $n_{CFC}$ and $n_{tile}$, in the sample of potential targets, and how many of them got assigned by the reference algorithm.
    We define $\mathcal{F}(n_{CFC}, n_{tile})$ as the ratio of such counts.
    By construction, $\mathcal{F}$ is an estimate of the (individual) assignment probability of a galaxy with a given $n_{CFC}$ and $n_{tile}$.    
\end{itemize}

Once we learned the $\mathcal{F}$ kernel, we can use it to emulate fiber assignment on any test sample of interest.
\begin{itemize}
    \item
    We process the test sample trough the same $n_{tile}$ finder algorithm used for the training sample: each galaxy in the test sample gets its own $n_{tile}$.
    \item
    We also apply the same FOF-CFC finder algorithm: each galaxy in the test sample gets its own $n_{CFC}$ and a link to its entangled partners. 
    \item
    We decide how many realisations, $N_{\rm bit}$, of the targeting we want to generate.
    Realistically, $N_{\rm bit}$ should be at least $\mathcal{O}(100)$ in order to have enough resolution for the anticorrelation post-processing stage, see below. 
    We randomly sample $N_{\rm bit}$ times each galaxy in the test sample, with its own individual probability, given by $\mathcal{F}(n_{CFC},n_{tile})$.
    \item
    We use the FOF information to add the desired anticorrelation on small scale, as described below.
\end{itemize}
The relation between anticorrelation and FOFs can be understood as follows.
Two galaxies compete for the same fiber only if they are both inside the patrol radius of that fiber, thereby establishing a characteristic exclusion scale (see also sections \ref{sec:estimator_level} and \ref{sec:model_level} for further considerations).
However, since the patrol radii overlap, it is formally possible to create chains of entangled galaxies that extend over scales larger than the exclusion one.
By construction, these larger entangled structures correspond to, or, more precisely, are contained in the (angular) FOF catalogue obtained using the exclusion scale as a linking length.
In other words, the anticorrelation induced by fiber assignment cannot travel distances larger than the size of the corresponding FOF group and, as a consequence, each group can be processed individually.

The anticorrelation algorithm works as follows.
For each galaxy inside the group we have $N_{\rm bits}$ realisations of the assignment, conveniently encoded as individual bitwise arrays, with the obvious correspondence between zeros/ones and discarded/assigned. 
By construction, any operation that keeps the number of zeros and ones in each single array conserved does not affect the individual assignment properties, but it can change the collective properties, specifically the assignment correlation.
With this in mind, we developed an algorithm that permutates ones and zeros, in each single bitwise array, in such a way that the anticorrelation of the pairs inside a FOF group gets maximized.
Details on how we build this maximum-anticorrelation state can be found in appendix \ref{app:antico}, in brief, the key requirement is for the number of assigned galaxies in a FOF group to vary the least possible from one targeting realisations to one another.
We can then tune the amount of anticorrelation by randomly swapping ones with zeros inside the bitwise arrays, effectively undoing part of the anticorrelation that we just imposed, while keeping all the individual properties untouched.
The fraction of randomly swapped bits, $f_{anti}$, is one of our free parameters.

There is a third free parameter, $\beta_{tilt}$, that enables us to adjust the $\mathcal{F}$ kernel by linearly changing its steepness along the $n_{CFC}$ direction.
While we have not utilized this parameter with any of the dark-time tracers examined here, we did employ it for the bright galaxy sample to address the issue of the pre-assignment mocks being less dense compared to the actual data. Consequently, we needed to suppress the high $n_{CFC}$ tail of the kernel, which would have assigned too many galaxies otherwise. 

The FFA approach immediately provides us with proper bitweights that we can use to implement the fiber-mitigation strategies described in section \ref{sec:estimator_level}.
In particular, we can obtain individual weights for each galaxy as $w_{IIP}=(c+1)/(K+1)$, where $c$ is the number of times the galaxy have been assigned and $K=N_{bits}$ is the number of realisations.
When dealing with FFA mocks we will sometimes refer to them as fiducial or weights, despite the fact that their definition formally differs from that of the completeness weights of data and altMTL mocks.

For all the FFA catalogues discussed in this work we trained the $\mathcal{F}$ kernel on actual data and set $\theta_{LL}=0.02\rm deg$, $f_{anti}=0.1$.
Finally, it is important to mention that in its current implementation, the emulator operates with one tracer at a time.
Extending it to a multi-tracer strategy is straightforward, but we leave that for future studies.

\section{Simulated catalogues}\label{sec:sims}

The simulation of DESI DR1 LSS samples (``mocks") is described fully in \cite{KP3s8-Zhao} and summarized in \cite{DESI2024.II.KP3}. Here, we provide some additional details on the simulations of DESI fiber assignment on these mocks.
We constructed three classes of fiber-assigned mocks, each with a different point of balance between level of realism and computational cost.
The three classes are obtained by coupling two sets of cosmological simulations, described below, with the two different strategies for simulating fiber assignment introduced in section \ref{sec:simFA},
\begin{itemize}
    \item 
    AbacusSummit + altMTLs; 25 mocks, one of which\footnote{Abacus mock n.11} with 128 independent  targeting realisations
    \item
    AbacusSummit + FFA; 25 mocks with 256 independent targeting realisations each
    \item
    EZmocks + FFA; 1000 mocks with 256 independent targeting realisations each
\end{itemize}
This large collection of mocks 
enables us to perform all sort of studies that require realistic survey geometry and completeness.

As a foundation for creating fiber-assigned mocks
we need lightcones with realistic distribution of the DESI targets, covering the same volume of DR1 data and having the same pre-fiber-assignment number density.
A set of 25 independent realisations of such catalogues has been extracted from the {\tt AbacusSummit} N-body suite \cite{abacus1, abacus2}.
It is worth noting, however, that not all features present in real observations, such as star contamination, have been simulated.
This high fidelity set is complemented by a larger set of 1000 realisations, obtained via approximate, faster methods ({\tt EZmocks}, \cite{EZmocks}).
Broadly speaking, the high fidelity set allows us to isolate systematic effects whereas the approximate set can be used to estimate covariance matrices.  
In this work, we focus on the Abacus set because it provides a more accurate description of the small-scale clustering, leading to a more realistic FA output. Furthermore, for the Abacus set we have samples processed with both the altMTL and FFA methods, facilitating a direct comparison of the two.
Unless specified otherwise, the results in this work are based on Abacus mock n.11.
All other mocks give consistent outcomes.

\subsection{Simulated DESI DR1 target samples}
{\tt AbacusSummit} is a suite of large, high-accuracy cosmological N-body simulations designed to meet (and exceed) the cosmological simulation requirements of the DESI survey \cite{abacus1, abacus2}. There are over 150 simulations with different cosmologies, containing $6912^3$ particles in a cubic volume of side 2 $h^{-1}$Gpc.
For the mocks on which we performed fiber assignment, the following different snapshots were employed: for LRGs, the snapshots used are \texttt{z0.500} for $z<0.6$ and \texttt{z0.800} for $z>0.6$; for ELGs, the snapshots used are \texttt{z0.950} for $z<1.1$ and \texttt{z.1325} for $z>1.1$; For QSOs, the snapshot used is \texttt{z1.400} for all redshifts. Simulated galaxies are placed within dark matter halos based on the results of analysis of early DESI clustering data by \cite{abacushod2,abacusHODELG}. These snapshots were then concatenated to produce mocks with all DESI dark-time tracers at the correct respective target densities.

The DR1 {\tt EZmocks} were produced at the same redshift snapshots as the {\tt AbacusSummit} mocks, with their clustering trained on the same {\tt AbacusSummit} mocks. They were produced within boxes of side 6 $h^{-1}$Gpc, which is large enough to simulate the entire DESI DR1 sample within each Galactic cap (GC). 2000 such simulations were produced at and concatenated across each of the same snapshots used for the {\tt AbacusSummit} mocks, which provided the North and South GCs. Full details are provided in \cite{KP3s8-Zhao}.

\subsection{Comparing altMTL and FFA mock catalogues}

\begin{figure}[htbp]
    \centering
    \includegraphics[width=0.49\linewidth]{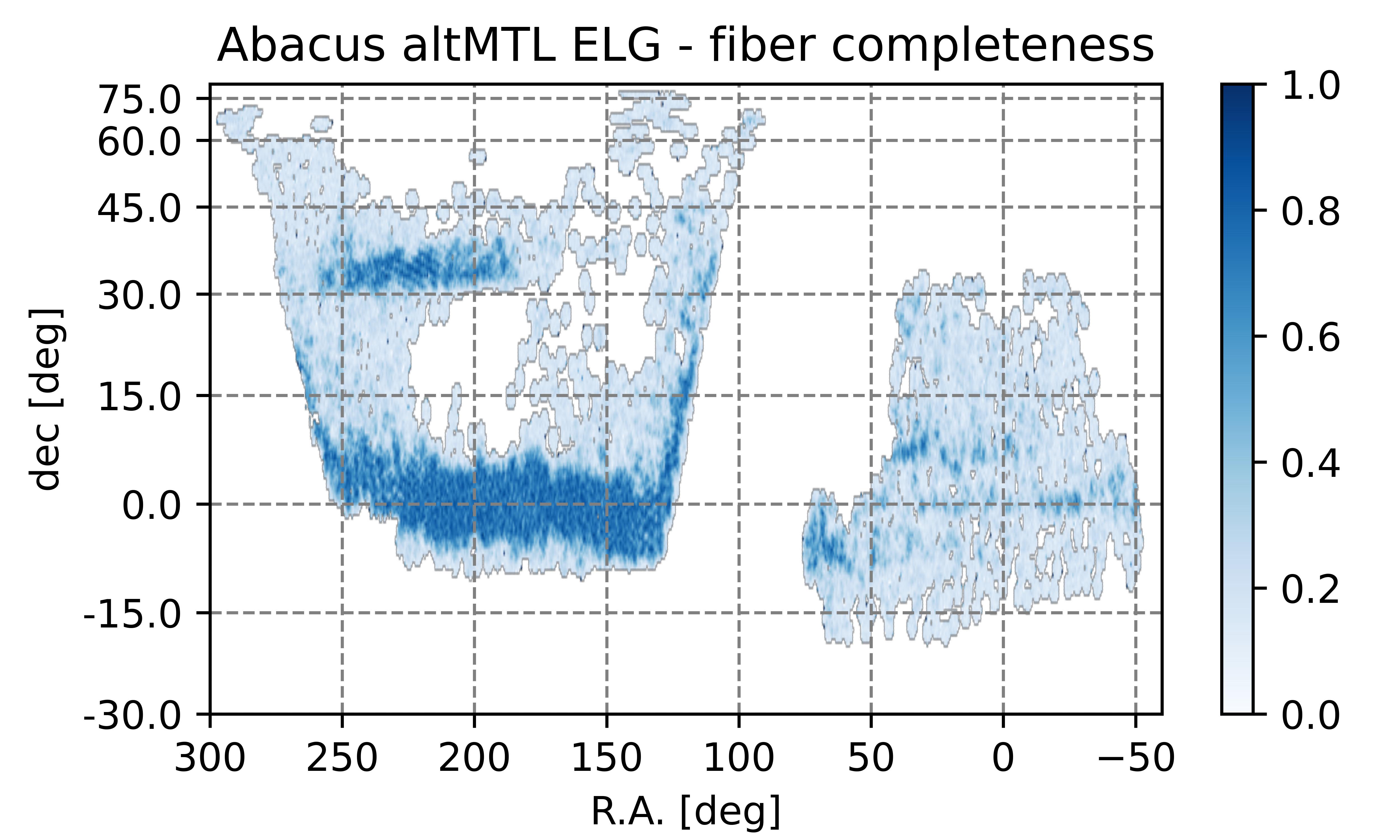}    \includegraphics[width=0.49\linewidth]{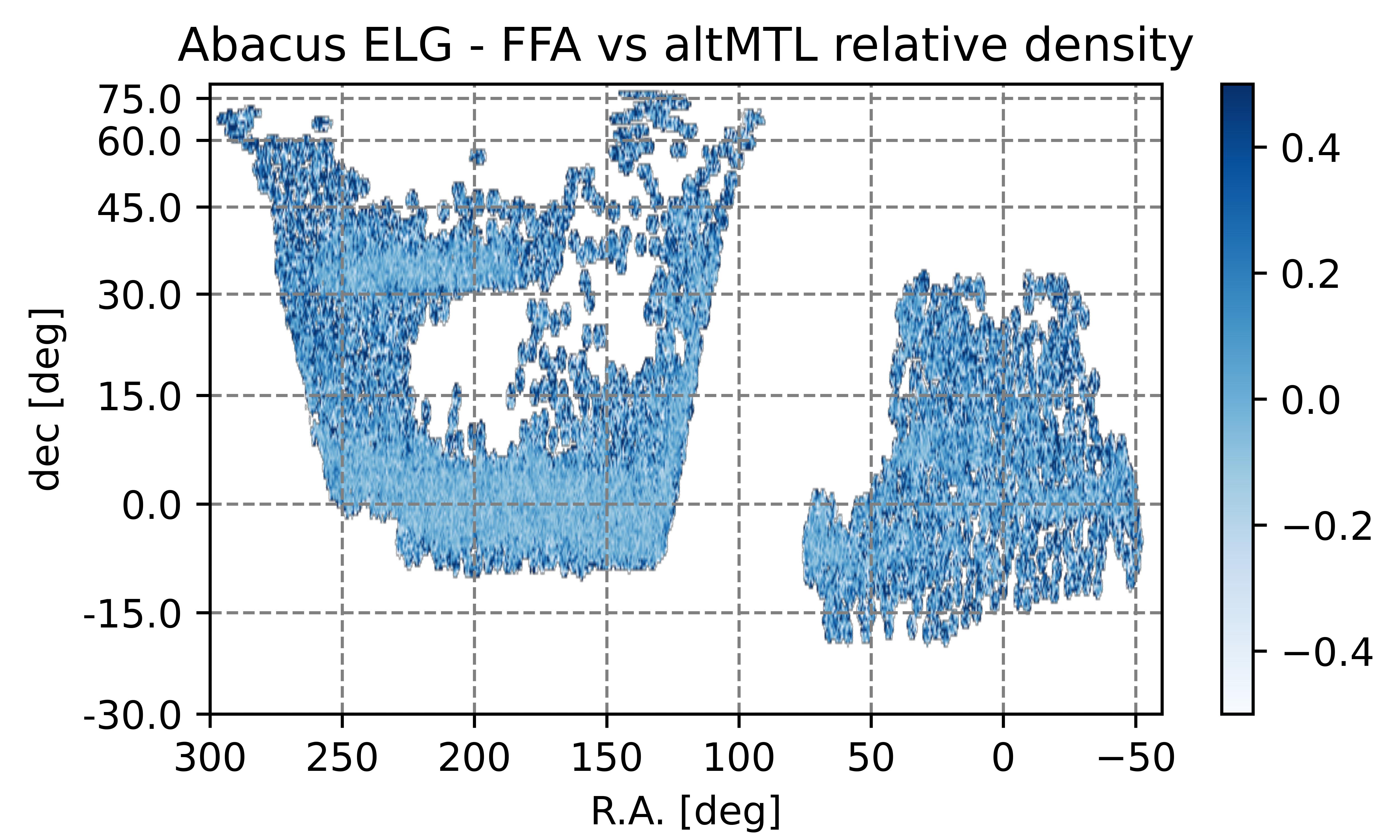}\\ \includegraphics[width=0.49\linewidth]{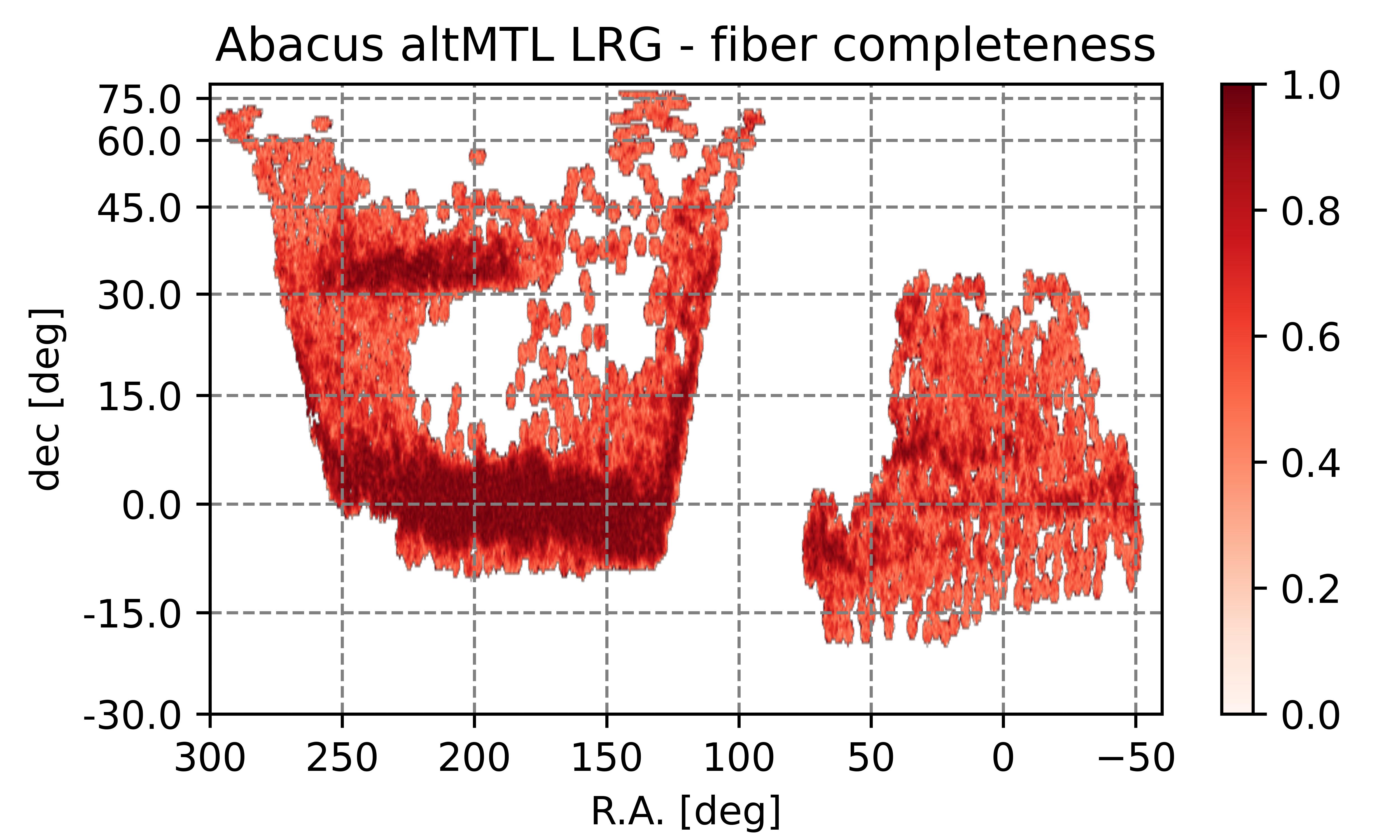}  \includegraphics[width=0.49\linewidth]{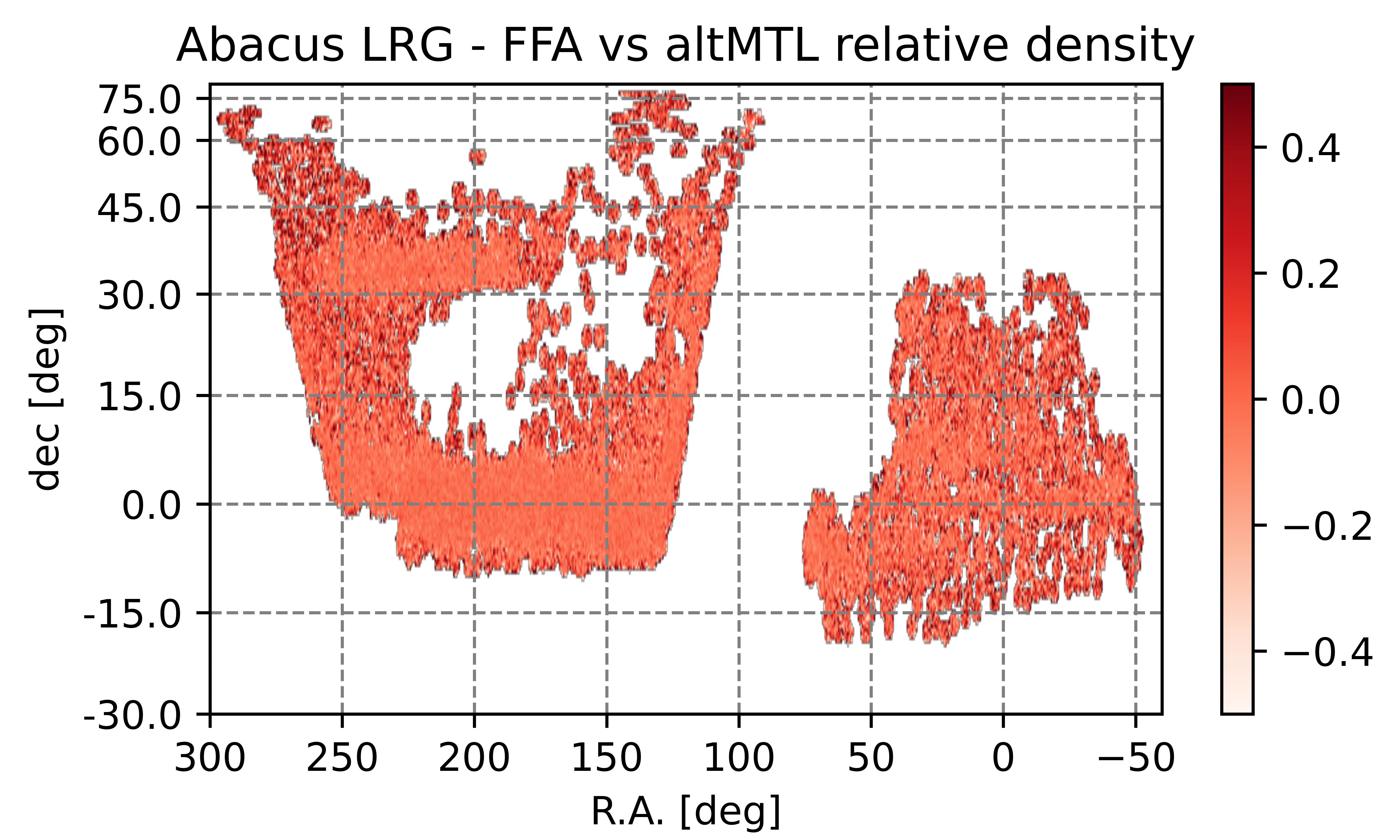}\\    \includegraphics[width=0.49\linewidth]{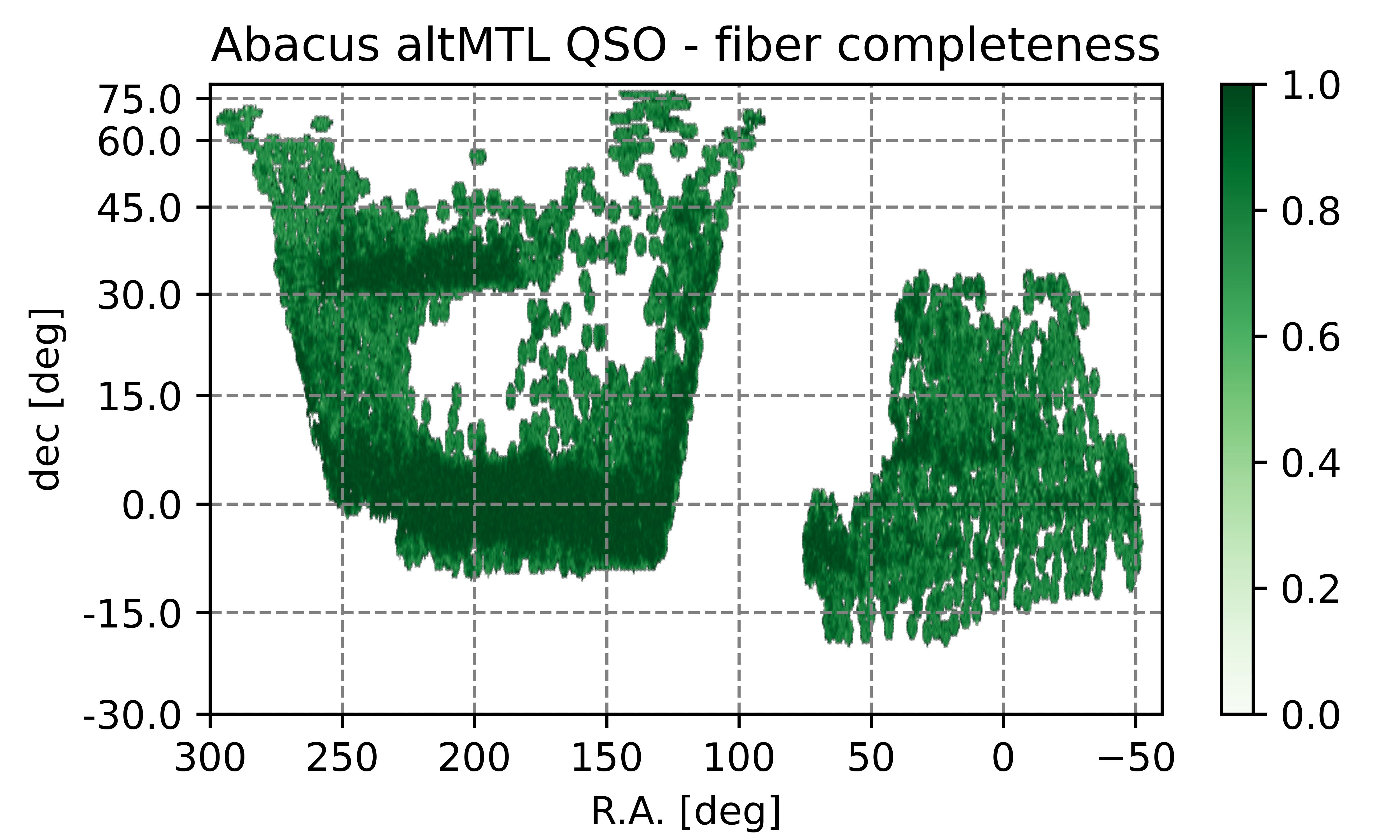}   \includegraphics[width=0.49\linewidth]{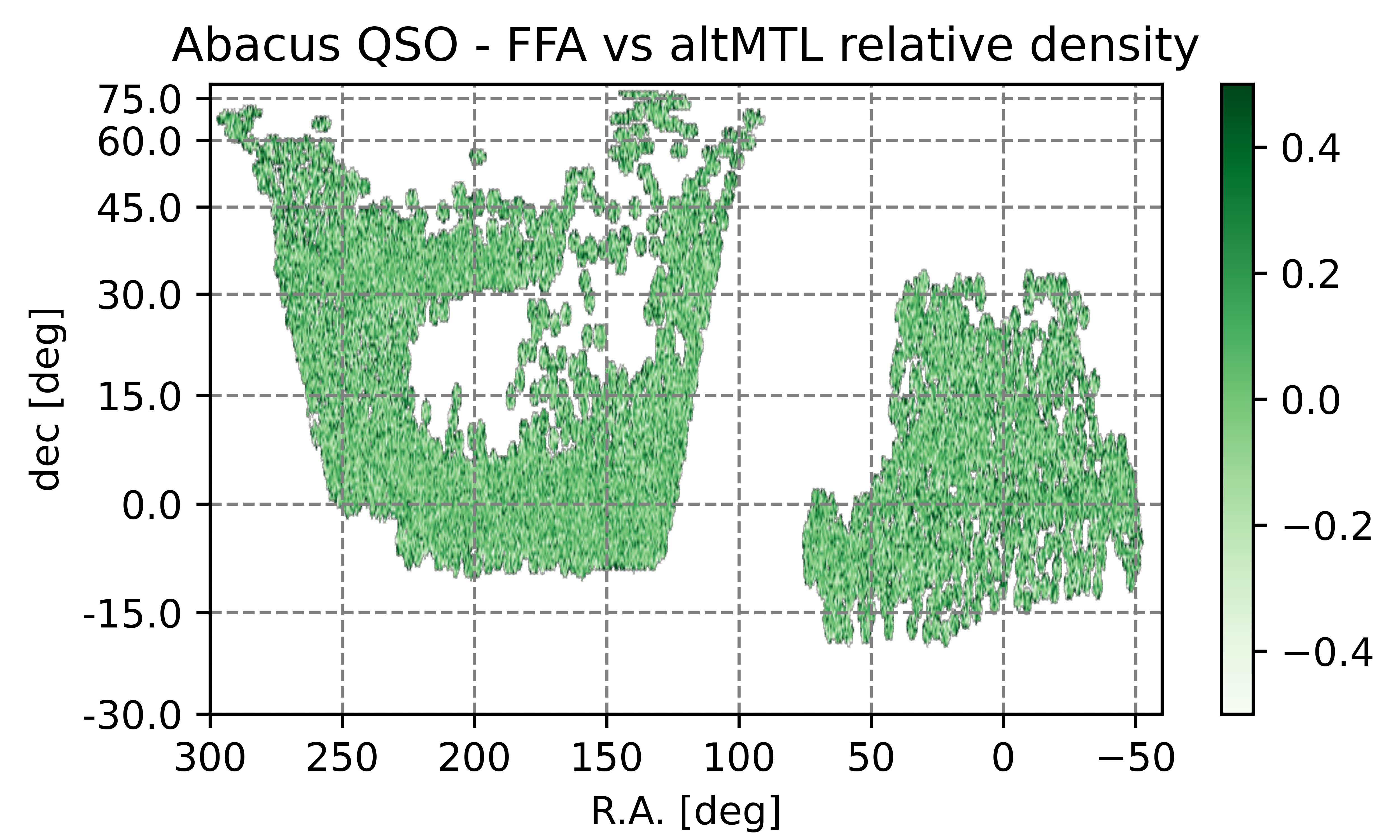}\\
    \caption{Left column: angular assignment completeness of the three dark-time tracers, ELGs, LRGs and QSOs, as labeled in the figure, measured from altMTL mocks. Right column: relative behaviour of FFA and altMTLs, $\rho_{\rm FFA} / \rho_{\rm altMTL} - 1$, where $\rho$ is the angular density of the assigned galaxies.}
    \label{fig:ang_2d_mocks}  
\end{figure}

In the left column of figure \ref{fig:ang_2d_mocks} we show the angular assignment completeness for the three dark-time tracers, ELGs, LRGs and QSOs, obtained by applying the altMTL approach.
The angular patterns, mostly driven by the tile coverage, are fully compatible with those of the data, shown in figure \ref{fig:ang_2d_data}.
In the right column of figure \ref{fig:ang_2d_mocks} we present the relative angular density, definded as 
$\rho_{\rm FFA} / \rho_{\rm altMTL} - 1$, of the FFA and altMTL samples of assigned galaxies, where $\rho_{\rm FFA}$ and $\rho_{\rm altMTL}$ are the respective densities.
This kind of comparisons can be interpreted as validation tests for FFA against the, more realistic, altMTL method, but, of course, finding good agreement between the two also reinforces our confidence in altMTLs.
As shown in the figure, the angular distributions are compatible among themselves, with the less complete regions revealing larger statistical fluctuations but no obvious signs of systematic trends.
For reference, the explicit values of the total number of objects in each catalogue pre and post fiber assignment is reported in table \ref{tab:n_obj}.
\begin{table}[htbp]
\centering
\begin{tabular}{rc|ccc}
\hline
 & & altMTL & FFA & Complete\\
\hline
ELG & n. of galaxies & $2.41 \times 10^6$ & $2.44 \times 10^6$ & $12.09 \times 10^6$\\
& fraction assigned & $0.20$ & 0.20 & 1 \\
LRG & n. of galaxies & $2.16 \times 10^6$ & $2.15 \times 10^6$ & $3.88 \times 10^6$\\
& fraction assigned & $0.56$ & 0.56 & 1 \\
QSO & n. of galaxies &$0.86 \times 10^6$ & $0.86 \times 10^6$ & $1.49 \times 10^6$\\
 & fraction assigned & $0.58$ & 0.58 & 1 \\
\hline
\end{tabular}
\caption{Total number of galaxies measured from the Abacus mocks pre and post fiber assignment with both the altLTM and FFA methods. We also report the fraction of assigned galaxies, which slightly differs from the definition of completeness used, e.g., for figure \ref{fig:ang_2d_data}, as here we show the ratio between clustering and complete sample. Values for all the three dark-time tracers, ELGs, LRGs and QSOs, are reported.}
\label{tab:n_obj}
\end{table}

\begin{figure}[htbp]\label{fig:z_dist}
    \centering
    \includegraphics[width=0.32\linewidth]{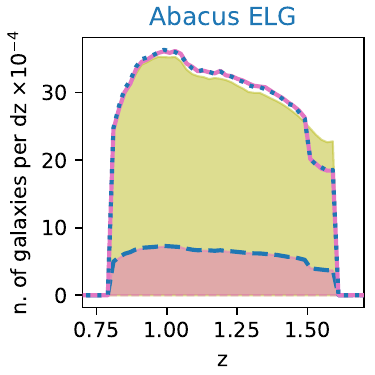}
    \includegraphics[width=0.32\linewidth]{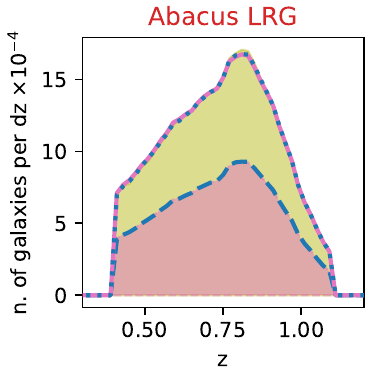}
    \includegraphics[width=0.32\linewidth]{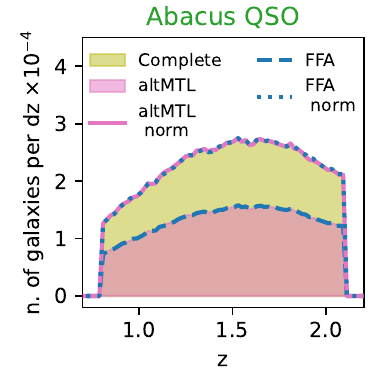}
    \caption{Comparison of the different redshift distribution ($dz = 0.02$) measured from the mocks without fiber assignment (complete sample, green shaded area) and with the two simulated assignment methods considered in this work: altMTLs (pink shaded area), FFA (dashed blue curve).
    The solid pink and dotted blue curves show the same altMTLs and FFA distributions rescaled by factor $N_{\rm complete}/N_{\rm assigned}$ to match the amplitude of the complete sample, where $N_{\rm complete}$ and $N_{\rm assigned}$ are the total number of objects in the complete and fiber-assigned samples, respectively. All three dark tracers are considered, as labeled in the figure.} 
    \label{fig:z_dist_mocks}
\end{figure}

The redshift distribution of the different tracers is shown in figure \ref{fig:z_dist_mocks}.
The altMTL and FFA mocks produce consistent results, represented by the pink shaded area and the dashed blue curve, respectively.
As expected, due to fiber assignment, both distributions are significantly suppressed compared to the complete sample (green shaded area). 
However, when 
rescaled by a factor $N_{\rm complete}/N_{\rm assigned}$, where $N_{\rm complete}$ and $N_{\rm assigned}$ are the total number of objects in the complete and fiber-assigned samples, respectively, 
the correct amplitude is consistently recovered across all tracer types (solid pink and dotted blue).
This confirms that the FA process does not significantly alter the shape of redshift distribution of the targets.
The only noticeable exception is given by the ELGs at $z>1.5$.
This discrepancy is not a result of fiber assignment; rather, it arises from redshift failures, which we intentionally incorporate into the clustering mocks, after the assignment process.
Specifically, the drop in the ELG distribution reflects the fact that, in real data, our ability to accurately measure redshifts above $1.5$ is limited by sky-line contamination (see figure \ref{fig:zdist_data}).
This also explains the slight excess of power at lower redshifts in the rescaled ELG distributions, as it reflects the ``conservation'' of the total number of galaxies.
It is worth stressing that the $N_{\rm complete}/N_{\rm assigned}$ rescaling serves solely as a visualization tool.
At no stage of the analysis of data and mocks we adopt such a rescaling or assume it can recover the ``true" redshift distribution.
Instead, we rely on redshift-failure weights, $w_{\rm zfail}$, as described in \cite{DESI2024.II.KP3} and reference therein.
 
\begin{figure}[htbp]
    \centering
    \includegraphics[width=0.32\linewidth]{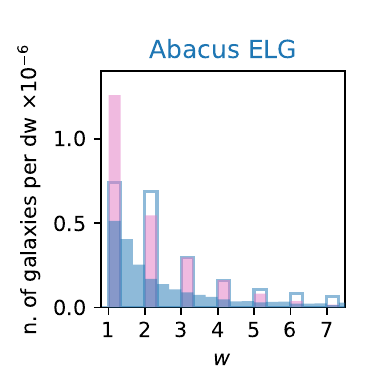}
    \includegraphics[width=0.32\linewidth]{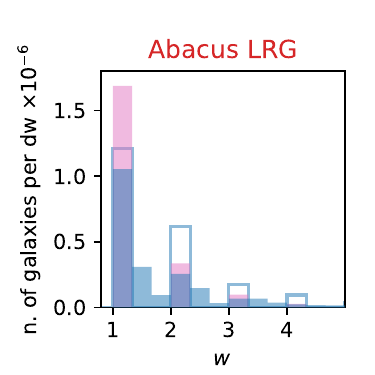}
    \includegraphics[width=0.32\linewidth]{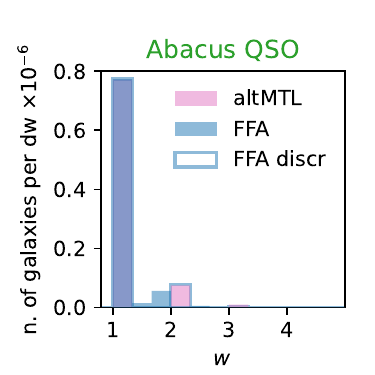}
    \caption{Histograms of the fiducial weights from the altMTL (blue shaded) and FFA (pink shaded) mocks ($dw=0.33$). The blue empty rectangles represent the same FFA weights rounded to the nearest integer. All the three dark-time tracers are shown, as labeled in the figure.}
    \label{fig:w_altMTL_FFA}
\end{figure}

Figure \ref{fig:w_altMTL_FFA} displays the histograms of the fiducial weights obtained through the altMTL and FFA approaches, represented by the pink shaded and blue shaded rectangles, respectively.
As explained in detail in section \ref{sec:estimator_level}, the two weight definitions are similar but formally distinct: the fiducial altMTL weights are derived from completeness considerations, while the fiducial FFA weights correspond to inverse probabilities, $w_{IIP}$.
In this section, we are not evaluating the performance of these weights as mitigation strategies; rather, we are examining how well the approximate FFA mocks replicate the key features of the more realistic altMTL mocks, including the behaviour of the corresponding default weights.
For this reason, we refer to both of them as ``fiducial".

Unlike the fiducial altMTL weights, which are obtained by counting the number of targets competing for the same fiber (see section \ref{sec:estimator_level} for the explicit definition), the FFA weights are not inherently constrained to take discrete integer values. Therefore, to facilitate comparison between the two, we include in the figure an illustrative histogram that represents the FFA weights rounded to the nearest integer (blue empty rectangles).
Overall, we observe a satisfactory agreement, which aligns with what can be reasonably expected considering the two different definitions.
More specifically, there is a general trend, more pronounced in the ELG case, for altMTLs to exhibit a shorter high-weight tail and a significantly larger proportion of weights equal to one.
The high-$w$ tail might influence the noise properties of the clustering statistics.
However, as the weight increases, the number of galaxies drops rapidly, making it difficult to anticipate whether the net effect will be significant.

\begin{figure}[htbp]
    \centering
    \includegraphics[width=0.32\linewidth]{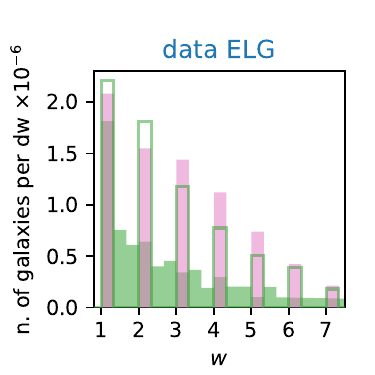}
    \includegraphics[width=0.32\linewidth]{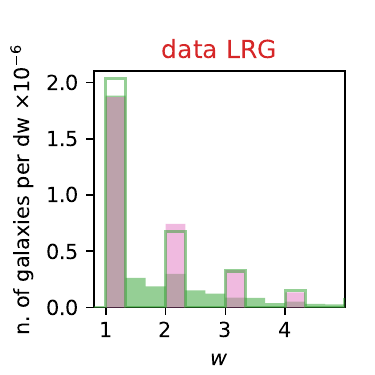}
    \includegraphics[width=0.32\linewidth]{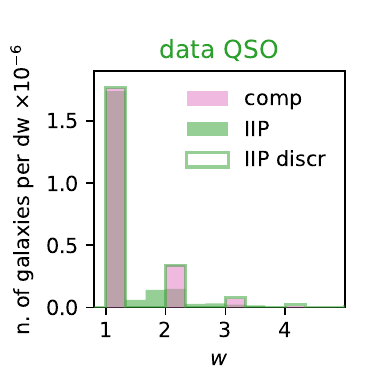}
    \caption{Histograms of fiducial (pink shaded rectangles) and IIP (green shaded rectangles) weights from the data ($dw=0.33$). The green empty rectangles represent the same IIP weights rounded to the nearest integer. All the three dark-time tracers are shown, as labeled in the figure.}
    \label{fig:comp_vs_IIP_aMTL}
\end{figure}

These discrepancies between altMTL and FFA weights may also arise from variations in how fiber assignment is implemented.
To separate the two potential contributions, we can leverage the fact that, for the data (and altMTL mocks), both weight definitions are available simultaneously.
Figure~\ref{fig:comp_vs_IIP_aMTL} shows the two histograms (shaded pink and green rectangles), together with the rounded version of $w_{IIP}$ (empty green rectangles).
The altMTL mocks yield almost indistinguishable results, not shown here for brevity, further reinforcing the reliability of the method.
The comparison is not entirely consistent, since, for practical reasons, these plots focus on the full sample of potential targets instead of that of the observed ones used for figure~\ref{fig:w_altMTL_FFA}.
Nevertheless, as illustrated in the figure, the two sets of weights show closer agreement when the fiber assignment strategy is fixed, suggesting that the main source of the discrepancies in figure \ref{fig:w_altMTL_FFA} is the assignment procedure itself, FFA versus altMTLs, at least for the $w=1$ peak.

\begin{figure}
    \centering
    \includegraphics[width=0.32\linewidth]{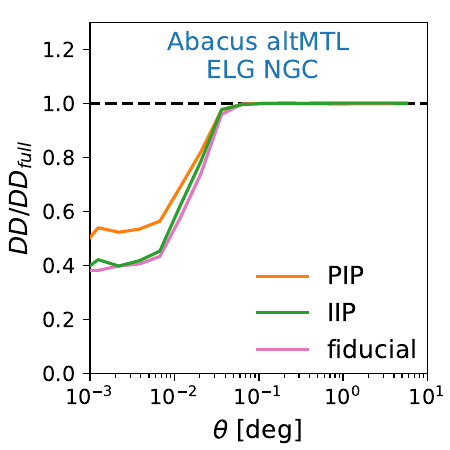}
    \includegraphics[width=0.32\linewidth]{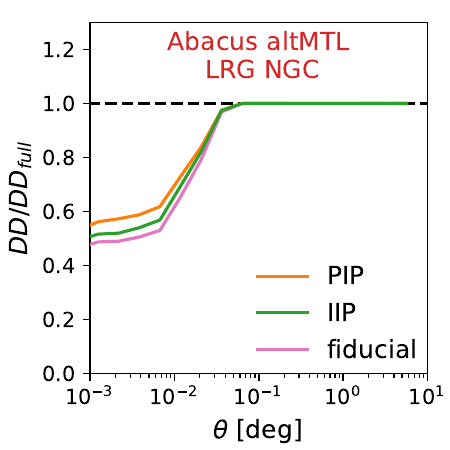}
    \includegraphics[width=0.32\linewidth]{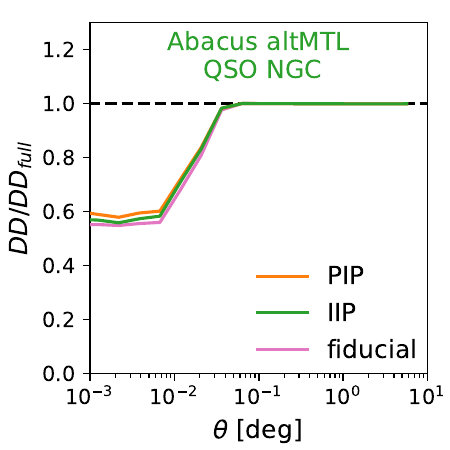}
    \caption{Normalised pair counts measured from altMTL mocks, as a function of the separation angle~$\theta$, with fiducial weights (solid pink), IIP weights (solid green) and PIP weights (solid orange), divided by the normalised pair counts from the full sample of potential targets. All three dark tracers are shown, as labeled in the figure.}
    \label{fig:ang_pair_counts_aMTL}
\end{figure}

\begin{figure}
    \centering
    \includegraphics[width=0.32\linewidth]{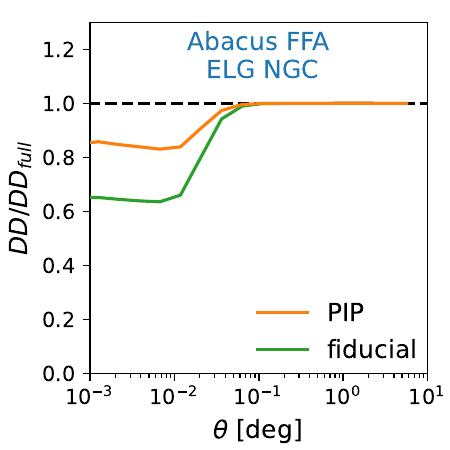}
    \includegraphics[width=0.32\linewidth]{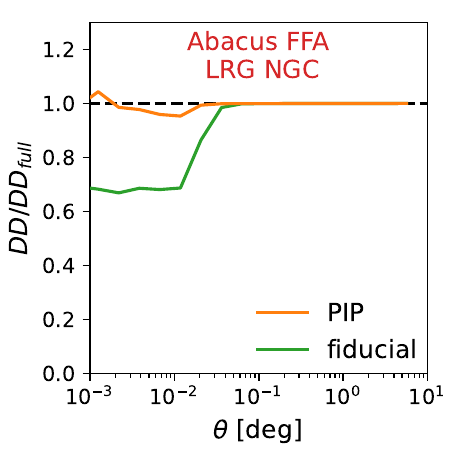}
    \includegraphics[width=0.32\linewidth]{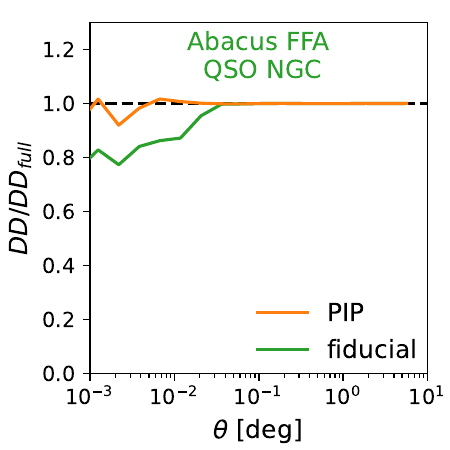}
    \caption{Same as figure \ref{fig:ang_pair_counts_aMTL} but for FFA mocks. For these mocks, by design, fiducial weights coincide with IIP weights (green solid). In contrast to altMTLs, for FFA we do not have access to a proper full sample, so we use the complete sample instead.}
    \label{fig:ang_pair_counts_FFA}
\end{figure}

Figures \ref{fig:ang_pair_counts_aMTL} and 
\ref{fig:ang_pair_counts_FFA}
show the collision windows extracted from the altMTL and FFA mocks, respectively.
By focusing on the fiducial weight curves, solid pink and solid green, repectively, we observe that while altMTLs yield results fully compatible with those from actual data (figure \ref{fig:ang_pair_counts_data}), FFA mocks display a less pronounced collision window, characterized by a slightly steeper transition and reduced overall depth, particularly for the QSOs.
It is nonetheless noteworthy the overall qualitative agreement, in particular for the angular size of the window, given the radical approximations involved.

\section{Estimator-level mitigation techniques}\label{sec:estimator_level}

In this section we focus on how to counteract the effect of fiber assignment at the estimator level, but it is worth mentioning that some of the techniques described here, namely $w_{comp}$ (or $w_{IIP}$), are required even when addressing the issue at the model level (see section \ref{sec:model_level}).
By {\it estimator-level} corrections we mean any weighting scheme applied to either individual galaxies or pairs of galaxies (or randoms) at the stage of measuring the clustering statistics of interest.
Such schemes are designed to statistically recover the ``true" clustering power of all the galaxies in the DESI volume, thus allowing us to ignore the effect of fiber assignment when modelling the statistic of interest for cosmological inference, except, of course, for the covariance part.   

We consider four different estimator-level corrections, 
\begin{enumerate}
    \item
    Completeness
    \item
    Individual inverse probability
    \item 
    Pairwise inverse probability
    \item
    Pairwise inverse probability with angular upweighting
\end{enumerate}
detailed in sections \ref{sec:wcomp}, \ref{sec:invprob}.
In appendix \ref{sec:collwind} we also discuss a fifth approach based on learning the collision window from the mocks and apply it to the data.

All configuration-space measurements in this work are obtained through the \cite{LandySzalay} estimator,
\begin{equation}\label{eq:xi_estimator}
  \xi({\bf s}) = \frac{DD({\bf s})}{RR({\bf s})} - 2 \frac{DR({{\bf s}})}{RR({\bf s})} + 1 \ ,
\end{equation}
where $DD$, $RR$ and $DR$ stand for data-data,
random-random and data-random (normalised) pairs counts, respectively, and {\bf s} is the separation. 
Fourier space measurements are obtained through the fast-Fourier-transform (FFT) implementation \cite{bianchi2015b, scoccimarro2015, Hand2017:1712.05834v1} of the \cite{yamamoto2006} estimator,
\begin{equation}\label{eq:P_estimator}
P_\ell({ k})= \ \frac{(2\ell+1)}{I}\int \frac{d\Omega_k}{4\pi}\, \left[ \int d^3 r_1\,\int d^3 r_2\,F({\bf r}_1)F({\bf r}_2) e^{i{\bf k}\cdot({\bf r}_1-{\bf r}_2)}\mathcal{L}_\ell(\hat{\bf k}\cdot\hat{\bf \eta})-S_\ell({\bf k})\right] \ ,
\end{equation}
where $\hat{\bf \eta} = \hat{\bf \eta}({\bf r}_1, {\bf r}_2)$ is the line of sight (LOS) of the pair formed by galaxies $1$  and $2$, $d\Omega_k$ the solid-angle element in $k$-space, $\mathcal{L}_\ell$ the $\ell-${th} order Legendre polynomial, $S_\ell$ the shot noise term (which is non-zero only for $\ell=0$) and $I$
the normalisation factor.
The density fluctuation is defined as $F ({\bf r})= n({\bf r}) - \alpha n_s({\bf r})$, where $n$ and $n_s$ are, respectively, the weighted number density of galaxies and randoms, the latter being $1/\alpha$ times more abundant.
The normalization is obtained as $I = \alpha \sum n n_s / dV$, with the sum performed over grid cells of fixed size $dV^{1/3} = 10 h^{-1} \rm Mpc$, as described in \cite{DESI2024.II.KP3}.
The shot noise term is given by $S_0 = \left( \sum n^2 + \alpha^2 \sum n_{s}^2 \right) / I$, where the sums are performed over galaxies and randoms.

\subsection{Completeness weights}\label{sec:wcomp}

This approach, originally developed for the SDSS survey \cite{bosslsscat}, relies on two separate sets of weights, described in detail in \cite{DESI2024.II.KP3, KP3s15-Ross}.
In brief,  we assign a completeness value to each target, $f_{\rm TLID}$, evaluated as the inverse of the total number of targets of the same class at the given tile and fiber,
and we define the corresponding completeness weight as:
\begin{equation}
w_{\rm comp} = 1/f_{\rm TLID}.
\label{eq:wcomp}
\end{equation}
$f_{\rm TLID}$ tells us how many targets were competing for the fiber and, consequently, represent a direct estimate of the assignment probability, in the case of a uniformly randomized process.
Being derived directly from galaxy counts, $w_{comp}$ is, by definition, an integer value, generally below 10, as shown in figures \ref{fig:w_altMTL_FFA} and \ref{fig:comp_vs_IIP_aMTL}.
$f_{\rm TLID}$ alone does not capture all incompleteness effects, such as, e.g., fibers assigned to standard stars or sky in order to meet a minimum threshold, or low-priority ELGs having no chance of being assigned in the proximity of an LRG (exclusion regions induced by QSO are already included in the veto mask).
These processes ultimately depend on random choices, made by the targeting algorithm, whose impact on the completeness is expected to be fairly uniform within any given group of overlapping tiles. 
For each of such groups we therefore introduce an additional completeness fraction, defined as
\begin{equation}
    f_{\rm tile} = \frac{N_{\rm assigned}+N_{\rm associated}}{N_{\rm tot}},
\end{equation}
where $N_{\rm tot}$ is the total number of targets of a given type and $N_{\rm associated}$ the number of not assigned targets at combinations of tile and fiber that got assigned to that target type. 
The numerator corresponds to the sum of the $w_{comp}$ weights over the assigned galaxies, thus if we divide each weight by $f_{\rm tile}$ their sum becomes $N_{\rm tot}$, by construction.
In practice, rather than introducing an additional weight $1/f_{\rm tile}$ for the data, we use $f_{\rm tile}$ as a weight for the randoms.
The procedure is well defined, as the same tile groupings exist in both the data and random samples, and it serves as an effective way to mitigate noise by utilizing a large number of tracers, since the number of randoms can be made much larger.
Throughout the manuscript we frequently refer to this weighting scheme as ``fiducial weights".

\subsection{Inverse probability weights}\label{sec:invprob}
Introduced by \cite{2017Bianchi} and further developed by \cite{bianchi2020}, inverse-probability 
(IP) weights are designed to counteract the effect of missing observations in an exact way.
Specifically, it was shown that unbiased $n$-point clustering measurements can be obtained by upweighting each galaxy $n$-plet by the inverse of its probability of being observed, provided there are no zero-probability $n$-plets.

In the case of 2-point statistics the appropriate inverse-probability correction is therefore pairwise in nature: each galaxy-pair requires its own specific weight, not necessarily expressible as the product of two individual weights.
We refer to this weighting scheme as pairwise inverse probability (PIP) and denote the corresponding pair weights as $w_{PIP}$.

However, if we can find some characteristic angular separation $\theta_{ind}$ above which the probability of a galaxy of being observed is (approximately) independent on whether its neighbours have been observed or not, for $\theta > \theta_{ind}$ we can safely use individual weights.
We refer to this weighting scheme as individual inverse probability (IIP) and denote the corresponding galaxy weights as $w_{IIP}$.

In summary, for both PIP and IIP, the correlation function is obtained by substituting in eq. (\ref{eq:xi_estimator}),
\begin{equation}\label{eq:DD_PIP}
    DD({\bf s}) = \sum_{{\bf x}_m - {\bf x}_n \approx {\bf s}} w_{mn} \ ,
\end{equation}
but the weights can be factorised as $w_{mn} = w^{IIP}_m  w^{IIP}_n$ only under the IIP assumption.
For the, more general, PIP scheme each pair weight is obtained via the bitwise weights introduced by \cite{2017Bianchi} and implemented for the DESI survey by \cite{KP3s7-Lasker}.
The symbol $\sum_{{\bf x}_m - {\bf x}_n \approx {\bf s}}$ is shorthand notation for a binned sum. 
$DR$ and $RR$ are always obtained using individual weights. 

The two approaches can be combined.
As a way to reduce the computational complexity of the pure PIP approach in Fourier space, \cite{bianchi2020} proposed an hybrid IIP-PIP procedure.
If $\theta_{ind}$ exists, this strategy yields results identical to those obtained by solely using PIP weights.
Therefore, we still categorize it as a PIP approach, for simplicity.
In this scenario, the power-spectrum (multipole) estimator becomes
\begin{align}\label{eq:P_PIP}
P^{\rm PIP}_\ell(k) &= P^{\rm IIP}_\ell(k) + \frac{(2\ell+1)}{I} \int \frac{d\Omega_k}{4\pi} \sum_{ij} A_{ij} e^{i{\bf k}\cdot({\bf r}_i - {\bf r}_j)} \mathcal{L}_\ell(\hat{\bf k}\cdot\hat{\bf \eta}_{ij}) \nonumber \\
&= P^{\rm IIP}_\ell(k)+ {(-i)}^\ell \frac{(2\ell+1)}{I} \sum_{ij} A_{ij} j_\ell(ks_{ij}) \mathcal{L}_\ell(\hat{\bf s}_{ij}\cdot\hat{\bf \eta}_{ij}) \ ,
\end{align}
where $P^{IIP}_\ell$ is obtained by supplementing eq. (\ref{eq:P_estimator}) with IIP weights, $j_\ell$ are spherical Bessel functions of the first kind and
\begin{equation}\label{eq:PIPco}
    A_{ij} = w^{PIP}_{ij} - w^{IIP}_i w^{IIP}_j \ .
\end{equation}
The selection probability of each object (galaxy or pair) and, most importantly, its inverse $1/p$, can be evaluated by running the targeting algorithm $K$ times and counting how many times, $c$, it gets selected (we follow the \cite{bianchi2020} notation and refer to $c$ as {\it recurrence}).
For a detailed description of how the procedure is implemented for DESI see \cite{KP3s7-Lasker}.
There is more than one, formally correct, way of combining recurrence and number of realisations into an inverse-probability estimator, which returns the true value of $1/p$ for $K \rightarrow \infty$. 
Given the limited amount of computational resources available, it is of crucial importance to use an estimator that converges the fastest possible with $K$.
The issue was studied by \cite{bianchi2020}, who recommended adopting the form
\begin{equation}
w^{eff} = \frac{K+1}{c+1} \ ,
\end{equation}
and named it {\it efficient} estimator\footnote{
In this scenario $K$ represents the number of replicas, the true realisation is not included in the count.}.
The authors also derived the exact distribution of the recurrence of a pair given the recurrences of the two individual galaxies, $i$ and $j$, when the latter are selected independently,
\begin{equation}\label{eq:dist_ind_vs_pair}
\mathcal{P}_K(c_{ij} | c_i, c_j) = \binom{c_i}{c_{ij}} \binom{K - c_i}{c_j -c_{ij}} {\binom{K}{c_j}}^{-1}  \ ,
\end{equation}
where $0 \le c_{ij} \le \min(c_i,c_j)$.
This allowed them to compute the expectation value,
\begin{equation}\label{eq:effpc}
g_{K}(c_i,c_j) =  \sum_{c_{ij}=0}^{\min(c_i,c_j)} \frac{K+1}{c_{ij}+1} \mathcal{P}_K(c_{ij} | c_i, c_j) \ ,
\end{equation}
of $w_{ij}^{eff}$, which they then used to obtain an accurate pair-count normalisation, free from unwanted sampling artefacts.  

For DESI we essentially follow the same strategy, but with an important refinement: rather than using $g_K$ for the normalisation, we include it in the pair-counting core.
Specifically, our definition of PIP weights becomes
\begin{equation}\label{eq:w_pip}
w^{PIP}_{ij} = w^{eff}_i w^{eff}_j  \frac{w^{eff}_{ij}}{g_K(c_i,c_j)} \ , 
\end{equation}
unless otherwise stated\footnote{
IIP weights are simply defined as $w_i^{IIP} = w_i^{eff}$, unless otherwise stated.}.
In other words, we factorise the weight into three terms: two purely individual, $w^{eff}_i$, and one purely pairwise, $w^{eff}_{ij}/ g_K(c_i,c_j)$, which incorporates the selection correlation.
The advantage over doing it at the normalisation level is that, with this pair-by-pair strategy, any residual scale dependent effect induced by, e.g., the clustering of low probability objects\footnote{
By low probability we mean probability roughly of the order of ten divided by the number of realisations.}, is removed.
Individual and pairwise recurrences $c$ are obtained, as usual, from the bitweights as ${\tt popcnt}(w^{\rm bit}_i)$ and ${\tt popcnt}(w^{\rm bit}_i {\tt \&} \ w^{\rm bit}_j)$, respectively, where ${\tt popcnt}$ is the standard bit-counting function and ${\tt \&}$ the bitwise ``and" operator.
$g_K$ is a universal function of the two integer variables $c_i$ and $c_j$, parameterized by $K$.
It only needs to be evaluated $K(K+1)/2$ times, corresponding to all possible (unordered) combinations of $c_i$ and $c_j$.
Once these values are stored, they can be used in the pair counts without additional computational expense.

\begin{figure}[htbp]
    \centering
    \includegraphics[width=0.57\linewidth]{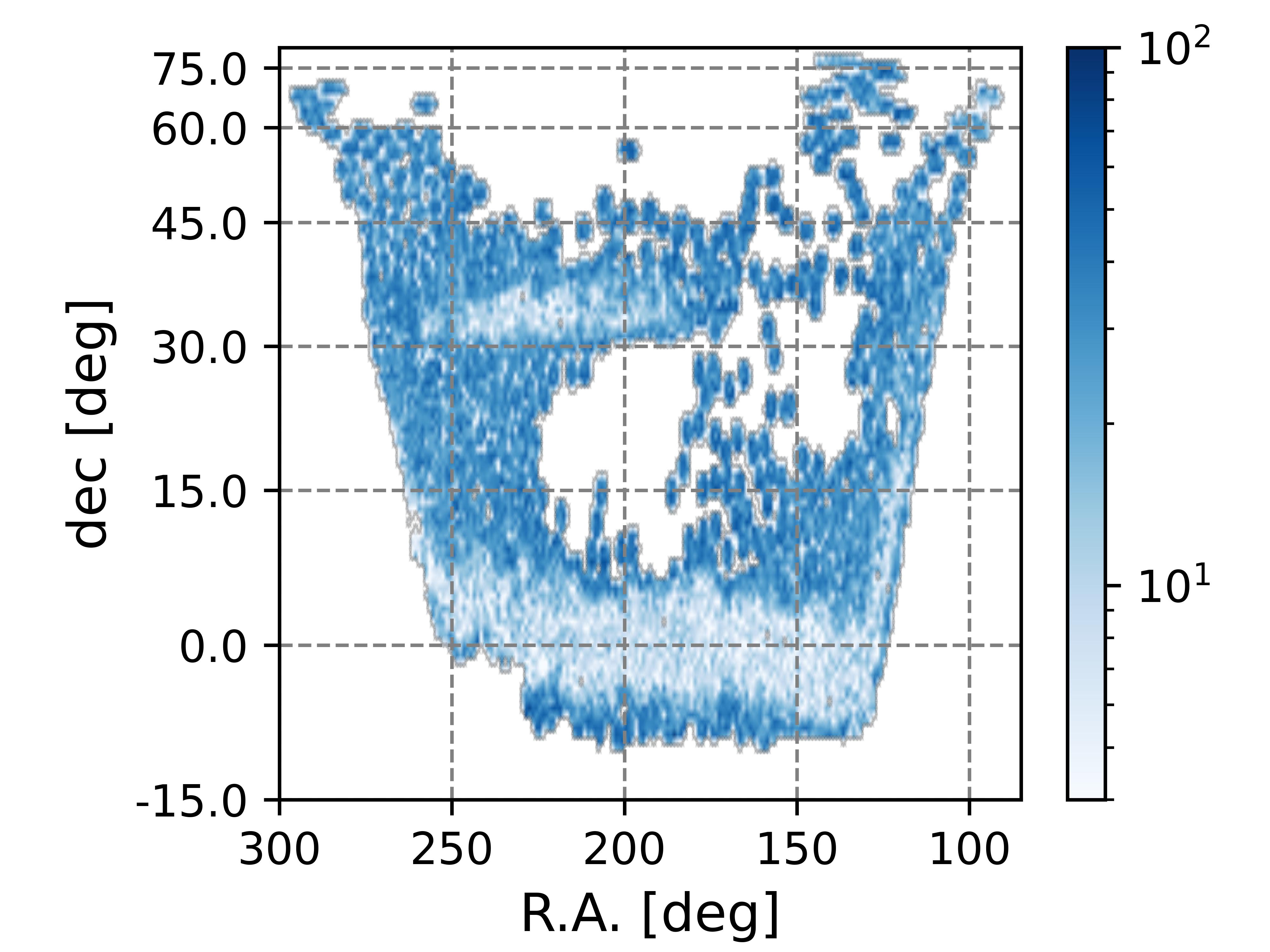}
    \includegraphics[width=0.42\linewidth]{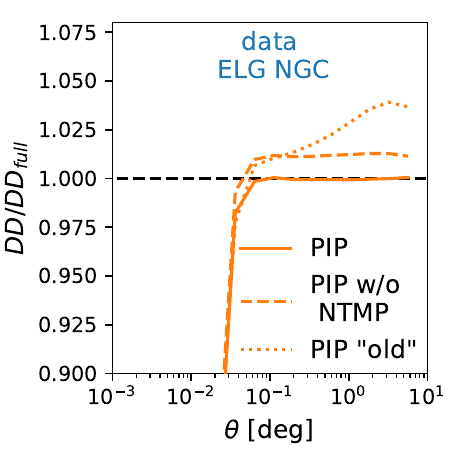}
    \caption{Left panel: angular number density of zero-probability ELGs in the north galactic cap (real data). Right panel: normalised pair counts as a function of the separation angle~$\theta$ obtained using PIP weights with and without the sampling correction described in eq. (\ref{eq:w_pip}), solid and dotted orange curves, respectively, and without $w_{NTMP}$, orange dashed, all divided by the normalised pair counts from the full sample of potential targets. Please note the different scales on the $y$ axes compared to the other pair count plots in the paper.}
    \label{fig:zeroprob_galaxies}
\end{figure}

As illustrated in figure \ref{fig:zeroprob_galaxies}, sampling artefacts become particularly significant for highly incomplete samples, such as the ELG catalogs.
The right panel of the figure compares normalized angular pair counts obtained with (solid) and without (dotted) the new definition introduced in eq. (\ref{eq:w_pip}).
The combined impact of low assignment probability and a limited number of realizations ($K=128$) causes the dotted curve to diverge significantly from the correct value, even on large angles.
If not corrected, the $5\%$ systematic effect in the figure (note the different ordinate scale compared to the previous pair count plots) would result in an arbitrarily large error on 2-point measurements when $DD/RR \rightarrow 1$, i.e. on large scale.
In contrast, the solid curve converges nicely to 1 for separations larger than the collision window (but see section \ref{sec:NTMP} below).
Higher probability samples like LRGs and QSOs (not shown in the figure) are significantly less affected by these sampling artefacts.

The effectiveness of the new definition of PIP weights can be explained as follows.
Based on mathematical considerations, \cite{bianchi2020} showed that the ratio of the estimated to the true value of the inverse probability is roughly characterized by two distinct regimes: for $p_{true} \gtrsim 10/K$ it remains virtually 1, whereas for smaller $p_{true}$ values, it decreases linearly with $p_{true}$.
If a large number of pairs belong to the second regime, any scale-dependent trend in the ``true'' probability leaves an imprint on the pair counts.
Such a trend naturally arise from the entanglement between clustering and probability and from the uneven distribution of $n_{tile}$, which mostly affect the large-scale behavior\footnote{Since the pair counts in figure \ref{fig:zeroprob_galaxies}, along with those in all other plots in this paper, are normalized, a scale-dependent suppression can appear as an enhancement over certain scale ranges.}.
Broadly speaking, eq. (\ref{eq:w_pip}) allows us to estimate the inverse probability of each pair with the accuracy we would have if its probability was that of the individual galaxies, which is typically one order of magnitude higher, thus effectively moving it from the biased to the unbiased regime.

\subsubsection{Inverse probability with angular upweighting}

By construction PIP weights return (on average) the ``true'' clustering of the pairs with non-zero probability in the parent sample.
The presence of zero-probability pairs manifests itself as power suppression on the scales where such pairs are more abundant, typically those corresponding to small separations. 
When it occurs, it is always possible to compensate through angular upweighting~\cite{2017Bianchi,percival2017}.
This comes with the implicit assumption that the clustering of the zero-probability pairs can be represented by the clustering, on the same angular scale, of the observed pairs.
Despite this being not necessarily true in general, it is in practice a valid assumption in most realistic scenarios.
Specifically, for DESI the main sources of zero-probabilty pairs are 1-pass regions and priority conflicts.
The fact itself that PIP weights alone do not fully correct for the collision window is direct proof of the presence of these pairs in real DESI data (figure~\ref{fig:ang_pair_counts_data}), altMTLs (figure~\ref{fig:ang_pair_counts_aMTL}) and, to a lesser extent, FFA mocks (figure~\ref{fig:ang_pair_counts_FFA}).

Angular upweighting operates by including an angular-separation dependent factor, $a_{DD}(\theta)$, in the evaluation of the galaxy pair counts of a clustering sample, such that eq. (\ref{eq:DD_PIP}) becomes
\begin{equation}
    DD({\bf s}) = \sum_{{\bf x}_m - {\bf x}_n \approx {\bf s}} a_{DD}(\theta_{mn}) \ w_{mn} \ ,
\end{equation}
and similarly for the galaxy-random pairs.
Typically, $a_{DD}$ takes the form of a ratio between angular pair counts, $DD_{\rm ref}/DD_{\rm dep}$, with the underlying idea that some reference sample, at the numerator, can be used to restore the missing power of a depleted sample, at the denominator.   

In the approach originally proposed by \cite{2017Bianchi} and implemented in the various works that followed, e.g. \cite{mohammad2020}, $DD_{\rm dep}$ and $DD_{\rm ref}$ were obtained, respectively, from the clustering sample itself and from the equivalent of our full sample, but restricted to the same redshift range of the clustering sample.  
In this work we take a different approach: both pair counts are obtained from the full sample, with no redshift cut. Specifically, for each tracer type, we define $DD_{\rm ref} = DD^{full}$ as the total count of all targets and $DD_{\rm dep} = DD^{full}_{PIP}$ as the PIP-weighted count of the targets that got assigned, regardless of whether they eventually obtained a redshift.   
In this approach, the upweighting factor represents the fraction of missing pairs caused solely by fiber assignment, making it unaffected by photometric redshift uncertainties or redshift failures.
Further details are reported in appendix \ref{app:AUW}.
Another distinction from previous works, e.g. \cite{KP3s7-Lasker}, is that we normalise the pair counts to the total number of pairs in the respective samples.
The advantage of using normalised pair counts over raw pair counts is mostly practical, as it allows us not to touch the normalisation factors\footnote{
This is desirable, e.g., if one wants to restrict angular upweighting to a limited range of separations, as a way to save computational resources. 
} of $\xi$ and, most importantly, of $P$.

We do in fact apply angular upweighting also to the power spectrum, for the first time.
Specifically, we modify eq. (\ref{eq:PIPco}) as follows,
\begin{equation}\label{eq:PIPcoAUW}
    A_{ij} = \frac{DD^{full}(\theta_{ij})}{DD^{full}_{PIP}(\theta_{ij})} \ w^{PIP}_{ij} - w^{IIP}_i w^{IIP}_j \ ,
\end{equation}
with $w^{PIP}_{ij}$ defined in eq. (\ref{eq:w_pip}).
A similar correction could in principle be applied to the data-random pairs; however, we found its effect to be negligible.

\subsubsection{Counteracting the effect of zero-probability galaxies: $w_{NTMP}$}\label{sec:NTMP}

The left panel of figure \ref{fig:zeroprob_galaxies} shows the distribution across the north galactic cap of the ELG targets that, based on the information encoded into the bitweights, have zero probability of assignment (about $2.5 \times 10^5$ objects).
Different tracers and regions exhibit similar behavior, albeit to a lesser degree.
It is evident that these zero-probability galaxies are not randomly distributed; instead, they follow a pattern that, not surprisingly, resembles the completeness pattern (see, for example, figure \ref{fig:ang_2d_data}) and hence correlates with $n_{tile}$.
As already said, by definition inverse probability weights cannot counteract the effect of zero-probability objects or, more precisely, clustered zero-probability objects.
The impact of these ``unobservable" galaxies on the angular pair counts is shown in the right panel of figure \ref{fig:zeroprob_galaxies} by the dashed orange curve.
We observe a $\sim 1\%$ systematic effect, which, as mentioned earlier, would lead to a severe large-scale error in 2-point measurements if not appropriately countered.

For this reason we define $f_{NTMP}$ as the sum of the IIP weights, $w_{IIP}$, of all the fiber-assigned galaxies in the full sample with a given $n_{tile}$ value divided by the total number of galaxies in the full sample\footnote{As in the case of angular upweighting, the full sample includes galaxies beyond the redshift range of the clustering sample and those with failed redshift measurements.} with the same $n_{tile}$ value, where NTMP stands for ``$n_{tile}$ missing power''\footnote{We could have simply counted the number of zero probability objects instead of using $\sum w_{IIP}$.
The primary benefit of the latter approach is that it would also account for unidentified systematic effects, should they exist.}. 
Since the expectation value of $\sum w_{IIP}$ is, by construction, the number of galaxies with a non-zero probability of assignment, $1 -f_{NTMP}$ estimates the fraction of zero-probability galaxies with a given $n_{tile}$.
We can therefore use it to compensate for the pattern shown in figure~\ref{fig:zeroprob_galaxies} by either defining and additional weight for the galaxies, $w_{NTMP} = 1/f_{NTMP}$, or for the randoms, $w^{\rm randoms}_{NTMP} = f_{NTMP}$.
Unless otherwise stated, for the measurements presented in this work we opted for the latter option, simply because the larger number density of the randoms helps to reduce the noise. 

Please note that we have deliberately formalized the problem in a way that resembles the definition of $f_{tile}$ in section \ref{sec:wcomp} to emphasize the similarities between the two quantities: the pairs $(w_{comp}, f_{tile})$ and $(w_{IIP}, f_{NTMP})$ play analog roles, with the first element representing an estimate of the inverse probability of assignment and the second element compensating for what the first does not capture.
Just as with the $f_{tile}$ definition, $f_{NTMP}$ could also be constructed from tile groups, rather than $n_{tile}$.
We defer this investigation to future work. 

As discussed in the previous section, angular upweighting is another effective countermeasure for zero-probability galaxies.
For example, with angular uwpweighting the ratio of the angular pair counts, right panel of figure \ref{fig:zeroprob_galaxies}, would be identically equal to one across the entire separation range.
The primary role of $w_{NTMP}$ is to bypass the angular upweighting step on large scale, where the assignment probabilities are independent.
This is essential to enable the use of PIP weights in Fourier space, and desirable, though not critical, in configuration space.

\section{Model-level mitigation techniques}\label{sec:model_level}

We classify as {\it model level} any fiber mitigation technique applied at the stage of cosmological inference, i.e. when evaluating the theoretical value of the clustering statistics, as either a function of compressed variables or actual cosmological parameters.
Note, however, that, despite their name, neither the method presented here nor those proposed in the literature are entirely model level, but rather hybrids, as they require some auxiliary estimator-level action, such as weighting and applying scale cuts.

For DR1 different model level corrections were explored, including the method proposed by \cite{hahn2017} and its natural extension to achieve the accuracy required by DESI: measuring the shape of the collision window from mocks, rather than assuming a top-hat shape.
This approach is essentially the model-level counterpart to the estimator-level technique presented in appendix \ref{sec:collwind} and shares with it good performances but also the limitation of being inherently mock dependent.
For the full-shape analysis we therefore opted for the more the agnostic $\theta$-cut approach, presented in a dedicated paper \cite{KP3s5-Pinon}, and summarized below.

\subsection{Removing small angular separations ($\theta$-cut)}

Once established that $\theta_{ind}$  exists, it becomes natural to simply exclude the scales with $\theta < \theta_{ind}$ from the analysis.
Specifically, we can pass to the clustering estimators only the pairs with angular separation larger than some $\theta_{cut}$, with $\theta_{cut} 
\gtrsim \theta_{ind}$, and model the resulting correlations and power spectra accordingly.
This is a valid fiber-mitigation approach as long as the following three conditions hold.
First, $\theta_{cut}$ must be small enough such that only a small fraction
of the information that we want to extract gets lost.
Second, we need a robust, unbiased correction for fiber incompleteness on scales larger than $\theta_{cut}$, with the advantage that, on such scales, we can rely on individual weights, by definition.
Third, the cut must be computationally tractable when applied to both the clustering estimator and the corresponding model.

A thorough analysis of $\theta$-cut approach can be found in \cite{KP3s5-Pinon}, where the above requirements are discussed in detail.
In brief, figure \ref{fig:ang_pair_counts_data} shows that by adopting $\theta_{cut} = 0.05 \rm deg$ we can safely exclude the scales where the assignment is intrinsically pairwise in nature.
The standard 2-point clustering analysis \cite{DESI2024.V.KP5} extracts information from linear and quasi-linear modes, corresponding to $k \lesssim 0.2 h {\rm Mpc}^{-1}$ or, equivalently, separations around $30 h^{-1} \rm Mpc$ and above.
As a reference, at $z=1$, adopting a Plank-like cosmology for the redshift-to-distance relation, a $0.05 \rm deg$ angular separation translates into a transverse comoving separation of approximately $3 h^{-1} \rm Mpc$, which suggests that most of the quasi-linear information is indeed preserved by the $\theta$-cut.
This is confirmed by direct tests on mocks, through the comparison of the posterior of cosmological parameters with and without $\theta$-cut, as shown in the companion paper. 

The effectiveness of the individual weighs, either $w_{comp}$ and $w_{IIP}$, on scales larger than $\theta_{cut}$ is demonstrated, at the level of angular clustering, by figure \ref{fig:ang_pair_counts_data} and confirmed for three-dimensional clustering by the results presented in section \ref{sec:results}.
The theoretical motivation of why $w_{IIP}$ weights yield unbiased clustering is presented in \cite{bianchi2020} and can be extended to $w_{comp}$ if we adopt the inverse-probability interpretation discussed in section \ref{sec:estimator_level}.    

Making the clustering estimators insensitive to small angular separations is straightforward in configuration space but not so trivial for the power spectrum, due to the non-local nature of the Fourier transform.
We address the problem by adopting the pair-count based solution introduced by \cite{bianchi2020} and summarised above, eq. (\ref{eq:P_PIP}), with the difference that, in the case of pure pair removal, eq. (\ref{eq:PIPco}) simply becomes $A_{ij} = - w_i w_j$.
The procedure is applied to the randoms, as well.
Lastly, since the impact of $\theta$-cut on the measured clustering is purely geometrical, it can be modelled along the line of what is usually done for survey footprint and selection effects \cite{wilson2017,BeutlerWindow}, i.e. as an additional contribution to the window matrix.
Specificaly, the expectation value of the $\ell$-th Legendre multipoles of the correlation function reads
\begin{equation}
\left\langle \widehat{\xi}_\ell^{cut}(s) \right\rangle = \sum_{\ell^{\prime}} W_{\ell \ell^{\prime}}^{cut}(s) \xi_{\ell^{\prime}}(s) \ ,
\end{equation}
whereas for the power spectrum multipoles, including a wide-angle expansion \cite{castorina2018} encoded by the index $n$, we get
\begin{equation}
\left\langle \widehat{P}_\ell(k) \right\rangle = 4 \pi \int {k^{\prime}}^{2 - n} dk^{\prime} \sum_{n, \ell^{\prime}} W_{\ell \ell^{\prime}}^{cut, (n)}(k, k^{\prime}) P_{\ell^{\prime}}(k^{\prime}) \ .
\end{equation}
The explicit expressions for the window matrices, $W_{\ell \ell^{\prime}}^{cut}(s)$ and $W_{\ell \ell^{\prime}}^{cut, (n)}(k, k^{\prime})$, are provided in \cite{KP3s5-Pinon}, together with the description of the numerical algorithm used to compute them efficiently. 

One potential issue with $\theta$-cut
is given by the extent of the window tails in   Fourier space, which couples long and short modes.
The resulting window matrix is very non-diagonal.
This means that in order to model the $\theta$-cut power spectrum on the scales of interest for standard cosmological analyses $k\lesssim 0.2 h \rm{Mpc}^{-1}$, one should formally know the correct amplitude of the theory power spectrum on smaller scales, where the theory is not valid anymore.
This effect was found to be negligible for DESI, as shown in \cite{KP3s5-Pinon}.
Nevertheless, a general strategy to compactify the window so that the impact of high theory modes gets suppressed was developed and successfully implemented by \cite{KP3s5-Pinon}.

\section{Comparison of results}\label{sec:results}

In this section we present the clustering measurements from the Abacus mocks processed trough the altMTL and FFA pipelines and we compare them to actual data measurements from DESI DR1.   
The goal is twofold: assessing the relative effectiveness of the different fiber incompleteness countermeasures and testing the soundness of the two strategies we have implemented to simulate fiber assignment.    
We do not show clustering measurements obtained with the $\theta$-cut approach as they are already presented and extensively tested against the corresponding analytical model in \cite{KP3s5-Pinon}. 

Given the high computational demands of the \texttt{fiberassign} code, we could only afford altMTL bitweights for one Abacus mock.
We arbitrarily chose mock 11 and ran $K = 128$ alternative realisations of the targeting in addition to the original one and use them to construct the corresponding bitweights, with $N_{bits}=128$.
The same number of alternative realisations/bits was used for the data sample.
With FFA we do not have such limitations, we therefore built $K = 256$ realisations for each of the 25 Abacus mocks.
To ensure a fair comparison among the different approaches, in this section we present clustering measurements for mock 11 alone, including all three dark-time tracers, ELGs, LRGs and QSOs.
An example of measurements averaged over a larger set of FFA mocks can be found in appendix~\ref{app:FFA_average}.
For compactness' sake we only show clustering measurements for the North Galactic Cap (NGC).
The results for the South Galactic Cap (SGC) are fully compatible but exhibit a lower signal-to-noise ratio due to a combination of higher incompleteness and a smaller area, see figures~\ref{fig:ang_2d_data} and \ref{fig:ang_2d_mocks}.

We measure the correlation functions using both linear bins of $4 h^{-1} \rm Mpc$ and logarithmic bins with centers given by $s_n = 10^{x_0+n\Delta x}$, where $x_0=-0.894$ and $\Delta x = 0.211$.
In all the figures we show a combination of the two, with logarithm bins (and abscissa scale) for $s<20 h^{-1} \rm Mpc$ and linear bins (and abscissa scale) on larger separations, as a way to compress all the relevant information into a single plot.
The numbers of galaxies in the NGC footprint are $N_{\rm ELG} = 1.8 \times 10^6$, $N_{\rm LRG} = 1.5 \times 10^6$, $N_{\rm QSO} = 5.6 \times 10^5$.
For each tracer we have multiple random samples spanning the same volume, with $N_{ran} = 1.2 \times 10^7$ particles each, which we can combine in different ways to reach a convenient balance between computational speed and mitigation of the noise.
For all the $\xi$ measurements presented here the pair counts supplied to eq. (\ref{eq:xi_estimator}) are obtained by merging 4 random samples.
More precisely, we use the full merged sample only for $s < 20^{-1}\rm Mpc$, whereas for the larger separations we split it into 6 randomly selected subsets and perform individual pair counts, which we then combine with the appropriate normalisation factors in order to obtain consistent all-scale estimates~\cite{SplitRandom}.
For the $\xi$ estimates that rely on angular upweighting we compute angular pair counts in logarithmic bins with centers given by $\theta_n = 10^{x_0+n\Delta x}$, where $x_0=-3.7$ and $\Delta x = 0.66$.
The upweighting is performed for angular separations smaller than $\theta_{max} = 0.2 \rm deg$.
Thanks to $w_{NTMP}$ weights, the choice of $\theta_{max}$ does not influence our results, as far as it is larger than the angular size of the collision window\footnote{Formally, a larger $\theta_{max}$ would have beneficial effects on the statistical error, as show in \cite{percival2017}. However, here we focus on systematics effects, using angular upweighting to address the presence of zero-probability pairs, and a lower value of $\theta_{max}$ allows us to reduce the overall computational time.}. 

The power spectra are estimated as described in section \ref{sec:estimator_level}, using a grid of $1024^3$ nodes, painted with cloud-in-cell assignment (with the appropriate deconvolution in Fourier space,~\cite{Jing2005:astro-ph/0409240}).
For all tracers, we obtain the density contrast by merging 2 random samples and, when computing FFTs, we mitigate the aliasing by performing one deinterlacing step, \cite{Sefusatti2015:1512.07295}.
As usual, the full 3-dim power spectrum is eventually compressed into Legendre multipoles, with linear $k$ bins of size $0.005 h \rm Mpc^{-1}$. 
We always subtract the shot noise contribution, as described in section \ref{sec:estimator_level}. 
For the $P$ estimates that rely on angular upweighting we compute angular pair counts in logarithmic bins with centers given by $\theta_n = 10^{x_0+n\Delta x}$, where $x_0=-5.7$ and $\Delta x = 0.64$.
Both PIP correction and angular upweighting (eqs. \ref{eq:PIPco}, \ref{eq:PIPcoAUW}) are limited to angular separations smaller than $\theta_{max} = 0.25 \rm deg$.
Similarly to the correlation function case, $\theta_{max}$ does not influence our results, as long as it exceeds the angular size of the collision window.

As explained in \cite{DESI2024.II.KP3}, for purely technical reasons, we created two separate versions of the clustering catalogs containing the same galaxies: one for use with fiducial weights and the other for use with IP weights. For all the measurements presented below it is intended that the appropriate catalog was used, except for the FFA mocks, where separate catalogs were unnecessary.

Below we present results for the four different estimator-level corrections discussed in section~\ref{sec:estimator_level}: fiducial, IIP, PIP and PIP weights plus angular upweighting.
The clustering catalogs that we used to obtain them feature a {\tt WEIGHT} column that specifies the total weight to be applied to galaxies and randoms. This weight incorporates factors that are not related to fiber assignment, as imaging systematics and redshift failures \cite{DESI2024.II.KP3}.
All the clustering measurements in this section are obtained by reading $w_{\rm galaxies}$ and $w_{\rm randoms}$ directly from these {\tt WEIGHT} columns.
For the fiducial weights, no further action is needed.
Similarly, the IP dedicated catalogues already include IIP weights in the {\tt WEIGHT} column so that no further operation would be required to use them.
However, this catalogues were created before of the introduction of $w_{NTMP}$ weights, which are uniquely used in this work.
We therefore had to, first, compute them and, second, redefine the random weights as $w_{\rm randoms}' = w_{\rm randoms} \times f_{NTMP}$ (see section~\ref{sec:estimator_level}).
For PIP weights we follow the same procedure, except that we also need to remove the IIP contribution by defining $w_{\rm galaxies}' = w_{\rm galaxies} / w_{IIP}$.
The power that is removed through this operation is ultimately restored (and enhanced) by the bitweights, $w_{\rm bit}$, which we obtain directly from {\tt BITWEIGHTS} column.
When we use PIP weights with angular upweighting we also need the full catalogue and the correspondent random sample.
The weights for galaxies, observed galaxies and randoms, are all obtained from the {\tt WEIGHT\_NTILE} columns, which account for variation of $w_{tot}/w_{IIP}$ with $n_{tile}$ \cite{DESI2024.II.KP3}.
To ensure consistency, the angular pair counts of the observed galaxies must be computed using bitweights.
In addition, at variance with the strategy adopted for the 3-dim pair counts from the clustering sample, for the angular counts we incorporate the NTMP correction into the (observed) galaxy weights, so that $w_{\rm observed} = w_{ntile} / f_{NTMP}$ plus bitweights, as we do not perform random-random pair counts.
We intentionally chose not to use FKP weights \cite{FKP} in order to keep our studies centered on fiber assignment issues; however, all the estimators employed in this work are fully compatible with FKP weights.

\begin{figure}
    \centering
    \includegraphics[width=0.32\linewidth]{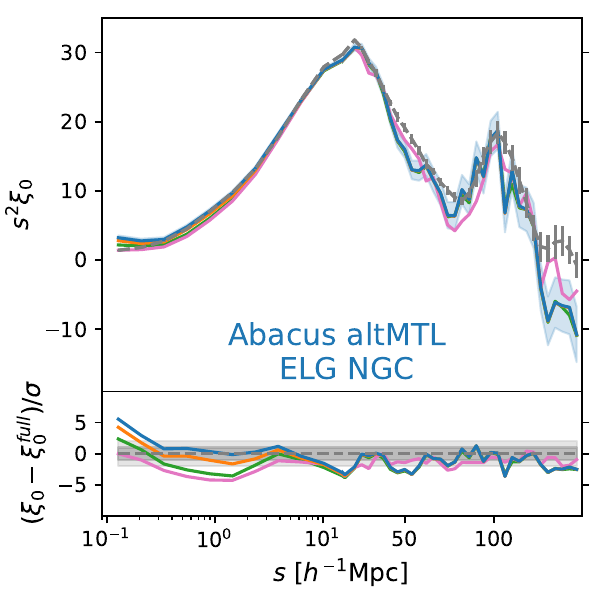}
    \includegraphics[width=0.32\linewidth]{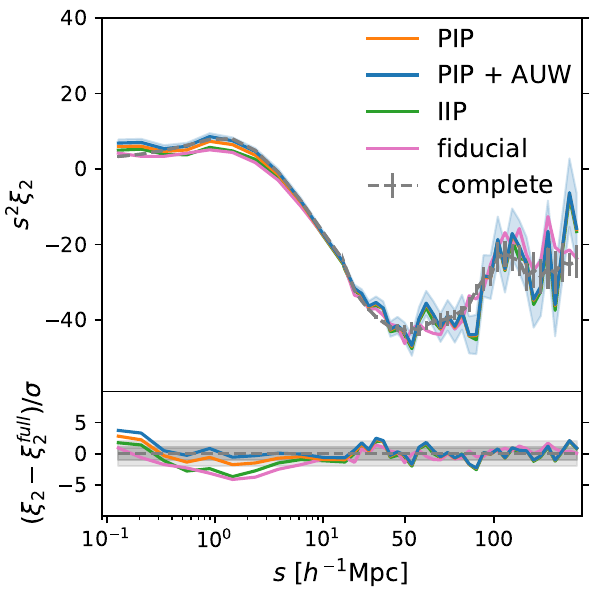}
    \includegraphics[width=0.32\linewidth]{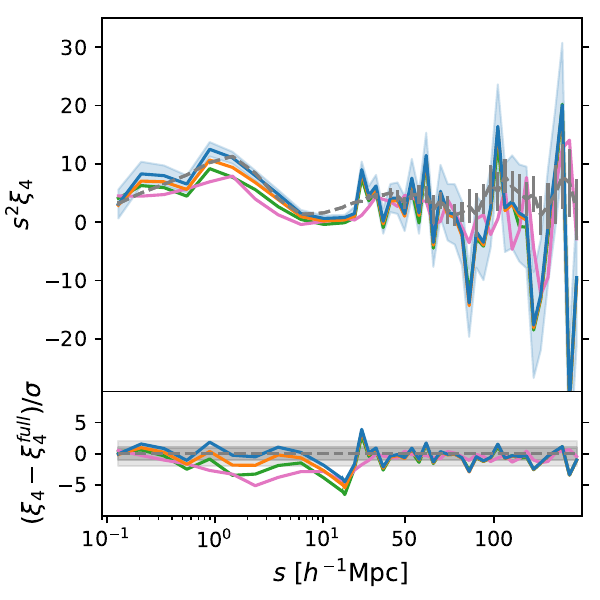} \\
    \includegraphics[width=0.32\linewidth]{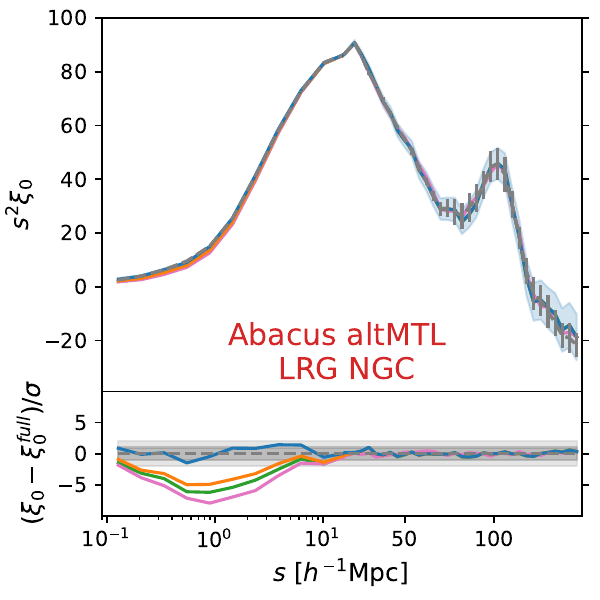}
    \includegraphics[width=0.32\linewidth]{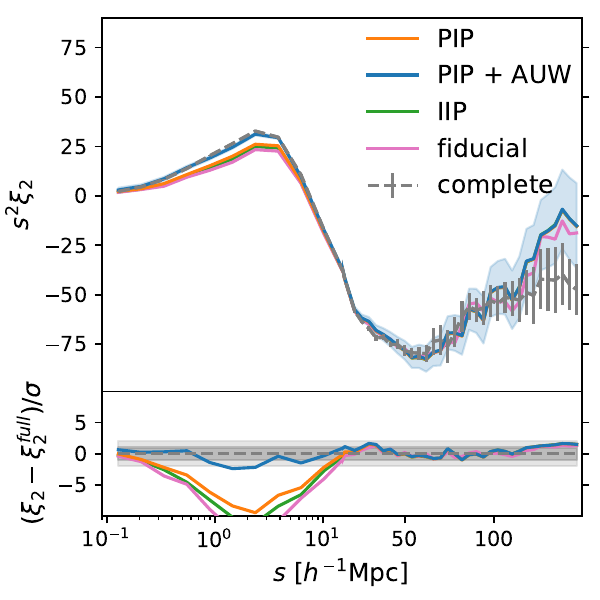}
    \includegraphics[width=0.32\linewidth]{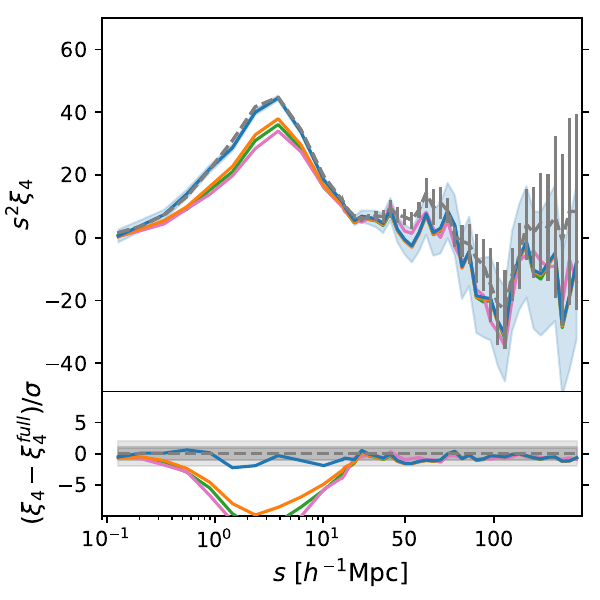} \\
    \includegraphics[width=0.32\linewidth]{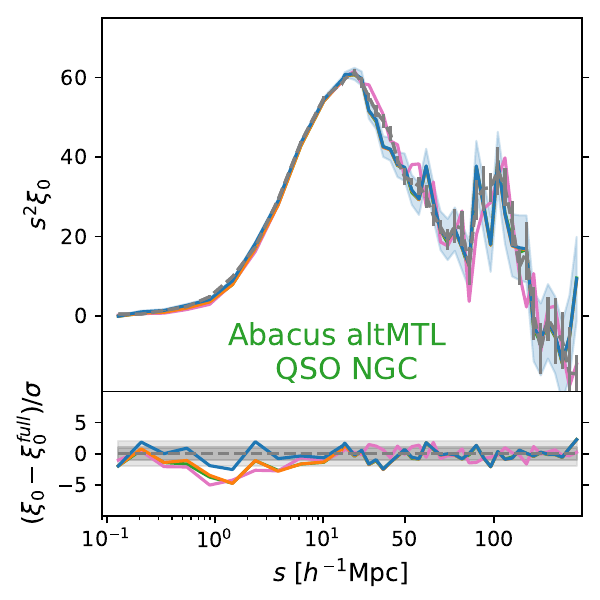}
    \includegraphics[width=0.32\linewidth]{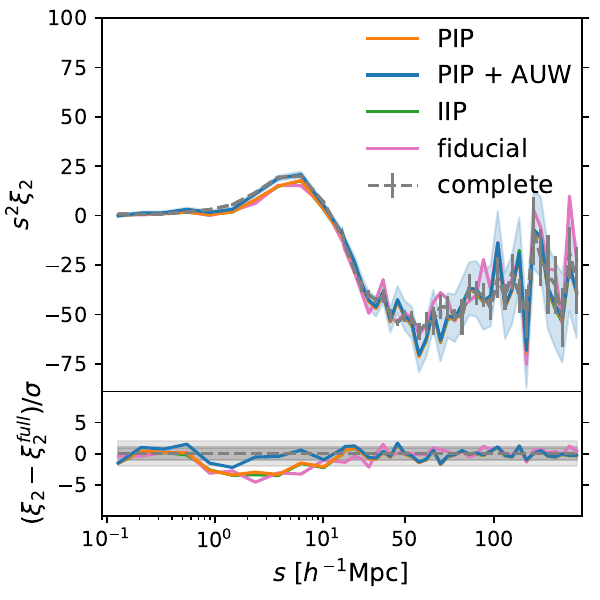}
    \includegraphics[width=0.32\linewidth]{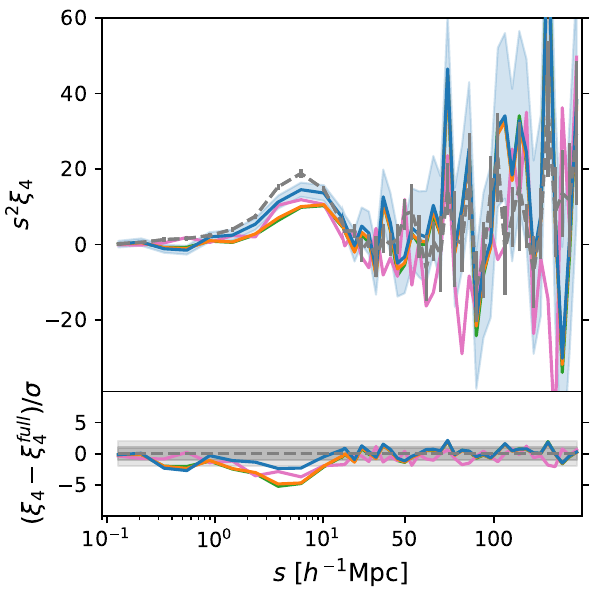}
    \caption{Multipoles of the correlation function $\xi_\ell$ as a function of the separation $s$, measured from the altMTL mocks (mock 11, NGC). For visualisation purpose, we follow the common practice of muliplying the amplitude of the multipoles by $s^2$. The columns correspond to $\ell=0,2,4$, from left to right, whereas the rows correspond to different dark-time tracers, ELGs, LRGs, QSOs, from top to bottom.
    The abscissa scale and the binning of correlation multipoles is logarithmic for $s<20 h^{-1}{\rm Mpc}$ and linear for larger separations in order to capture all the relevant fiber-assignment-related features at once. The grey dashed curves correspond to measurements obtained from the complete sample with no additional weights and represent a reference target for all the other measurements.
    The corresponding error bars represent the scatter between different mocks.
    The remaining curves are obtained from the fiber-assigned samples using: fiducial weights (solid pink); IIP weights (solid green); PIP weights (solid orange); PIP weights plus angular upweighting (solid blue), with the blue shaded area indicating the corresponding $1\sigma$ error.
    In the insets at the bottom of each panel we show the difference between fiber-assigned and complete samples in units of standard deviation.}
    \label{fig:xi_ASaMTL}
\end{figure}

In Figure \ref{fig:xi_ASaMTL} we show the correlation function multipoles measured from altMTL mocks. 
The shaded areas represent the standard deviation extracted from the fiducial DESI covariance matrices, which, for the correlation function, are obtained through a semi-analytical approach, implemented as the \texttt{RascalC} code \cite{KP4s7-Rashkovetskyi}.
Since on small scale this approach is not reliable, we complete the semi-analytic covariance with a purely empirical one.
Specifically, for $s < 20 h^{-1} \rm Mpc$, we use as error bars the standard deviation of the correlation functions measured with PIP weights plus angular upweighting from 7 FFA mocks.
To reduce noise caused by the small number of mocks, we interpolate it with a power law.
This approach clearly limits the reliability of the small-scale error and should be seen more as a convenient method for normalizing and visualizing the systematic trends of the different mitigation techniques, rather than providing an exact quantification of their confidence intervals.

For the altMTL mocks we can test all the four different estimator-level corrections considered in this work: fiducial weights (pink  solid), IIP weights (green solid), PIP weights (orange solid), PIP weights plus angular upweighting (blue solid).
Their effectiveness can be assessed by comparing them with the ``true" correlation, directly measured from the complete sample (grey dashed), i.e. the set of all potential targets of a given tracer class, ELG, LRG or QSO. 
For the complete samples we also show the error bars obtained from the variance of 7 mocks, as a rough measure of the intrinsic scatter of the clustering itself, regardless of fiber assignment.   

Figure \ref{fig:xi_ASaMTL} clearly shows that all the various weighting schemes produce fully consistent estimates at scales greater than $20 h^{-1} \rm Mpc$, and, most importantly, they all successfully recover the expected clustering, i.e. that of the complete sample.
At smaller scales, the impact of the collision window is distinctly visible, especially in the insets at the bottom of each plot, as a suppression of the power.
The effect becomes more pronounced and extends to larger scales for higher order multipoles.
The behavior of fiducial and IIP weights is similar, which is expected, since the former can be seen as a limiting case of the latter, as discussed in section \ref{sec:estimator_level}.
More precisely, the overall trend across different tracers and multipoles seems to indicate that IIP weights are slightly more effective in counteracting the collision window, but not in a significant way.  
PIP weights provides a further boost against the small-scale suppression, but the adjustment is still not sufficient to restore the "true" clustering.
This is not surprising, given the substantial number of zero-probability pairs, as illustrated in figure~\ref{fig:ang_pair_counts_aMTL}.
To compensate for these pairs we supplement PIP weights with angular upweighting.
As demonstrated in the figure, this combination produces estimates that are closer to the ``truth" than the others, with no discernible trend in the residuals.

For what concerns the statistical error, the figure clearly shows that, for the ELG sample, the fiducial weights produce clustering estimates with less noise on large scale.
A quantitative assessment of this effect, based on the scatter between 60 jacknife samples extracted from the ELG data sample, is presented in \cite{DESI2024.II.KP3}.
The test reveals that the variance obtained with IIP weights is approximately $60\%$ greater than that obtained with fiducial weights.
This result cannot be straightforwardly extrapolated to the measurements presented in this work, as they are obtained with different settings, the most significant being the inclusion of the $w_{NTMP}$ weights discussed above, but the trend seems confirmed, at least qualitatively (top row of figure \ref{fig:xi_ASaMTL}).

It is important to note that the excess variance is significant only for the ELG sample, which is highly incomplete (see table \ref{tab:n_obj}) and characterized by the lowest assignment probabilities, with a large portion of the survey footprint covered by just one pass of the instrument.
In the next data release, dramatic improvements in all these areas are expected, and any variance-related issues should be highly suppressed.

We can identify two primary underlying causes for the excess variance.
Firstly, low probabilities (more precisely, high inverse probabilities) are inherently more difficult to sample, where by low probability we roughly mean $p \lesssim 10/N_{bits}$.
With only 128 realisations, a relatively small number of bits compared to previous studies in the literature, the variance of the weights is directly impacted by sampling noise, with the most unlikely targets being more exposed to this effect. 

Secondly, even if the probabilities were perfectly measured ($N_{bits} \rightarrow \infty$) the resulting shot noise properties would not necessarily align with those of the fiducial weights.
IP weights are designed to be agnostic, meaning they do not rely on specific assumptions about the targeting algorithm or the underlying clustering.
This is accomplished by compensating for each individual assignment choice made by the algorithm, such that the contribution of each object or pair is exactly 1, when averaged over all possible realisations.
As a consequence, in extreme scenarios, some pairs might be given an inflated weight or, more precisely, some pairs might obtain different weights despite being statistically equivalent.

One possible cause of this effect is chance alignments with higher-priority targets, though it is not immediately clear why the fiducial weights would be less susceptible to this issue. 
It is nonetheless true that $w_{comp}$ effectively acts as a local smoothing of the probabilities over areas the size of the fiber patrol radius, suggesting that averaging over (inverse) probabilities could indeed have an impact.

Additionally, the fiducial weights include the $f_{tile}$ term, which compensates for missing power not captured by $w_{comp}$ on the tile-group angular scale. In contrast, the equivalent correction for the IP weights, $w_{NTMP}$, follows a simpler $n_{tile}$ angular pattern.
The $f_{tile}$ correction is more accurate, in the sense that it adjusts for fluctuations on a more localized scale, potentially helping to reduce variance.
There are straightforward countermeasures to address all the aforementioned potential causes of variance-related issues, such as increasing the number of realisations to mitigate sampling noise and redefining $w_{NTMP}$ based on tile groups, which we leave for future studies.

\begin{figure}
    \centering
    \includegraphics[width=0.32\linewidth]{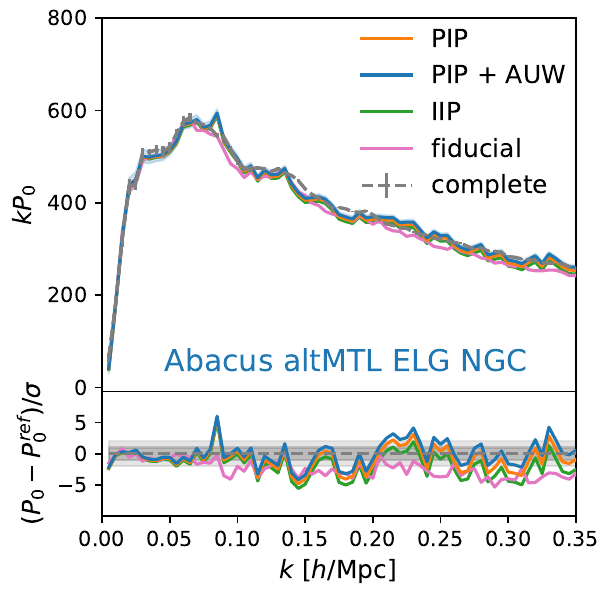}
    \includegraphics[width=0.32\linewidth]{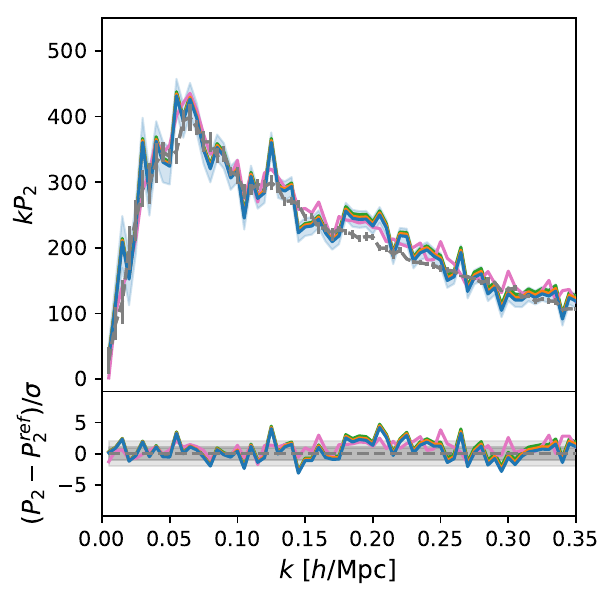}
    \includegraphics[width=0.32\linewidth]{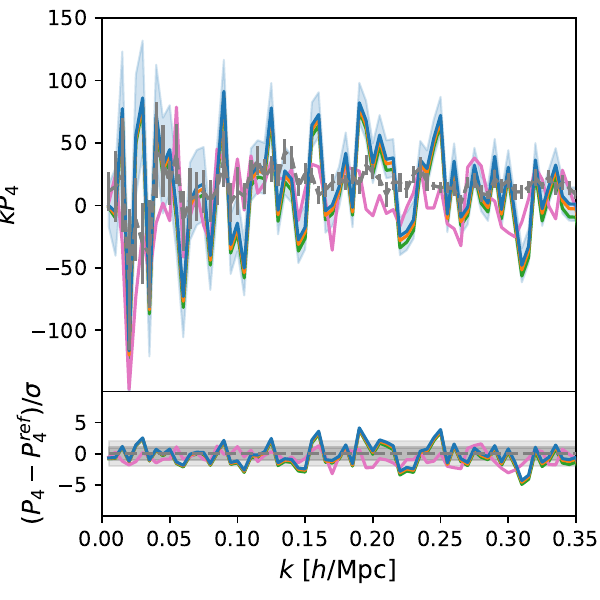} \\
    \includegraphics[width=0.32\linewidth]{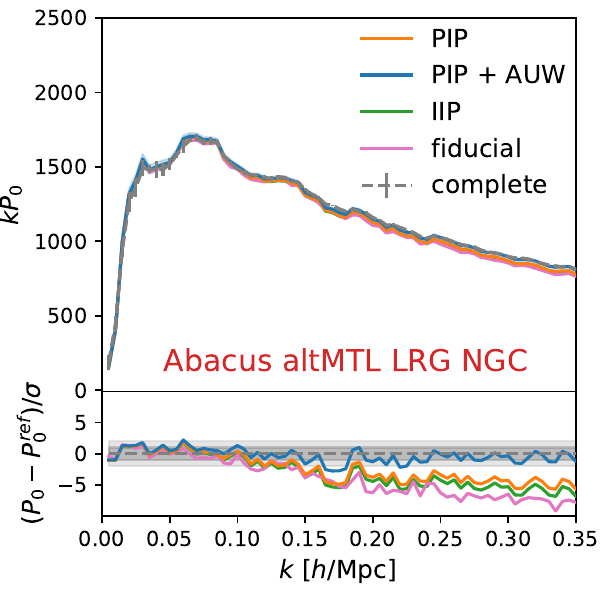}
    \includegraphics[width=0.32\linewidth]{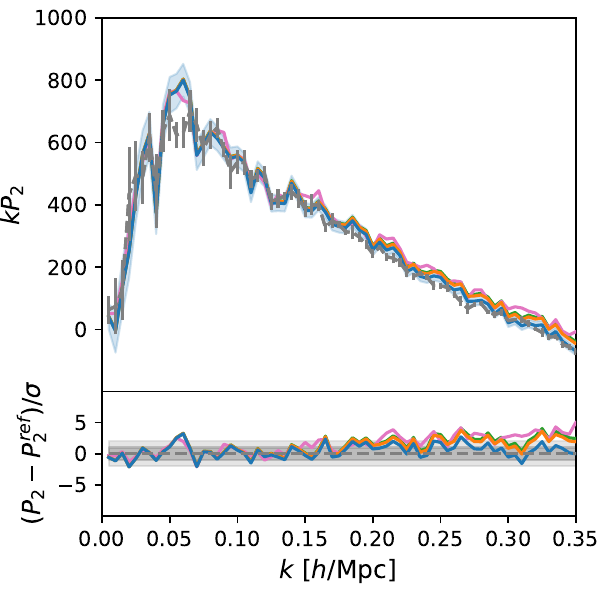}
    \includegraphics[width=0.32\linewidth]{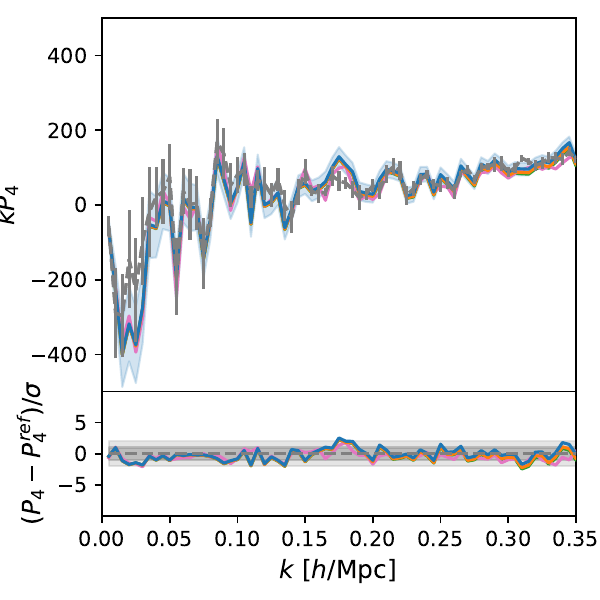} \\
    \includegraphics[width=0.32\linewidth]{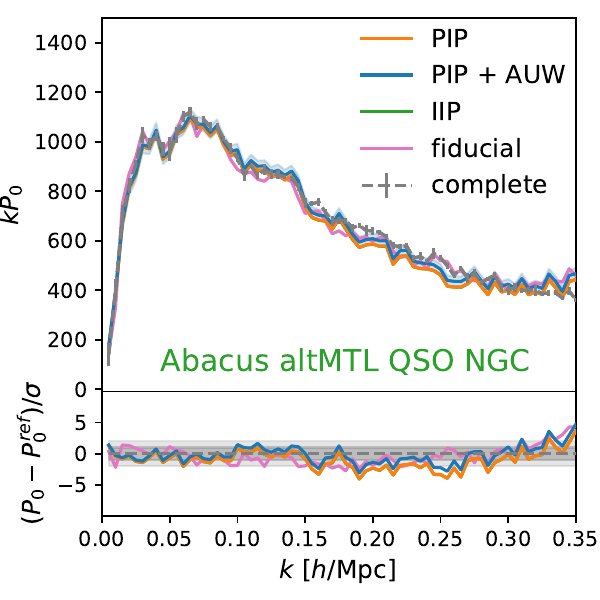}
    \includegraphics[width=0.32\linewidth]{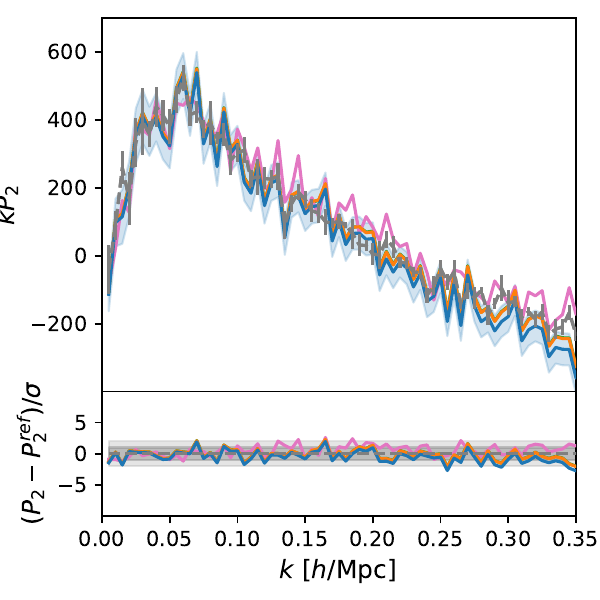}
    \includegraphics[width=0.32\linewidth]{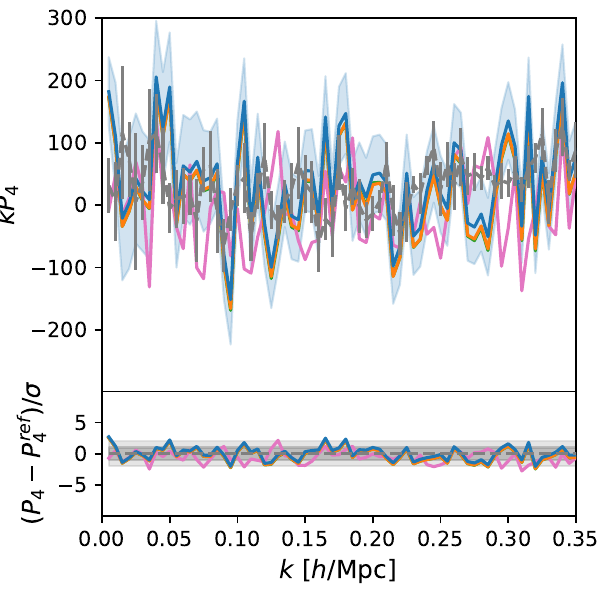}
    \caption{Multipoles of the power spectrum $P_\ell$ as a function of the wave number $k$, measured from the altMTL mocks (mock 11, NGC). For visualisation purpose, we follow the common practice of muliplying the amplitude of the multipoles by $k$.
    The columns correspond to $\ell=0,2,4$, from left to right, whereas the rows correspond to the different dark-time tracers, ELGs, LRGs, QSOs, from top to bottom. The grey dashed curves correspond to measurements obtained from the complete sample with no additional weights and represent a reference target for all the other measurements.
    The corresponding error bars represent the scatter between different mocks.
    The remaining curves are obtained from the fiber-assigned samples using: fiducial weights (solid pink); IIP weights (solid green); PIP weights (solid orange); PIP weights plus angular upweighting (solid blue), with the blue shaded area indicating the corresponding $1\sigma$ error. In the insets at the bottom of each panel we show the difference between fiber-assigned and complete samples in units of standard deviation.}
    \label{fig:P_ASaMTL}
\end{figure}

Figure \ref{fig:P_ASaMTL} shows the multipoles of the power spectrum obtained from the same altMTL samples.
We adopt the same color coding conventions as in the previous figure.
The shaded areas represent the standard deviation extracted from the fiducial DESI covariance matrices. For the power spectrum, these are measured from the scatter of 1000 FFA EZmocks and subsequently scaled in amplitude using the semi-analytic $\xi$ covariance mentioned earlier, as detailed in \cite{DESI2024.II.KP3}.

In essence, the same considerations made for the correlation function also hold for the power spectrum, with the key difference being that in Fourier space, the influence of fiber assignment affects significantly larger scales.
This is expected, due to the non local nature of the Fourier transform, which produces extended tails in the collision window.
The relative behaviour of the different weighting schemes is confirmed, with PIP weights plus angular upweghting recovering the clustering of the complete sample with no indication of systematic error.     
The only significant exception is the monopole of the QSO sample, which shows an excess of power at large $k$, consistent across all curves. Given that this sample has the lowest Nyquist frequency due to its large volume, and no similar trends are observed in the corresponding configuration space measurements, we argue this is likely a result of aliasing.

\begin{figure}
    \centering
    \includegraphics[width=0.32\linewidth]{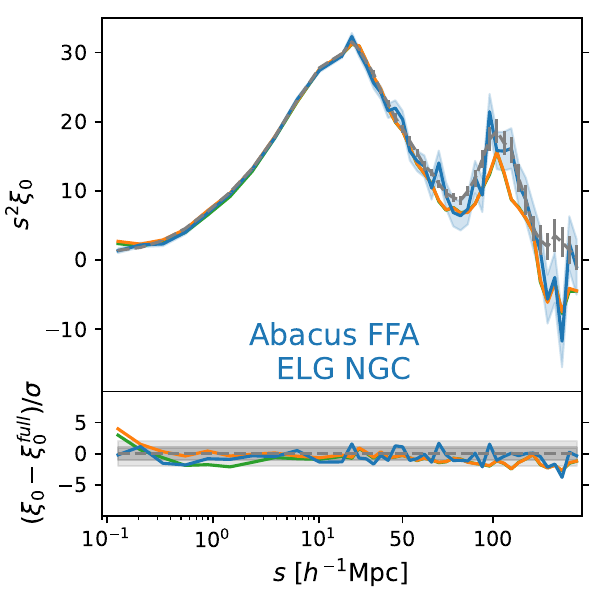}
    \includegraphics[width=0.32\linewidth]{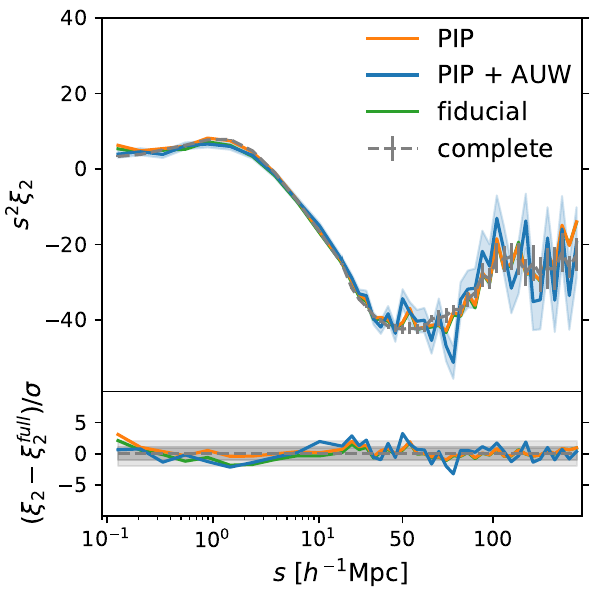}
    \includegraphics[width=0.32\linewidth]{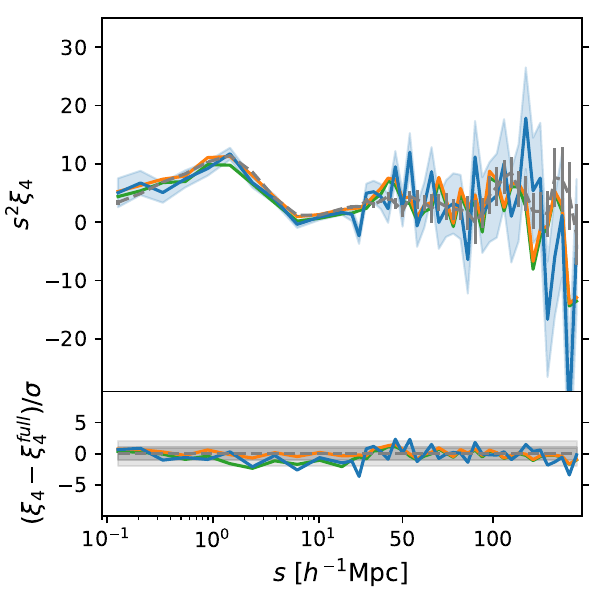} \\
    \includegraphics[width=0.32\linewidth]{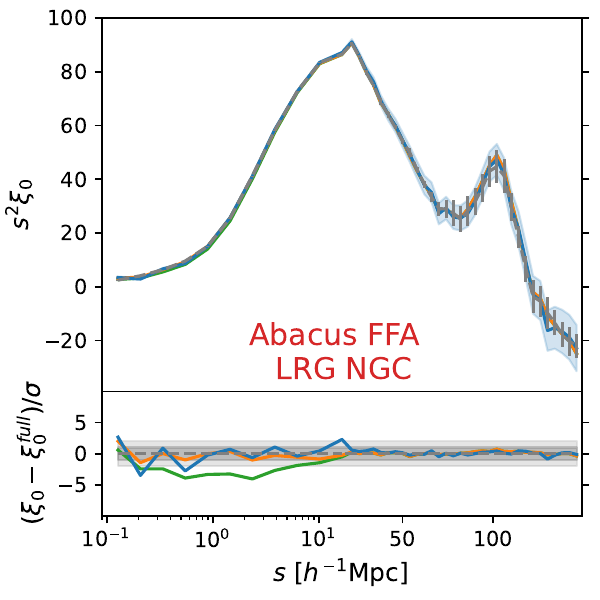}
    \includegraphics[width=0.32\linewidth]{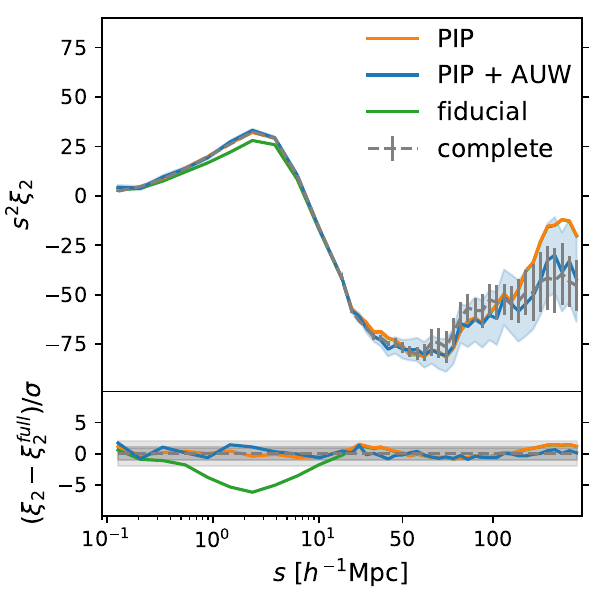}
    \includegraphics[width=0.32\linewidth]{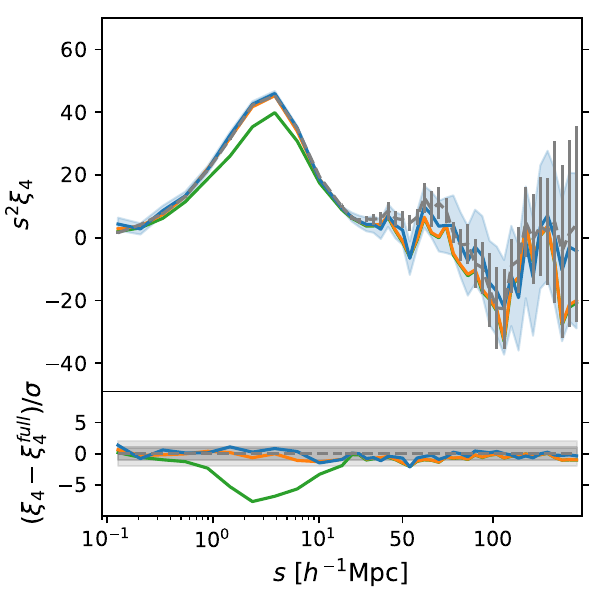} \\
    \includegraphics[width=0.32\linewidth]{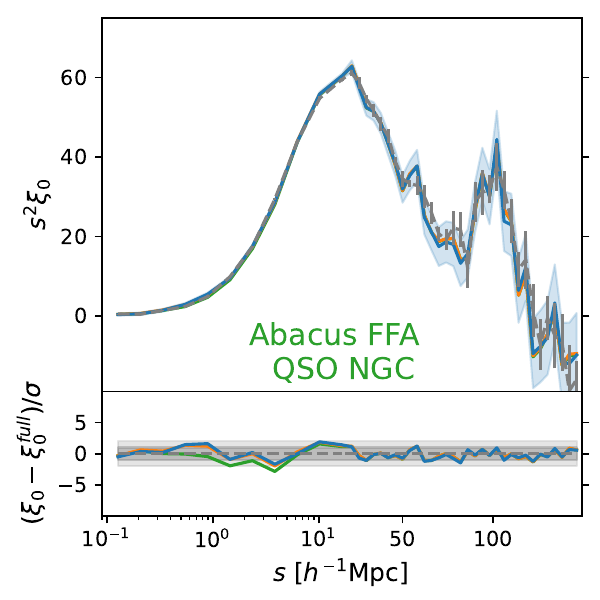}
    \includegraphics[width=0.32\linewidth]{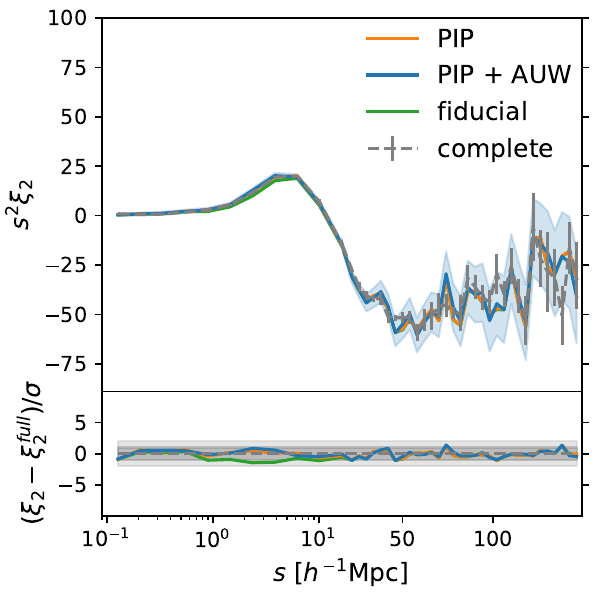}
    \includegraphics[width=0.32\linewidth]{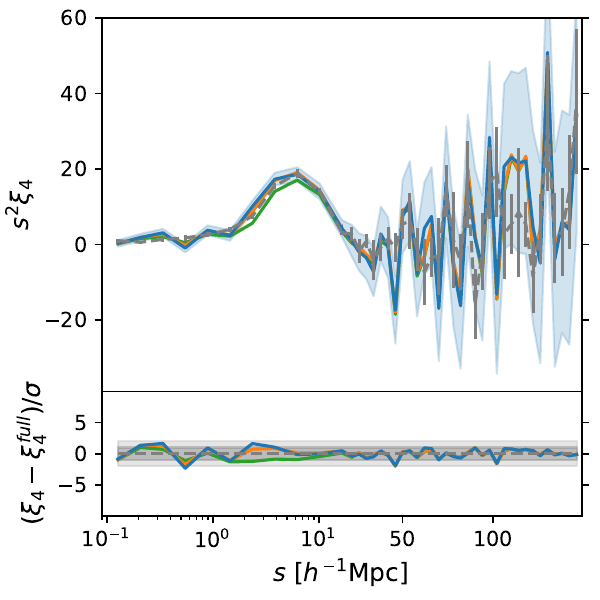}
    \caption{Same as figure \ref{fig:xi_ASaMTL} but for the FFA mocks (mock 11, NGC). Unlike altMTL mocks, the fiducial weights here (solid green) are, by design, IIP weights.}
    \label{fig:xi_ASffa}
\end{figure}

\begin{figure}
    \centering
    \includegraphics[width=0.32\linewidth]{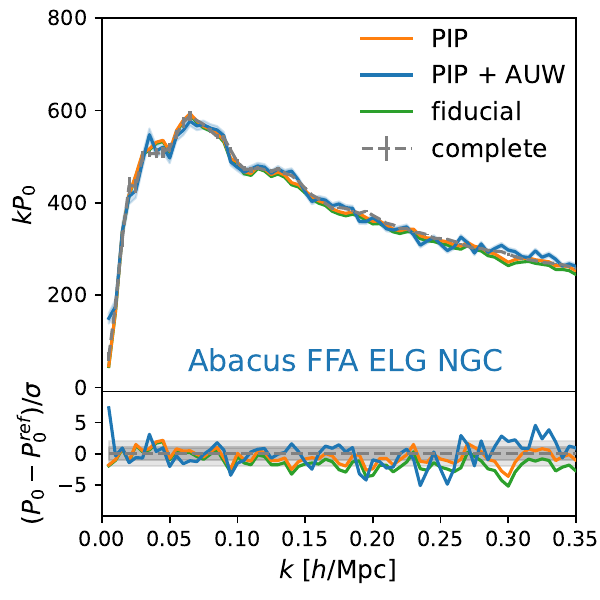}
    \includegraphics[width=0.32\linewidth]{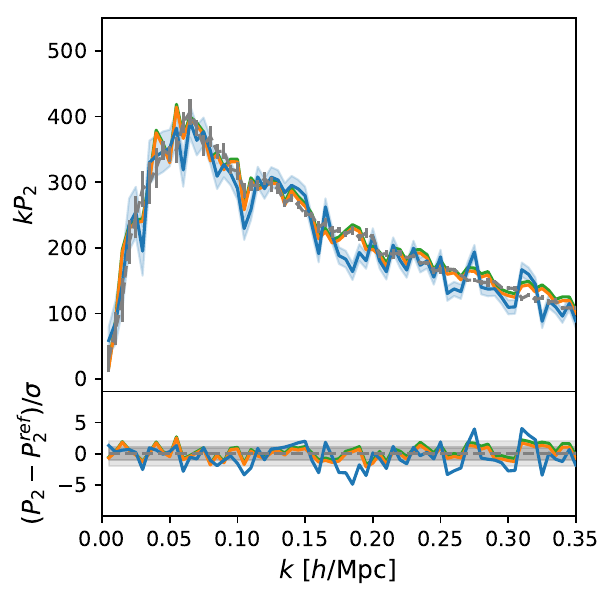}
    \includegraphics[width=0.32\linewidth]{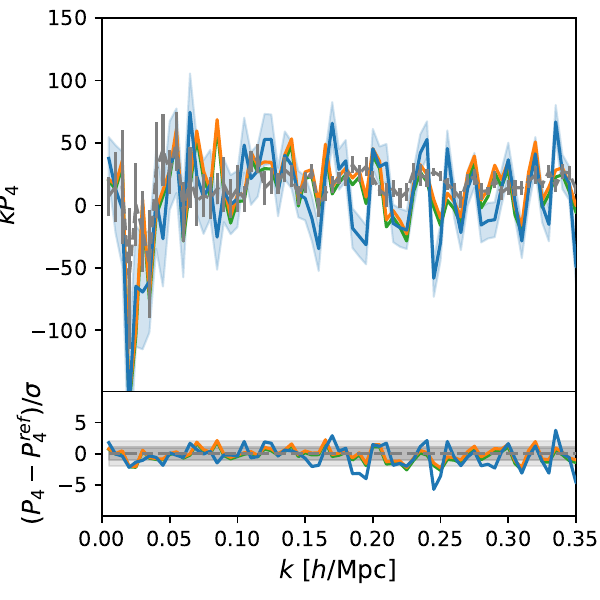} \\
    \includegraphics[width=0.32\linewidth]{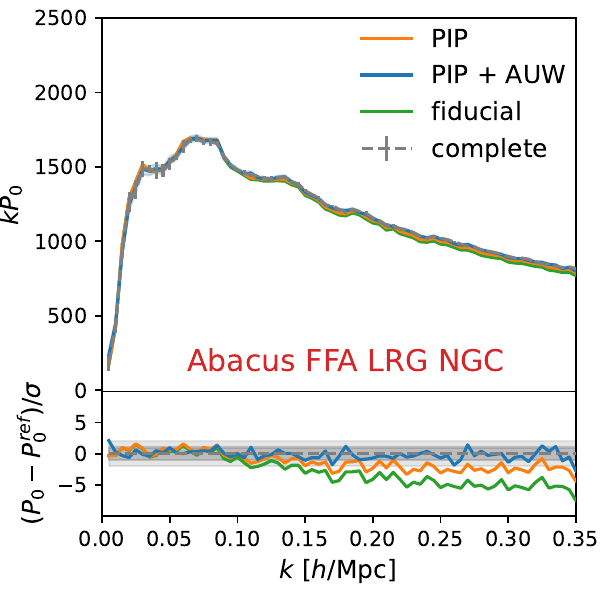}
    \includegraphics[width=0.32\linewidth]{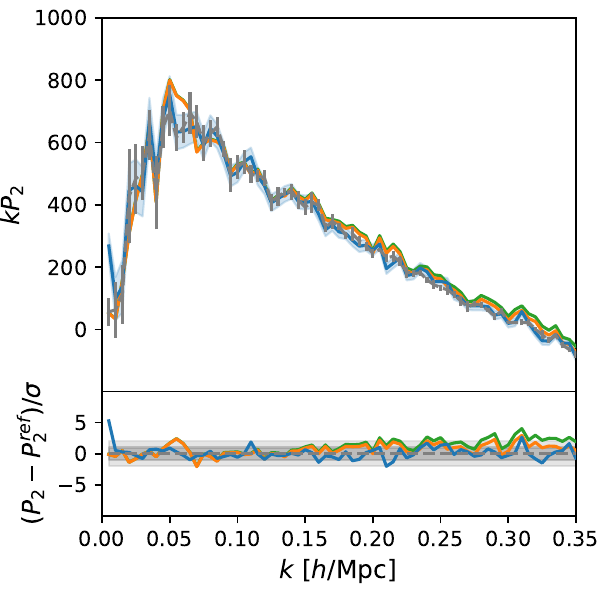}
    \includegraphics[width=0.32\linewidth]{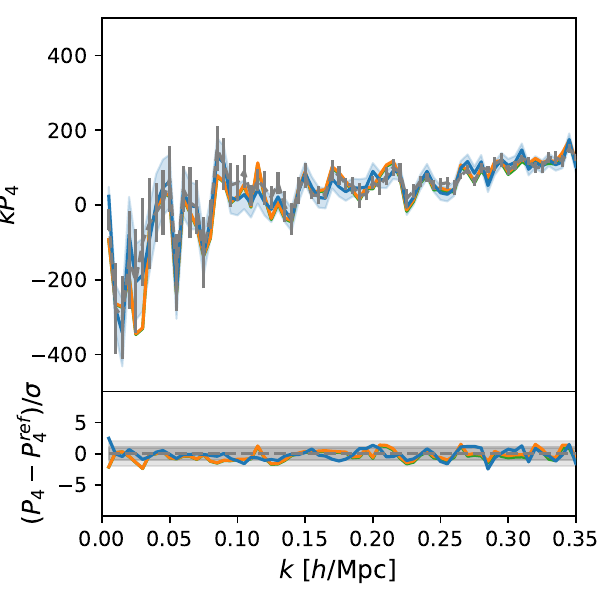} \\
    \includegraphics[width=0.32\linewidth]{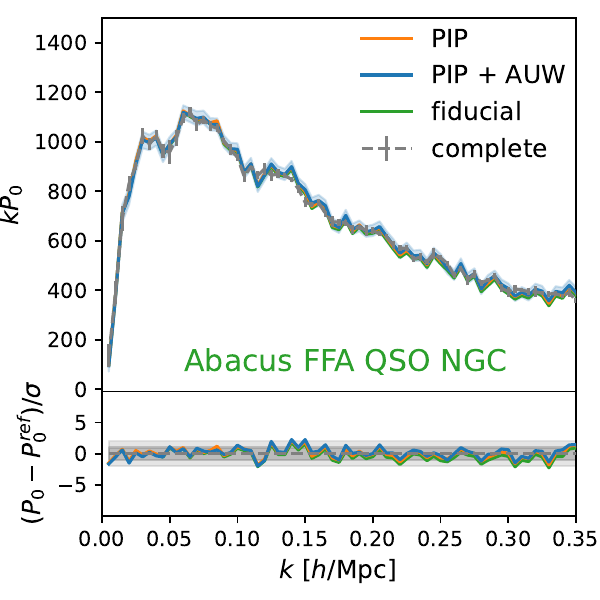}
    \includegraphics[width=0.32\linewidth]{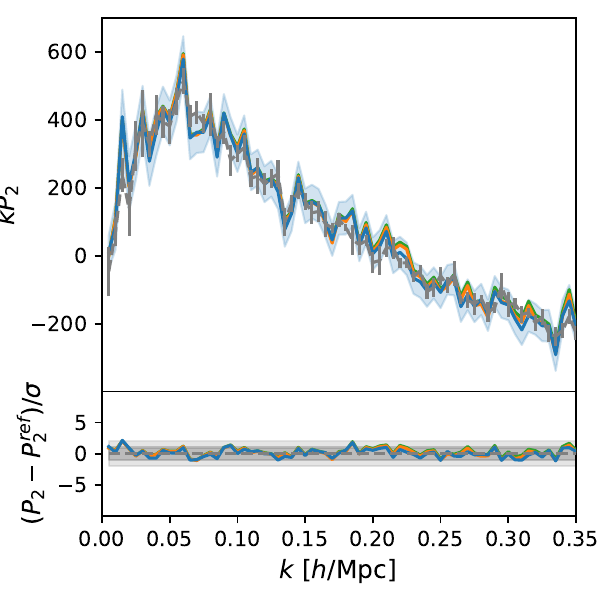}
    \includegraphics[width=0.32\linewidth]{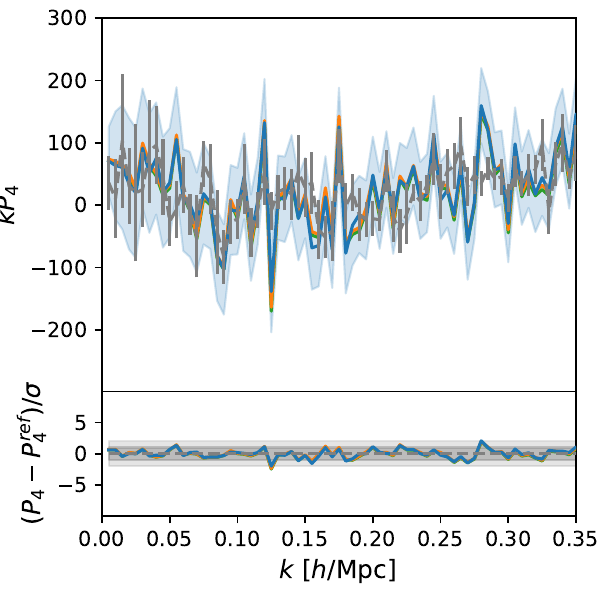}
    \caption{Same as figure \ref{fig:P_ASaMTL} but for the FFA mocks (mock 11, NGC). Unlike altMTL mocks, the fiducial weights here (solid green) are, by design, IIP weights.}
    \label{fig:P_ASffa}
\end{figure}

Using the same settings adopted for the altMTL mocks, we measured both the correlation function and power spectrum from the FFA mocks.
The results are shown in figures \ref{fig:xi_ASffa} and \ref{fig:P_ASffa}, respectively. 
With this set of mocks, we can compare three fiber mitigation techniques, fiducial weights (green solid), PIP weights (orange solid) and PIP weights with angular upweighting (blue solid), to the reference provided by the complete sample (gray dashed).
It is important to note that, by design (see section \ref{sec:FFA}), the fiducial weights in the FFA case are actually IIP weights.
Furthermore, unlike the altMTL catalogs, the FFA mocks were not processed to generate a proper full sample for angular upweighting.
To work around this limitation, we use the complete sample instead, which is, by construction, cut to the same redshift range of the tracer under investigation.
This scenario is not entirely realistic because, in a real situation, we would not have access to all the (spectroscopic) redshifts required to make such a cut, specifically the redshifts of galaxies that were not observed.
As a result, in these tests, angular upweighting may be more effective at mitigating systematic effects than it would in a real-world application, at least in principle.
In addition, the complete sample does not include $w_{ntile}$ weights, so for the clustering sample we cannot apply the factorization $w_{\rm galaxies}' = w_{\rm galaxies} / w_{IIP}$ descibed above (or use FKP weights), as that would interfere with angular upweighting.
Consequently, the PIP plus angular upweighting measurements show a greater statistical error than they ideally would.
Finally, since the complete sample does not incorporate the veto mask of the clustering sample, angular upweighting would technically require normalizing the angular pair counts with the corresponding random samples.
For all these reasons, the comparison with the other two weighting schemes is not entirely fair. 

In figure \ref{fig:xi_ASffa} the enhanced statistical noise is clearly noticeable, e.g. when comparing the $\xi$ multipoles measured with PIP to those measured with PIP plus angular upweighting, especially for the ELG sample in the top raw.
Nevertheless, PIP weights with angular upweighting provide unbiased estimates, as expected. Interestingly, PIP weights alone also successfully recover the correct clustering across all scales. This is not surprising, as FFA mocks contain significantly fewer zero-probability pairs compared to altMTL mocks and real data (see figures 
\ref{fig:ang_pair_counts_data}, \ref{fig:ang_pair_counts_aMTL}, \ref{fig:ang_pair_counts_FFA}), making angular upweighting unnecessary, at least for addressing systematic effects.

Similar considerations apply to the power spectrum multipoles, shown in figure \ref{fig:P_ASffa}.
The most noticeable difference, compared to the correlation function, is arguably the behavior of the PIP measurements from the LRG sample.
Specifically, the PIP power spectrum exhibits a slight systematic suppression at high $k$, which is not readily noticeable in its configuration space counterpart (as seen in the bottom insets of the central row in both figures).
This is likely due to the combined effects of the collision window's impact being more pronounced in Fourier space and the fact that we use two different prescriptions for the standard deviations.
As mentioned earlier, the small-scale variance of the correlation is not sufficiently reliable for a quantitative comparison of confidence intervals. In fact, a closer look at the LRG's PIP correlation monopole in figure \ref{fig:xi_ASffa} reveals indications of power suppression at this scale, roughly of the order of $\sigma$.
Setting aside the definitional issue mentioned earlier, the combination of PIP and angular upweighting consistently produces unbiased estimates for all the FFA samples.

The direct comparisons between figures \ref{fig:xi_ASaMTL} and \ref{fig:xi_ASffa} and between figures \ref{fig:P_ASaMTL} and \ref{fig:P_ASffa} provide valuable insights into the reliability of the two assignment techniques we developed.
Overall, we observe good agreement between the more realistic altMTL method and the computationally simpler FFA emulator. 
The most significant difference is the reduced impact of the collision window in the FFA case, though it remains evident for the fiducial weights.
Additionally, due to a significantly lower number of zero-probability pairs, PIP weights alone are (almost) sufficient to restore the ``true" clustering without the need for angular upweighting.
Both the effects are not surprising, given how the collision window is handled by the FFA algorithm (section \ref{sec:FFA}), and they are perfectly consistent with the behaviour of the angular pair counts shown in figure \ref{fig:ang_pair_counts_FFA}. 
Possible improvements to enhance the realism of the FFA algorithm involve optimizing its hyperparameters, adapting it for multi-tracer emulation, and developing strategies to actively manipulate the shape of the collision window.
We leave these investigations to future work.

\begin{figure}
    \centering
    \includegraphics[width=0.32\linewidth]{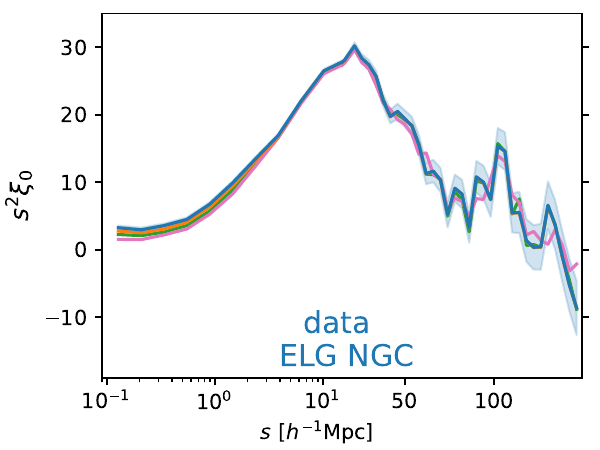}
    \includegraphics[width=0.32\linewidth]{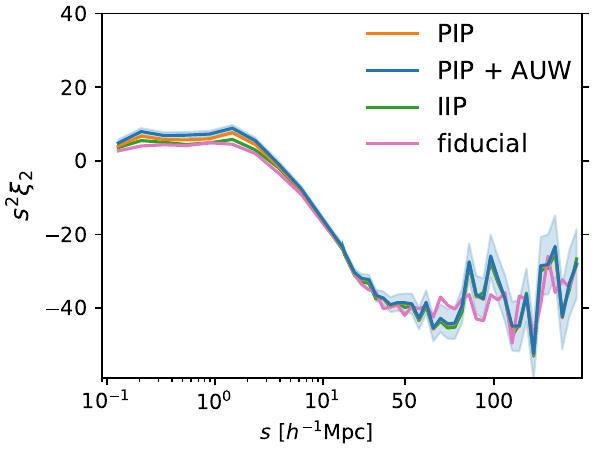}
    \includegraphics[width=0.32\linewidth]{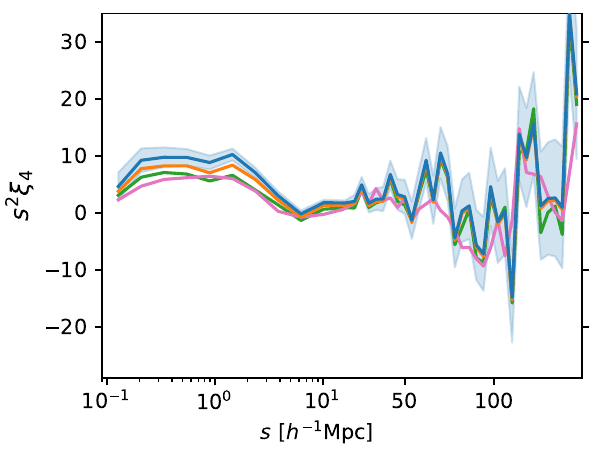} \\
    \includegraphics[width=0.32\linewidth]{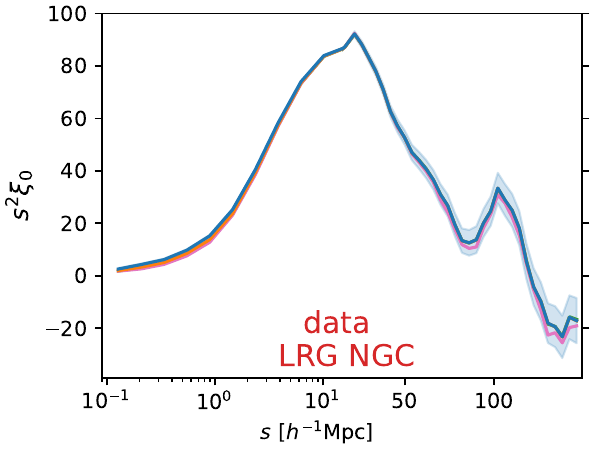}
    \includegraphics[width=0.32\linewidth]{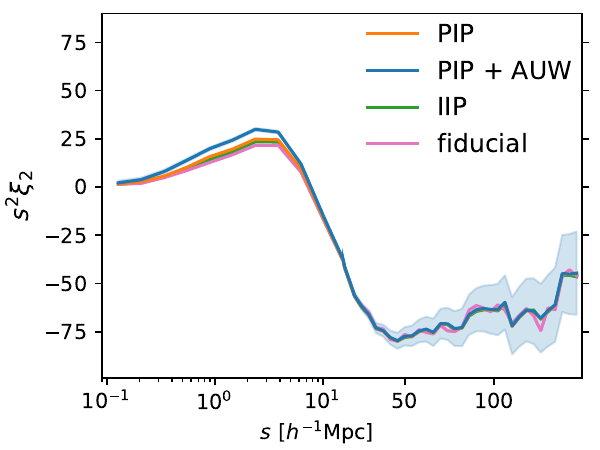}
    \includegraphics[width=0.32\linewidth]{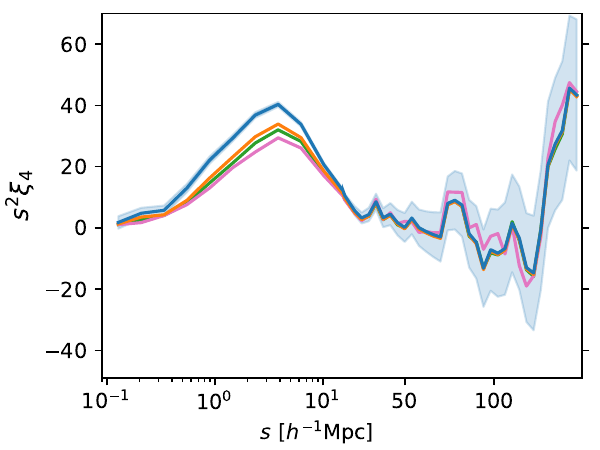} \\
    \includegraphics[width=0.32\linewidth]{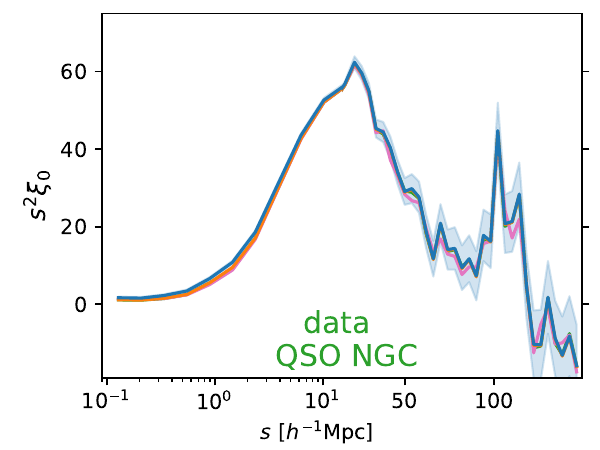}
    \includegraphics[width=0.32\linewidth]{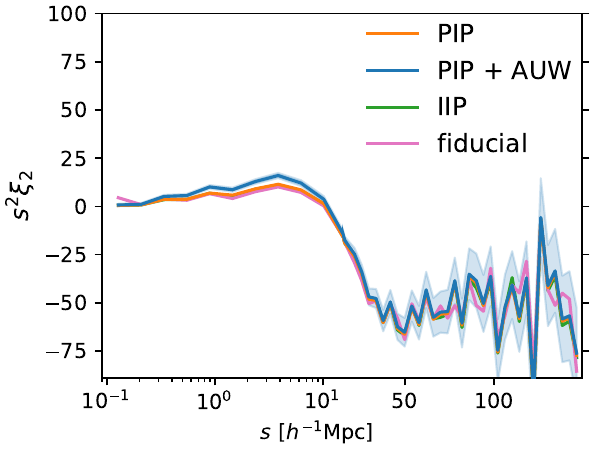}
    \includegraphics[width=0.32\linewidth]{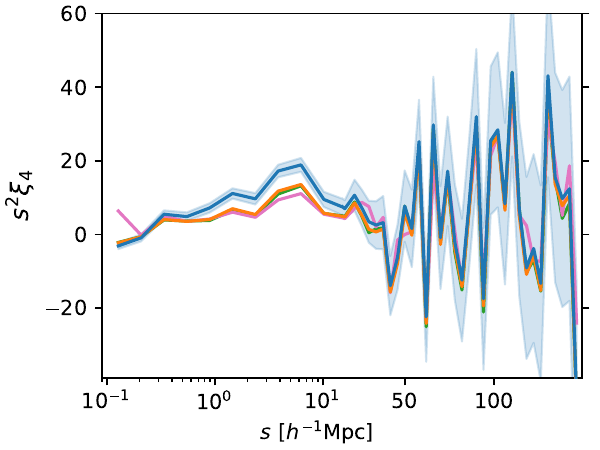}
    \caption{Same as figure \ref{fig:xi_ASaMTL} but applied to the DESI DR1 catalogues, with the key difference that no reference sample is available for comparison in this case.}
    \label{fig:xi_data}
\end{figure}

\begin{figure}
    \centering
    \includegraphics[width=0.32\linewidth]{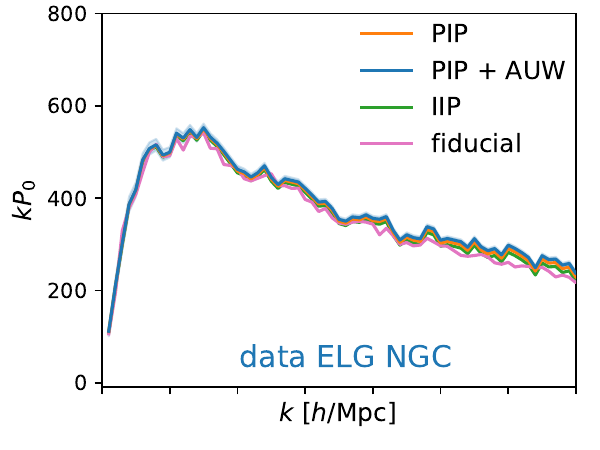}
    \includegraphics[width=0.32\linewidth]{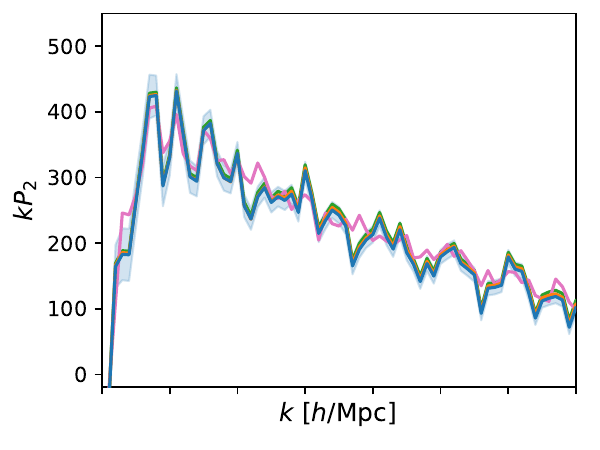}
    \includegraphics[width=0.32\linewidth]{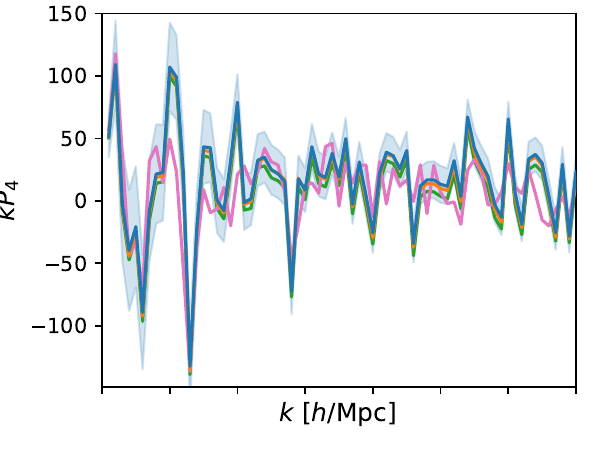} \\
    \includegraphics[width=0.32\linewidth]{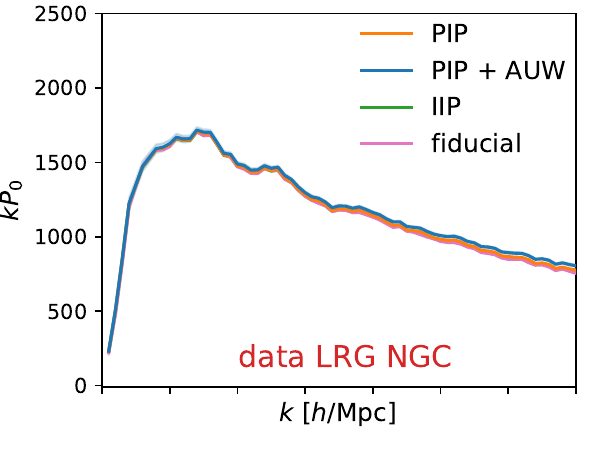}
    \includegraphics[width=0.32\linewidth]{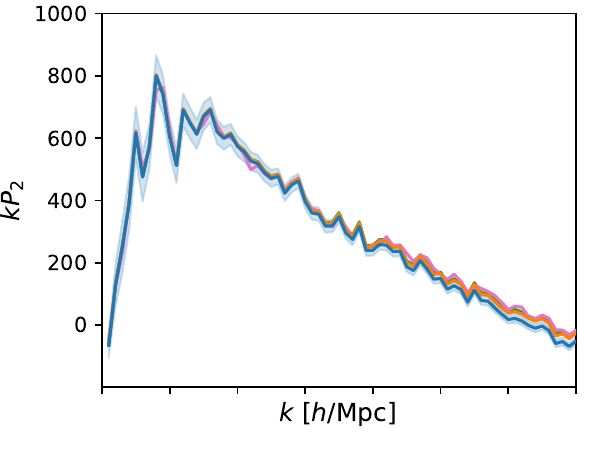}
    \includegraphics[width=0.32\linewidth]{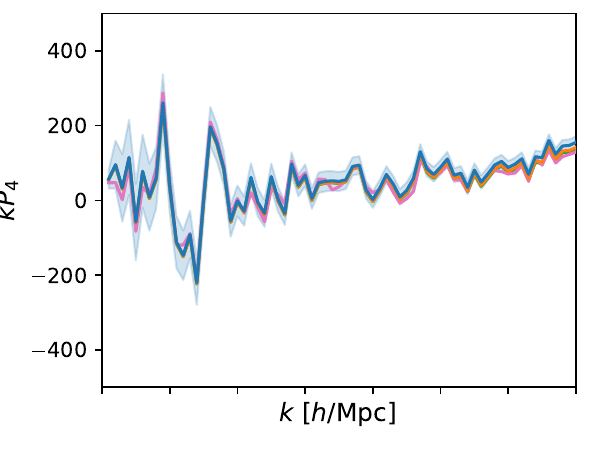} \\
    \includegraphics[width=0.32\linewidth]{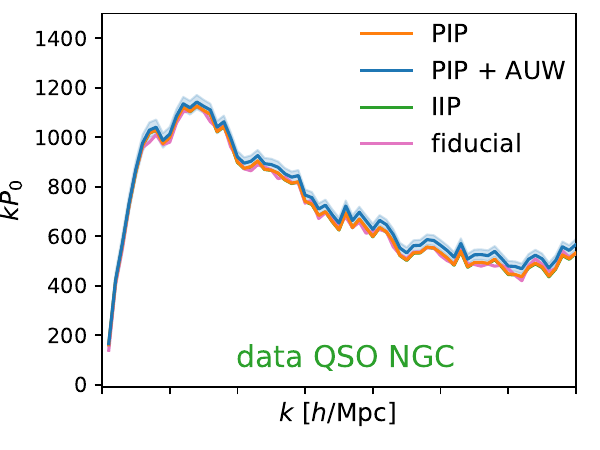}
    \includegraphics[width=0.32\linewidth]{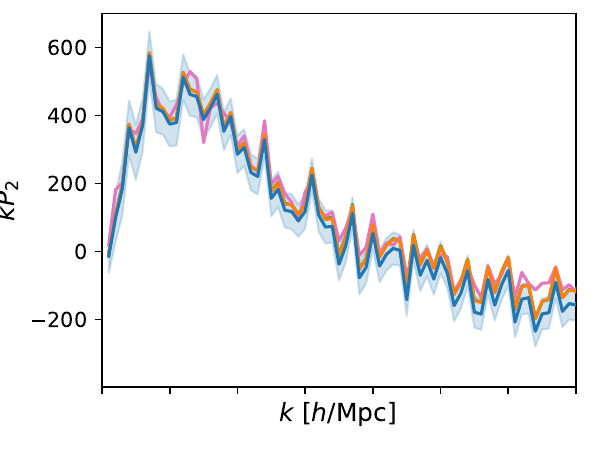}
    \includegraphics[width=0.32\linewidth]{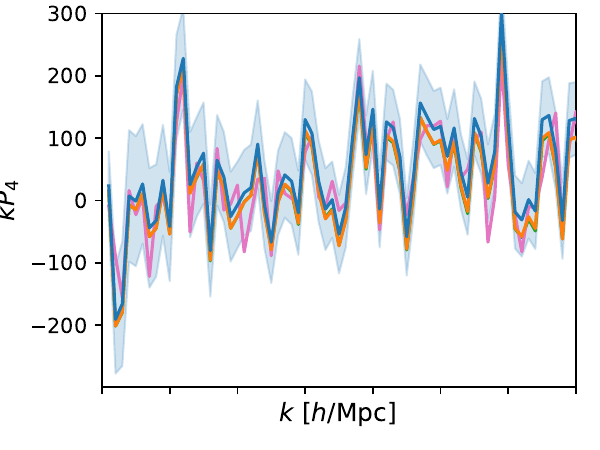}
    \caption{Same as figure \ref{fig:P_ASaMTL} but applied to the DESI DR1 catalogues, with the key difference that no reference sample is available for comparison in this case.}
    \label{fig:P_data}
\end{figure}

Finally, figures \ref{fig:xi_data} and \ref{fig:P_data} display the multipoles of the correlation function and power spectrum, respectively, as measured directly from the DESI DR1 samples.
These calculations were performed using the same settings as for the altMTL mocks, and the same color coding conventions are followed.
To the best of our knowledge, this is the first time PIP weights have been applied to measure the power spectrum from real data. 
Although we lack a ``true" sample for comparison, it is evident that the behavior of the various mitigation schemes aligns well with the altMTL results and, more broadly, with expectations.
On one hand, this further reinforces the validity of altMTLs, and to a lesser extent, FFA, as realistic fiber assignment methods.
On the other hand, it confirms the effectiveness of the mitigation techniques, or at least their consistency across different datasets, with no unexpected trends emerging.

\section{Conlusions}\label{sec:conclusions}

We have presented a detailed analysis of the impact of fiber assignment incompleteness on 2-point statistics for DESI DR1, discussed the countermeasures implemented to address it, and described the methods developed to simulate the fiber assignment process on synthetic galaxy catalogs.

This work is part of a larger effort to characterize the DESI fiber incompleteness, with one of its main goals being to provide a coherent overview of this effort.
For a more detailed description of the altMTL assignment method and the $\theta$-cut fiber mitigation, we refer the reader to \cite{KP3s7-Lasker} and \cite{KP3s5-Pinon}, respectively, whereas for a comprehensive explanation of the process used to build the catalogues and obtain the 2-point statistics for DR1 please see \cite{DESI2024.II.KP3}.

We have conducted systematic studies on the impact of fiber assignment on the clustering of all the three DESI dark-time tracers, ELGs, LRGs, QSOs, across the entire DR1 volume, unsing both real and simulated data.
As expected, we have found clear evidence of fiber incompleteness patterns in the DESI DR1 footprint and, most importantly, in the 2-point clustering.
These effects are successfully removed on large and intermediate scale with the use of proper weights, applied to the galaxy and random samples.
However, these weights alone do not address those aspects of fiber assignment that are intrinsically pairwise in nature.
As a consequence, the angular pair counts show a large power suppression on scales $\lesssim 0.02 \rm deg$ for all the tracer types and sky areas considered, reaching $\sim 80\%$ for the ELGs in south Galactic cap.
Properly addressing the impact of  this ``collision window'' on power spectrum and correlation function estimates is the primary goal of the fiber mitigation techniques presented in this work.

We have explained what concepts motivate the fast-fiber-assignment procedure (FFA) and how the algorithm is practically implemented.
We have tested its performance against altMTL mocks and real data, finding reasonably good agreement considering the approximations involved.
The most notable differences we observed are the reduced number of zero-probability pairs and a less pronounced collision window, both of which are fully explained by the FFA algorithm's procedures for assigning probabilities and imposing small-scale anticorrelation.
The primary reason for developing FFA was the necessity to run fiber assignment on thousands EZmocks to extract realistic covariance matrices for cosmological inference.
This is how we obtained the power-spectrum covariance used for the DR1 full-shape analysis, \cite{DESI2024.V.KP5, DESI2024.VII.KP7B}.
However, as discussed in \cite{DESI2024.II.KP3} and \cite{KP4s6-Forero-Sanchez}, we have indications that this covariance might have roughly $10\%$ less power than it should.
We therefore applied a final rescaling in amplitude, as described in \cite{DESI2024.II.KP3}.  
FFA is a potential reason for this missing variance, alongside with pre-existing discrepancies in the 3- and 4-point functions.
These aspects will be further investigated in future data releases.

We have assessed the reliability of the altMTL mocks, which are more realistic but computationally demanding. They have successfully passed all tests against the data, with no apparent signs of any issues.
For future data releases, we are looking into ways to accelerate the altMTL process and enhancing the accuracy of FFA by incorporating more realism into the procedure, such as integrating a multi-tracer kernel \cite{KP3s11-Sikandar}.

We have conducted a comprehensive analysis of the different fiber-mitigation techniques, including fiducial weights derived by matching the local completeness of different tracer types, individual inverse probability (IIP) weights, pairwise inverse probability (PIP) weights, and their integration with angular upweighting.
To achieve the high level of accuracy required by DESI and address limitations like the limited number of targeting realizations and the significant incompleteness in DR1, we have used inverse probability weights in a novel manner.
In particular, we have introduced a new definition for the PIP weights that enhances their robustness against scale-dependent variations in assignment probability. Additionally, we have defined a new set of weights, $w_{NTMP}$, to correct for the irregular distribution of the numerous zero-probability galaxies found in DR1.  

These new developments allowed us not only to obtain reliable estimates of the correlation function down to the smallest scale for DR1 but also to successfully apply, for the first time ever, PIP weights (plus angular upweighting) to power spectrum measurements from real data.
Since these innovations, especially the $w_{NTMP}$ weights, were finalized only recently, this is the only DR1 paper where they have been utilized.  

In summary, we found that PIP weights with angular upweighting deliver the most accurate estimates of the ``true" clustering, in all scenarios considered, both in configuration and Fourier space, and for all types of tracers.
These results complement those in a parallel paper, \cite{KP3s5-Pinon}, which shows that similar performances can be achieved with completeness weights, provided that small scales are removed from both the estimator and the model of the statistics of interest ($\theta$-cut method).

So far, PIP weights have been employed in various analyses of DESI data, from both SV3 \cite{DESI2023a.KP1.SV, DESI2023b.KP1.EDR} and DR1, to recover the correct small-scale clustering (e.g., \cite{pearl2024, abacusHODELG, abacushod2, yu2024, gao2024}).
For the full-shape analysis, \cite{DESI2024.V.KP5, DESI2024.VII.KP7B}, the completeness weights with $\theta$-cut method was chosen, whereas the BAO analysis \cite{DESI2024.III.KP4, DESI2024.VI.KP7A} relied solely on completeness weight, as it was shown that no further correction was necessary.  
However, for those studies that, like BAO, depend on reconstruction, the $\theta$-cut approach is less effective, due to the transfer of small-scale incompleteness to larger scale.  
This work opens the opportunity to use PIP weights with angular upweighting for all these analyses in future data releases, thus allowing for the preservation of small-scale information and providing a more clearly defined framework for reconstruction.

\appendix

\section{Adding small-scale anticorrelation to the FFA catalogues}\label{app:antico}

Once a group of entangled galaxies has been identified through the FOF algorithm, we set it into a maximum anticorrelation state, by manipulating the corresponding bitweights.
Specifically, we allow displacements of bits from one realisation to one another, inside each bitweight, but not from one bitweight to one another.  
As a consequence, the procedure conserves individual probabilities while arbitrarily changing the pairwise ones.
By maximum anticorrelation state we mean bitweights such that the corresponding pair count (and any $t$-plet count with $t \ge 2$), averaged over all the realisations, is the lowest compatible with the given numbers of bits per bitweight or, in other words, compatible with the given individual probabilities. 

Two preliminary considerations: (i) for a given group there are many of such states; (ii) all galaxies in a FOF group are treated equally, i.e. we ignore their spatial locations and relative separations.
Regarding (i), our procedure is such that the maximum anticorrelation state is fully determined by the order with which the galaxy are processed, which can be easily randomised to obtain alternative states.
Regarding (ii), it would be technically possible to add scale dependency to the procedure but, realistically, that would come at the cost of more computational complexity/time, which is precisely what we want to avoid with FFA. 

\begin{figure}
    \centering
    \includegraphics[width=1.0\linewidth]{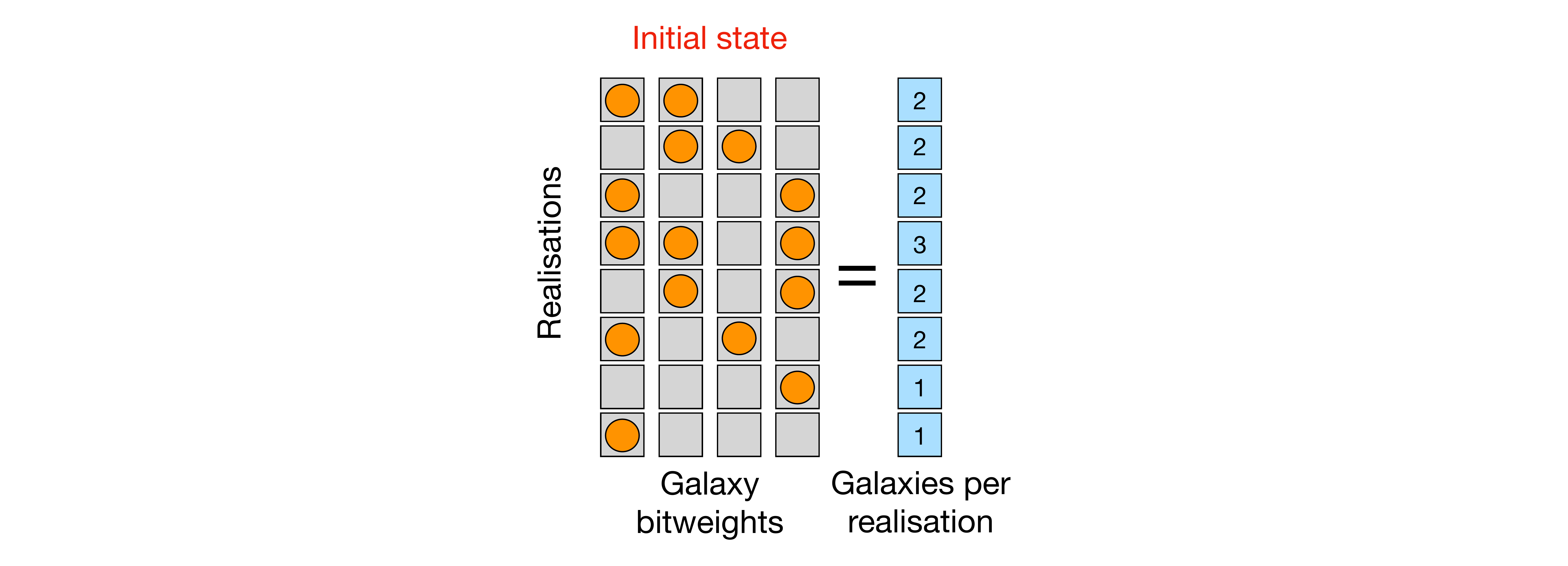}\\
    \includegraphics[width=1.0\linewidth]{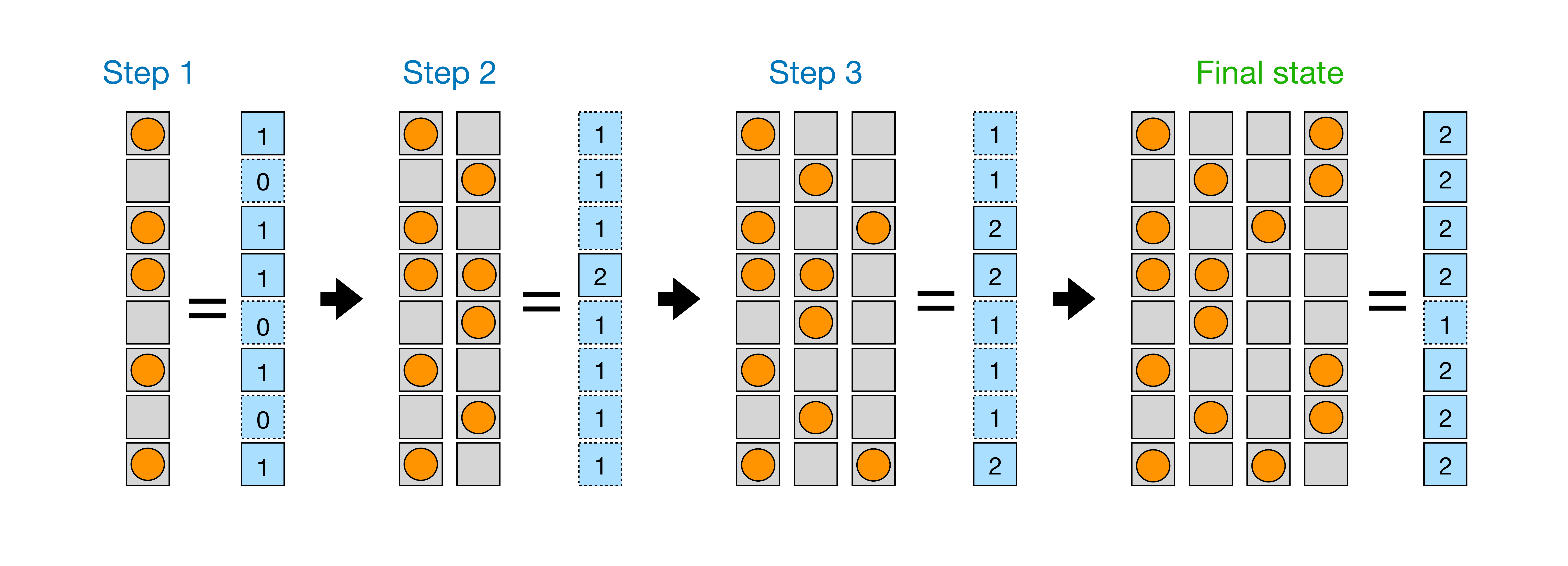}
    \caption{Scheme of the procedure used to find a maximum anticorrelation state for the bitweights, as decribed in the text.}
    \label{fig:anticorr_scheme}
\end{figure}

The actual algorithm works as sketched in figure \ref{fig:anticorr_scheme}.
In brief, given a group of $N_{FOF}$ ordered galaxies, each with its own bitweight, we initialise the process with galaxy 1, for which we count the total number of bits (i.e. galaxies) in each realisations.
Clearly, such number can only be 1 or 0 a this stage.   
We then move to galaxy 2 and permutate its bits randomly, but giving priority to the ``holes" left by galaxy 1, i.e. those realisations with no galaxies yet.
We then iterate the process to all the remaining $N_{FOF}-2$ galaxies.
With this approach the sum of galaxies per realisation at each step will be either $a_n$ or $a_n-1$, where $a_n$ is an integer which varies with the iteration number $n$. 
What we call ``holes" at the $n$-th step are those realisations with $a_n-1$ total bits.

Since the number of pairs in each realisation grows with the square of the number of objects and since the number of objects is non negative, it follows that no further bit permutation of the so obtained final state can decrease the total number of pairs (and $t$-plets with $t \ge 2$), i.e. it is a maximum anticorrelation state.

\section{Collision kernel from the mocks as a mitigation strategy}\label{sec:collwind}
The method described here builds on the individual weighting schemes discussed in section~\ref{sec:estimator_level}, either $w_{comp}$ or $w_{IIP}$, for a first-stage correction.
When these weights are applied, the residual bias is of pure pairwise nature and confined to the $\theta < \theta_{ind}$ separations, inside the collision window.
As described throughout the paper, PIP weights statistically extract this window from the data, whereas $\theta$-cut removes it and adjusts the theory model accordingly.  

A third option consists of assuming a given shape for the window, e.g., a top hat function~\cite{hahn2017}, or the actual shape measured from a set of high-fidelity mock catalogues.
This shape can be used inside either a model-stage or an estimator-stage mitigation. 
Here we discuss the latter option, with shape measured from the mocks. 
 
We are not interested in applying this method to configuration-space statistics, as the net effect would be to leave the large-scale correlation of the data virtually unchanged, while making mock-dependent the small-scale one, with little control on the correspondent error.
Instead, we want to use it as way to compensate for the large-scale ($k \lesssim 0.2$) tails of the collision window in Fourier space.

With the expression ``measuring the window" we mean that we can measure from mocks the following collision kernel,
\begin{equation}\label{eq:windco_pairs}
\mathcal{K}(s_\parallel, s_\perp) = \frac{DD_{complete}(s_\parallel, s_\perp)}{DD_{IIP}(s_\parallel, s_\perp)} - 1 \ ,
\end{equation}
where $DD_{complete}$ represents the pair counts from the complete sample and $DD_{IIP}$ the pair counts computed with IIP weights from the fiber-assigned sample,
as function of the parallel and perpendicular separations, $s_\parallel$ and $s_\perp$.
An equivalent expression can be written for the fiducial weights.
We can then measure the data power spectrum using the estimator defined in eq.~(\ref{eq:P_PIP}), but with
\begin{equation}\label{eq:windco}
    A_{ij} = w^{IIP}_i w^{IIP}_j \mathcal{K}(s_\parallel^{ij}, s_\perp^{ij}) \ .
\end{equation}
 
In the idealised case in which the mocks are a perfect representation of the (unknown) clustering of the data, this approach returns the exact pairwise correction.
In a more realistic scenario, the unavoidable discrepancies between mocks and data are mitigated by the differential nature of the $\mathcal{K}$ kernel.  
In other words, $\mathcal{K}$ can be seen as an impulse-response device, designed to capture the effect of fiber assignment alone, regardless of the underling clustering.
More precisely, this is a first order description of such a device, as the true targeting mechanism cannot be fully captured by a multiplicative factor: the fraction of observed pairs must drop for increasing number density and eventually saturate.
Lastly, despite the fact that formally the kernel depends on the $s_\parallel$ variable, its variability along the parallel direction turned out to be negligible. 
We therefore decided to integrate $s_\parallel$ and keep only the $s_\perp$ dependence, as a way to compress the information an minimize the noise. 

\begin{figure}
    \centering
    \includegraphics[width=0.32\linewidth]{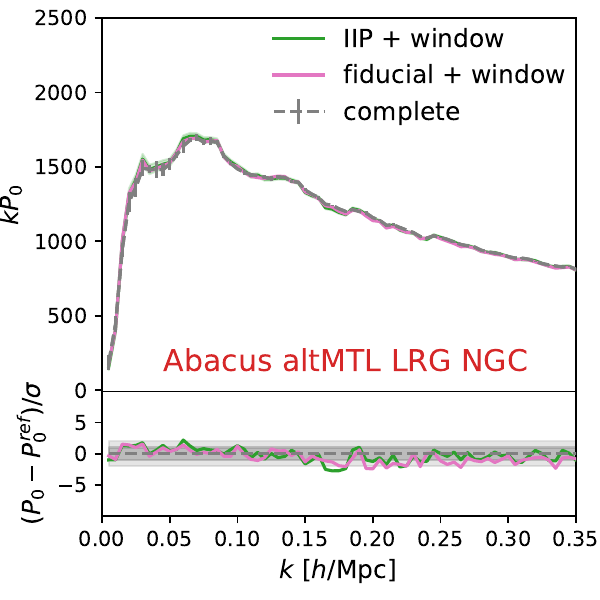}
    \includegraphics[width=0.32\linewidth]{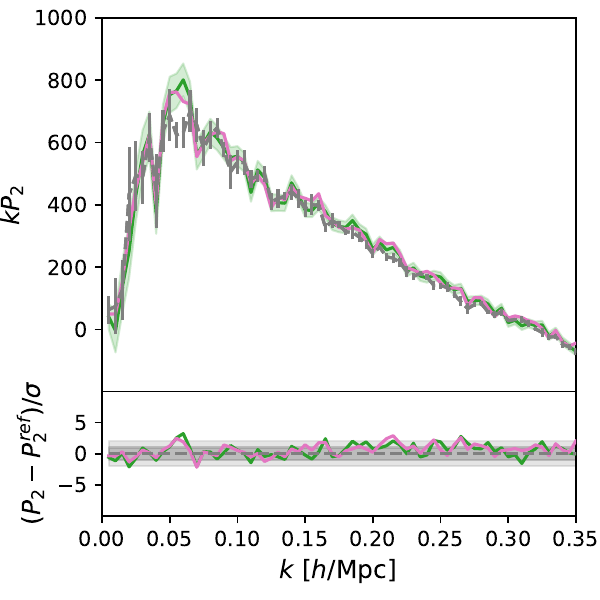}
    \includegraphics[width=0.32\linewidth]{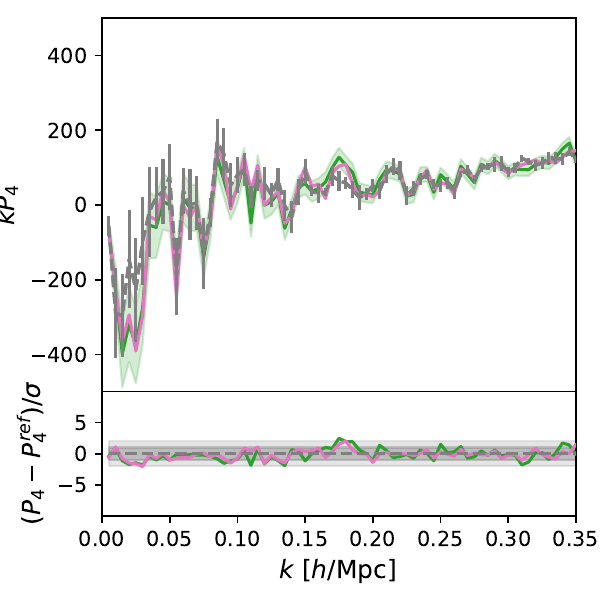}
    \caption{Multipoles of the power spectrum $P_\ell$ as a function of the wave number $k$, measured from the altMTL mocks (mock 11, LRGs NGC). For visualisation purpose, we follow the common practice of muliplying the amplitude of the multipoles by $k$. The gray dashed curves correspond to measurements obtained from the complete sample with no additional weights and error bars indicating the scatter between different mocks. They represent a reference target for the other two measurements, obtained from the fiber-assigned samples using the transverse collision window coupled to fiducial weights (solid pink) and IIP weights (solid green with shaded green area for the corresponding $1\sigma$ error), as described in the text. In the insets at the bottom of each panel we show the difference between fiber-assigned and complete samples in units of standard deviation.}
    \label{fig:P_ASaMTL_windco}
\end{figure}

Figure \ref{fig:P_ASaMTL_windco} shows the LRG power spectrum multiploes measured from the altMTL mocks using eq. (\ref{eq:windco}).
We considered both fiducial and IIP weights as the individual weights required by the equation, represented by the pink and green solid curves, respectively.
Since the complete and clustering samples do not share the same mask we have normalised the pair counts of eq. (\ref{eq:windco_pairs}) with those of the corresponding random samples.
For this test the $\mathcal{K}$ function is derived from the same mock used to measure the power spectrum.
By doing so we are unrealistically providing the estimator with nearly all the missing information, except for the integration along $s_\parallel$.
The results displayed in the figure should therefore be viewed as proof of concept rather than a definitive test.
In a more realistic scenario one should take the average of $\mathcal{K}$ measured over the 25 altMTL mocks and apply it to the data.
That said, the figure shows the effectiveness of this mitigation strategy in recovering the ``true'' clustering, with a minor preference for IIP weights.
As discussed in the text, we eventually opted for more agnostic approaches, but this is a valid way to restore the correct amplitude of $P$ on large and intermediate scale, as far as the mock catalogues provide a good representation of fiber assignment process and underlying clustering.

\section{Clustering measurements averaged over multiple FFA mocks}\label{app:FFA_average}

\begin{figure}
    \centering
    \includegraphics[width=0.32\linewidth]{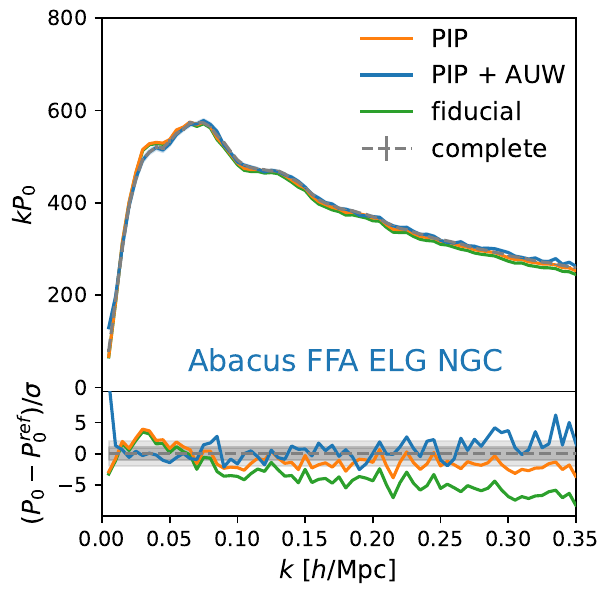}
    \includegraphics[width=0.32\linewidth]{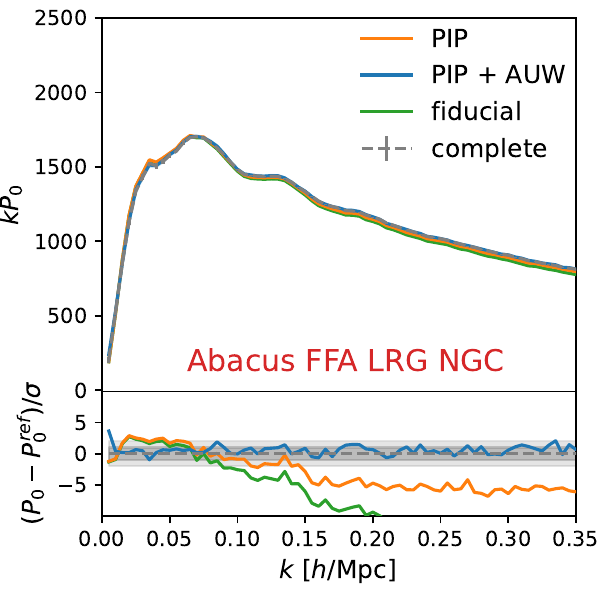}
    \includegraphics[width=0.32\linewidth]{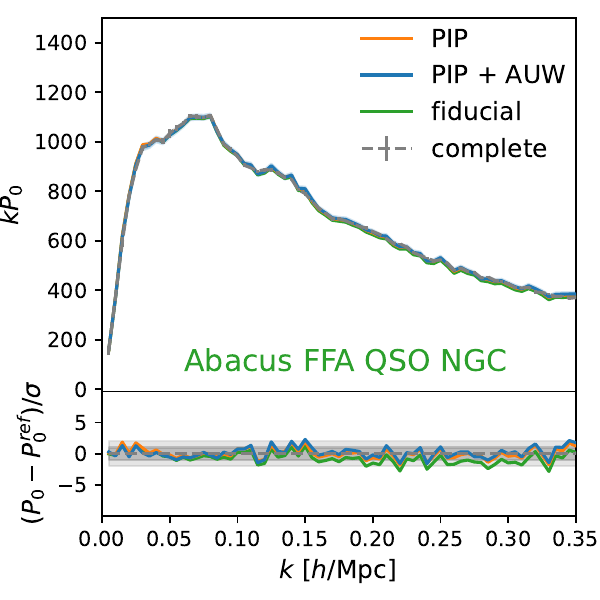}
    \caption{Monopole of the ELG, LRG, and QSO power spectrum averaged over 7 FFA mocks (mocks 11-17, NGC). The grey dashed curves indicate the measurements obtained from the complete sample with no additional weights and error bars indicating the scatter between different mocks. The remaining curves are obtained from the fiber-assigned samples using: fiducial (IIP) weights (solid green); PIP weights (solid orange); PIP weights plus angular upweighting (solid blue), with the blue shaded area indicating the corresponding $1\sigma$ error. The bottom insets show the difference between fiber-assigned and complete samples in units of the standard deviation of the mean.}
    \label{fig:averaged_P}
\end{figure}

Figure \ref{fig:averaged_P} shows the monopole of the power spectrum averaged over 7 FFA mocks, with the variance rescaled accordingly.
The trends discussed in section \ref{sec:results} are confirmed and, to some extent, enhanced.
For instance, the systematic suppression of the power spectrum measured from the ELG sample with fiducial weights emerges more clearly.
As previously noted, with FFA mocks we could not apply angular upweighting in a fully consistent way because the catalogs lack some of the necessary components.
This is at the origin of the low $k$ peak observed in the bottom insets of both the ELG and, to a lesser extent, the LRG samples.
It also explains why the estimates obtained with PIP plus angular upweighting are noisier than the others. 

\section{Additional information on the  catalogues and robustness of the mitigation schemes against pothometric errors and redshift failures.}\label{app:AUW}

\begin{figure}
    \centering
    \includegraphics[width=0.32\linewidth]{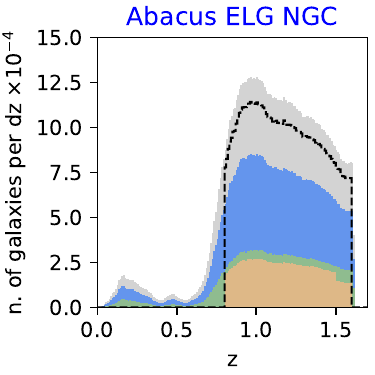}
    \includegraphics[width=0.32\linewidth]{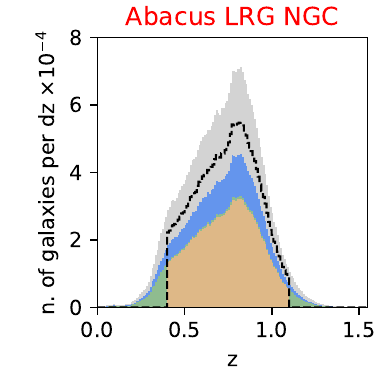}
    \includegraphics[width=0.32\linewidth]{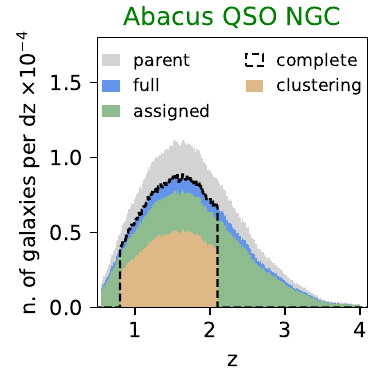}
    \caption{Redshift distributions measured from the altMTL mocks (NGC), with $dz=0.01$, for the five catalogues discussed in the text: parent (gray area), full (blue area), assigned (green area), complete (black dashed curve), clustering (gold area). Each panel displays one of the dark-time tracers, as labeled in the figure. The parent catalogue, which formally includes all the potential targets, is divided accordingly.}
    \label{fig:zdist_parent}
\end{figure}

Organizing the data into various catalogs, each tailored for a specific purpose, is essential for simulating fiber assignment, developing mitigation strategies, and testing them.
The four catalogue classes employed in this work are those introduced in section \ref{sec:data} and explained in detail in \cite{DESI2024.II.KP3, KP3s15-Ross}.
Here we provide additional clarifications to assist the reader in navigating the various definitions and to further support the robustness of the fiber mitigation strategies we adopted.
To aid the discussion, we explicitly define a fifth catalog, the {\it assigned} sample, which does not physically exists in the DESI database but is instead extracted ``on the fly" from the full sample whenever required.
Below is a summary of the properties of the catalogues.
Their redshift distributions, measured from the altMTL mocks, are shown in figure \ref{fig:zdist_parent}.
\begin{itemize}
    \item{\bf Parent.}
    The parent sample (gray area) is the collection of all potential targets across all tracer types, serving as the input for the fiber assignment algorithm for both mocks and real data.
    Its $z$-distribution, determined by the color cuts, can be estimated from the data through photometry and closely reproduced in the mocks.
    Its sky area is a collection of ``circles", each representing the area covered by the DESI focal plane in one of the DR1 pointings.
    To allow for a direct comparison with the other samples, which are all defined for individual tracer types (ELG, LRG, or QSO), in the figure the parent sample has also been divided by tracer types.

    \item{\bf Complete.}
    The complete sample (black dashed curve) is similar to the parent sample but is cut to the exact angular footprint solely including the targets that were potentially reachable by a (functioning) fiber, including those that actually have zero probability of being assigned due to, e.g., priority issues.
    Most importantly, the complete sample is an non-observable catalogue, defined only in the mocks, where we have exact knowledge of the ``true" redshift of all the tracers.
    It represents the ``true" clustering that we want to recover through our fiber mitigation strategies and, as such, it must be cut to match the volume of the clustering sample under examination.

    \item{\bf Full.}
    The full sample (blue area) is also cut to the exact survey footprint but also includes the veto mask.
    It neither depends on nor provides any explicit redshift information, as it is derived from the parent sample through processes that depend exclusively on angular information.
    It is available for both real data and mocks and serves as the ``reference" sample that enters the definitions of angular upweighting factor and $w_{NTMP}$ given in section \ref{sec:estimator_level}. 
    
    \item{\bf Assigned.}
    The assigned sample (green area) is the subset of the targets of the full sample that obtained a fiber, regardless of whether the redshift measurement was successful or not.
    It is used as the ``depleted" sample that enters the definitions of angular upweighting factor and $w_{NTMP}$ given in section \ref{sec:estimator_level}.

    \item{\bf Clustering.}
    The clustering sample (gold area) is the set of all the galaxies allowed by the veto mask that obtained a reliable spectroscopic redshift.
    Similarly to the complete sample, in the DESI database it is only defined within the fiducial redshift range of the different tracer types, described in section \ref{sec:data}. 
    Any arbitrary additional cut to construct one or more redshift bins is formally allowed on this catalogue and should be accompanied by corresponding cuts in the complete sample.
\end{itemize}
From the definitions we just provided, it follows that, within the redshift interval common to all the samples, i.e. where the complete and clustering samples are defined, we should observe a gradual loss of amplitude progressing from parent to complete, to full, to assigned, and finally to the clustering sample.
This behavior is clearly illustrated in
figure \ref{fig:zdist_parent}, with the primary causes being:
\begin{itemize}
    \item 
    Parent to complete: angular cut to the exact survey footprint.
    \item
    Complete to full: veto mask.
    \item
    Full to assigned: fiber assignment.
    \item
    Assigned to clustering: redshift failures.
\end{itemize}
As depicted in the figure, there are two important features that we incorporated into the mocks to accurately replicate what is observed in the real data.
First, the full and assigned samples, which are used to define the angular upweighting factor, extend well beyond the redshift range of the clustering sample.
Second, the clustering sample is impacted by redshift failures, with a non-uniform dependence on redshift in the ELG case.

We emphasize these two aspects because they strongly support the robustness of our mitigation strategies against photometric uncertainties and redshift failures.
If, for example, we add a random displacement along the line of sight to simulate photometric errors, the angular upweighting factor does not change, by construction, as we are already including all possible redshifts, and similarly for $w_{NTMP}$ and $w_{comp}$.
In this picture, the clustering and complete samples remain unchanged, as they are defined, within spectroscopic uncertainties, by the ``true" redshifts.
The essential point is that with our approach the full sample is not required to share the same angular clustering and/or redshift range as the clustering sample.
In fact, the very fact that they are defined over different redshift ranges implies that their angular correlations differ.
However, our tests (section \ref{sec:results}) demonstrate that the ``true" clustering can still be effectively recovered, as the critical factor is obtaining an accurate estimate of the fraction of zero-probability pairs for a given target type, as a function of (angular) scale.
Properly estimating this ratio only requires that the numerator and denominator are defined over the same volume, without the need for precise knowledge of the volume itself, at least for a survey with the characteristics of DESI.

Similar considerations apply to redshift failures: under our definition, the angular upweighting factor is unaffected by redshift failures and the same holds for $w_{NTMP}$ and $w_{comp}$.
The clustering sample, however, requires proper corrections, particularly if redshift failures are unevenly distributed.
Strictly speaking, this is not a fiber-assignment issue and has been addressed separately, with two dedicated works \cite{krolewski2025, yu2025}, summarized in \cite{DESI2024.II.KP3}.
We incorporated the corrections outlined in those papers into all the measurements presented in this work, including those based on mocks.

Finally, if the parent sample itself includes fluctuations of non-cosmological origin, usually denoted as imaging systematics, our fiber mitigation strategies treat them as if they were ``true", effectively restoring their pre-assignment amplitude.
As a consequence, these spurious fluctuations can impact our clustering estimates and require dedicated countermeasures.
Once again, this is technically not a fiber-assignment issue and has been investigated separately~\cite{Rezaie23, Chaussidon21QSOsys}. The fiducial mitigation scheme for DR1 imaging systematics, specific to each tracer type, is detailed in \cite{DESI2024.II.KP3}.
We used that scheme for all the measurements from real data (imaging systematics are not present in DR1 mocks) in this work.

\section*{Data availability}
Data from the plots in this paper will be available on Zenodo as part of DESI's Data Management Plan. The data used in this analysis will be made public along the Data Release (details in
\url{https://data.desi.lbl.gov/doc/releases/}).

\acknowledgments


This material is based upon work supported by the U.S. Department of Energy (DOE), Office of Science, Office of High-Energy Physics, under Contract No. DE–AC02–05CH11231, and by the National Energy Research Scientific Computing Center, a DOE Office of Science User Facility under the same contract. Additional support for DESI was provided by the U.S. National Science Foundation (NSF), Division of Astronomical Sciences under Contract No. AST-0950945 to the NSF’s National Optical-Infrared Astronomy Research Laboratory; the Science and Technology Facilities Council of the United Kingdom; the Gordon and Betty Moore Foundation; the Heising-Simons Foundation; the French Alternative Energies and Atomic Energy Commission (CEA); the National Council of Humanities, Science and Technology of Mexico (CONAHCYT); the Ministry of Science, Innovation and Universities of Spain (MICIU/AEI/10.13039/501100011033), and by the DESI Member Institutions: \url{https://www.desi.lbl.gov/collaborating-institutions}. Any opinions, findings, and conclusions or recommendations expressed in this material are those of the author(s) and do not necessarily reflect the views of the U. S. National Science Foundation, the U. S. Department of Energy, or any of the listed funding agencies.

The authors are honored to be permitted to conduct scientific research on Iolkam Du’ag (Kitt Peak), a mountain with particular significance to the Tohono O’odham Nation.



\bibliographystyle{JHEP}
\bibliography{DESI2024_updated21Nov, KP3_references, biblio}







\end{document}